\documentclass[twocolumn, times, trackchanges]{aastex63_bren}

\usepackage[ruled,vlined]{algorithm2e}
\usepackage{graphicx}
\usepackage{bbm}
\usepackage{float}
\usepackage{xspace}
\usepackage[hypertexnames=false]{hyperref}
\usepackage{subfiles} % Best loaded last in the preamble
\hypersetup{linkcolor=red,citecolor=green,filecolor=cyan,urlcolor=magenta}
\defcitealias{baron19}{B19}
\defcitealias{baron24}{B24}

\newcommand{\halpha}{H$\alpha$\xspace}
\newcommand{\hbeta}{H$\beta$\xspace}
\newcommand{\oiiifull}{$\text{[O\,{\sc iii}]}\lambda \, 5007\mathrm{\AA}$\xspace}
\newcommand{\oiii}{$\text{[O\,{\sc iii}]}$\xspace}
\newcommand{\oifull}{$\text{[O\,{\sc i}]}\lambda \, 6300\mathrm{\AA}$\xspace}
\newcommand{\oi}{$\text{[O\,{\sc i}]}$\xspace}
\newcommand{\niifull}{$\text{[N\,{\sc ii}]}\lambda \, 6584\mathrm{\AA}$\xspace}
\newcommand{\nii}{$\text{[N\,{\sc ii}]}$\xspace}
\newcommand{\siifull}{$\text{[S\,{\sc ii}]}\lambda\lambda \, 6717\mathrm{\AA}+6731\mathrm{\AA}$\xspace}
\newcommand{\sii}{$\text{[S\,{\sc ii}]}$\xspace}

\newcommand{\oiiihbeta}{$\log (\text{[O\,{\sc iii}]}/\text{H}\beta)$\xspace}
\newcommand{\niihalpha}{$\log (\text{[N\,{\sc ii}]}/\text{H}\alpha)$\xspace}
\newcommand{\siihalpha}{$\log (\text{[S\,{\sc ii}]}/\text{H}\alpha)$\xspace}
\newcommand{\oihalpha}{$\log (\text{[O\,{\sc i}]}/\text{H}\alpha)$\xspace}

\newcommand{\mic}{$\mathrm{\mu m}$\xspace}

\shorttitle{PAH band - optical line ratios relation}
\begin{document}

\title{PHANGS-ML: the universal relation between PAH band and optical line ratios across nearby star-forming galaxies}

\author[0000-0003-4974-3481]{Dalya Baron}
\email{dalyabaron@gmail.com}
\altaffiliation{Carnegie-Princeton Fellow}
\affiliation{The Observatories of the Carnegie Institution for Science. 813 Santa Barbara Street, Pasadena, CA 91101, USA}
\affiliation{Kavli Institute for Particle Astrophysics \& Cosmology (KIPAC), Stanford University, CA 94305, USA}

\author[0000-0002-4378-8534]{Karin M. Sandstrom}
\affiliation{Department of Astronomy \& Astrophysics, University of California, San Diego. 9500 Gilman Drive, La Jolla, CA 92093, USA}

\author[0000-0002-9183-8102]{Jessica Sutter}
\affiliation{Whitman College, 345 Boyer Avenue, Walla Walla, WA 99362, USA}

\author[0000-0002-8806-6308]{Hamid Hassani}
\affiliation{Dept. of Physics, University of Alberta, 4-183 CCIS, Edmonton, Alberta, T6G 2E1, Canada}

\author[0000-0002-9768-0246]{Brent~Groves}
\affiliation{International Centre for Radio Astronomy Research, University of Western Australia, 35 Stirling Highway, Crawley, WA 6009, Australia}

\author[0000-0002-2545-1700]{Adam~K.~Leroy}
\affiliation{Department of Astronomy, The Ohio State University, Columbus, Ohio 43210, USA}

\author[0000-0002-3933-7677]{Eva Schinnerer}
\affiliation{Max Planck Institute for Astronomy, K\"{o}nigstuhl 17, D-69117, Germany}

\author[0000-0003-0946-6176]{M\'ed\'eric Boquien} 
\affiliation{Universit\'e C\^ote d'Azur, Observatoire de la C\^ote d'Azur, CNRS, Laboratoire Lagrange, 06000, Nice, France}

\author{Matilde Brazzini}
\affiliation{Dipartimento di Fisica, Università di Trieste, Sezione di Astronomia, Via G.B. Tiepolo 11, I-34143 Trieste, Italy}

\author[0000-0002-5235-5589]{J\'er\'emy Chastenet}
\affiliation{Sterrenkundig Observatorium, Universiteit Gent, Krijgslaan 281-S9, 9000 Gent, Belgium}

\author[0000-0002-5782-9093]{Daniel~A.~Dale}
\affiliation{Department of Physics and Astronomy, University of Wyoming, Laramie, WY 82071, USA}

\author[0000-0002-4755-118X]{Oleg~V.~Egorov}
\affiliation{Astronomisches Rechen-Institut, Zentrum f\"{u}r Astronomie der Universit\"{a}t Heidelberg, M\"{o}nchhofstra\ss e 12-14, D-69120 Heidelberg, Germany}

\author[0000-0001-6708-1317]{Simon C.~O.\ Glover}
\affil{Universit\"{a}t Heidelberg, Zentrum f\"{u}r Astronomie, Institut f\"{u}r Theoretische Astrophysik, Albert-Ueberle-Str 2, D-69120 Heidelberg, Germany}

\author[0000-0002-0560-3172]{Ralf S.\ Klessen}
\affiliation{Universit\"{a}t Heidelberg, Zentrum f\"{u}r Astronomie, Institut f\"{u}r Theoretische Astrophysik, Albert-Ueberle-Str.\ 2, 69120 Heidelberg, Germany}
\affiliation{Universit\"{a}t Heidelberg, Interdisziplin\"{a}res Zentrum f\"{u}r Wissenschaftliches Rechnen, Im Neuenheimer Feld 225, 69120 Heidelberg, Germany}
\affiliation{Harvard-Smithsonian Center for Astrophysics, 60 Garden Street, Cambridge, MA 02138, U.S.A. \label{CfA}}
\affiliation{Elizabeth S. and Richard M. Cashin Fellow at the Radcliffe Institute for Advanced Studies at Harvard University, 10 Garden Street, Cambridge, MA 02138, U.S.A. \label{Radcliffe}}

\author[0000-0003-2721-487X]{Debosmita Pathak}
\affiliation{Department of Astronomy, The Ohio State University, Columbus, Ohio 43210, USA}
\affiliation{Center for Cosmology \& Astro-Particle Physics, The Ohio State University, Columbus, Ohio 43210, USA}

\author[0000-0002-5204-2259]{Erik Rosolowsky}
\affiliation{Dept. of Physics, University of Alberta, 4-183 CCIS, Edmonton, Alberta, T6G 2E1, Canada}

\author[0000-0003-0166-9745]{Frank Bigiel}
\affiliation{Argelander-Institut f\"ur Astronomie, Universit\"at Bonn, Auf dem H\"ugel 71, 53121 Bonn, Germany}

\author[0000-0002-5635-5180]{M\'{e}lanie~Chevance}
\affiliation{Astronomisches Rechen-Institut, Zentrum f\"{u}r Astronomie der Universit\"{a}t Heidelberg, M\"{o}nchhofstra\ss e 12-14, 69120 Heidelberg, Germany}
\affiliation{Cosmic Origins Of Life (COOL) Research DAO, coolresearch.io}

\author[0000-0002-3247-5321]{Kathryn~Grasha}
%\altaffiliation{ARC DECRA Fellow}
\affiliation{Research School of Astronomy and Astrophysics, Australian National University, Canberra, ACT 2611, Australia} 

\author[0000-0002-9181-1161]{Annie~Hughes}
\affiliation{IRAP/OMP/UPS, 9 Av. du Colonel Roche, BP 44346, F-31028 Toulouse cedex 4, France}

\author[0000-0002-6972-6411]{J. Eduardo M\'endez-Delgado}
\affiliation{Astronomisches Rechen-Institut, Zentrum f\"{u}r Astronomie der Universit\"{a}t Heidelberg, M\"{o}nchhofstra\ss e 12-14, D-69120 Heidelberg, Germany}

\author[0000-0003-3061-6546]{J\'er\^ome Pety}
\affil{IRAM, 300 rue de la Piscine, 38400 Saint Martin d'H\`eres, France}
\affil{LERMA, Observatoire de Paris, PSL Research University, CNRS, Sorbonne Universit\'es, 75014 Paris}

\author[0000-0002-0012-2142]{Thomas~G.~Williams}
\affiliation{Sub-department of Astrophysics, Department of Physics, University of Oxford, Keble Road, Oxford OX1 3RH, UK}

\author[0000-0001-9628-8958]{Stephen Hannon}
\affiliation{Max-Planck-Institut f\"ur Astronomie, K\"onigstuhl 17, D-69117, Heidelberg, Germany}

\author[0000-0002-4781-7291]{Sumit K. Sarbadhicary}
\affiliation{Department of Physics, The Ohio State University, Columbus, Ohio 43210, USA}
\affiliation{Center for Cosmology \& Astro-Particle Physics, The Ohio State University, Columbus, Ohio 43210, USA}
\affiliation{Department of Astronomy, The Ohio State University, Columbus, Ohio 43210, USA}

\suppressAffiliations



\begin{abstract}

The structure and chemistry of the dusty interstellar medium (ISM) are shaped by complex processes that depend on the local radiation field, gas composition, and dust grain properties. Of particular importance are Polycyclic Aromatic Hydrocarbons (PAHs), which emit strong vibrational bands in the mid-infrared, and play a key role in the ISM energy balance. We recently identified global correlations between PAH band and optical line ratios across three nearby galaxies, suggesting a connection between PAH heating and gas ionization throughout the ISM. In this work, we perform a census of the PAH heating -- gas ionization connection using $\sim$700,000 independent pixels that probe scales of 40--150 pc in nineteen nearby star-forming galaxies from the PHANGS survey. We find a universal relation between $\log$PAH(11.3 \mic/7.7 \mic) and $\log$([SII]/H$\alpha$) with a slope of $\sim$0.2 and a scatter of $\sim$0.025 dex. The only exception is a group of anomalous pixels that show unusually high (11.3 \mic/7.7 \mic) PAH ratios in regions with old stellar populations and high starlight-to-dust emission ratios. Their mid-infrared spectra resemble those of elliptical galaxies. AGN hosts show modestly steeper slopes, with a $\sim$10\% increase in PAH(11.3 \mic/7.7 \mic) in the diffuse gas on kpc scales. This universal relation implies an emerging simplicity in the complex ISM, with a sequence that is driven by a single varying property: the spectral shape of the interstellar radiation field. This suggests that other properties, such as gas-phase abundances, gas ionization parameter, and grain charge distribution, are relatively uniform in all but specific cases.

\end{abstract}


\section{Introduction}\label{sec:intro}

The mid-infrared emission spectra of most Galactic and extragalactic sources are dominated by several prominent emission features at 3.3, 6.2, 7.7, 8.6, 11.3, and 12.7 \mic (see reviews by \citealt{tielens08, li20}). These prominent features are broad (full width at half maximum; FWHM of$\sim$0.5--1 \mic) and show complex spectral shapes, and their total emission amounts to $\lesssim$20\% of the total infrared radiation of the Milky Way and other star-forming galaxies (e.g., \citealt{smith07}). They are believed to arise from vibrational modes of polycyclic aromatic hydrocarbon (PAH) molecules excited via stochastic heating by starlight (\citealt{sellgren83, leger84, allamandola85}). 

PAHs are a type of hydrocarbon molecule consisting of tens to hundreds of carbon atoms, arranged in aromatic structures (see reviews by \citealt{tielens08, li20}). They are believed to be ubiquitous in the interstellar medium (ISM) of star-forming galaxies, constituting $\sim$5\% of the dust mass (e.g., \citealt{draine07, tielens08, sutter24}). Their significant emission in the mid-infrared suggests that they are an important absorber of starlight \citep{joblin92, cecchi_pestellini08, mulas13}, and their presence has been linked to the prominent extinction bump at 2175 $\mathrm{\AA}$ \citep{joblin92, li01, steglich10, gordon24}. As a result, they have been used as tracers of the integrated star formation rate in local and high-redshift galaxies (e.g., \citealt{genzel98, lutz07, pope08, belfiore23, leroy23, gregg24}).

PAHs are believed to have a significant impact on the phase structure of the ISM and on the ion-molecule chemistry responsible for simple gas-phase species. By providing a large number of photoelectrons and by being an effective recombination channel for singly-ionized carbon atoms, they dominate the heating of the neutral gas in the diffuse ISM and the ionization balance in molecular clouds (e.g., \citealt{bakes94, weingartner01, lepage03}). The photoelectric heating efficiency and ionization balance depend on the charge and size distribution of PAHs, and thus mapping these properties across different environments may help to better constrain the physical processes taking place in the ISM (see reviews by \citealt{tielens08, draine11, klessen16}.

Observations of PAH band ratios can constrain the PAH charge and size distributions. Laboratory measurements and theoretical calculations show that ionized and neutral PAHs have different band strengths, with ionized PAHs having stronger 6.2 and 7.7 \mic band strengths compared to their 3.3 and 11.3 \mic bands, and neutral PAHs having stronger 3.3 and 11.3 \mic band strengths (e.g., \citealt{defrees93, allamandola99}). As a result, numerous studies have been using the 11.3 \mic/7.7 \mic PAH band ratio as a tracer of the ionized PAH fraction (e.g., \citealt{hony01, kaneda05, flagey06, smith07, galliano08, diamond_stanic10, vega10, peeters17, lai22, chastenet23b, dale23}). To constrain the PAH size distribution, studies have been using short-to-long wavelength PAH band ratios such as e.g., 3.3 \mic/7.7 \mic, 3.3 \mic/11.3 \mic, 6.2 \mic/7.7 \mic, 11.3 \mic/17 \mic, and more (e.g., \citealt{smith07, diamond_stanic10, sales10, chastenet23b, dale23, lai23, ujjwal24, whitcomb24}). For a given radiation field, since smaller PAHs have smaller heat capacities, single photon absorption raises their peak temperature to higher values, resulting in an overall stronger emission in shorter wavelength bands such as 3.3 \mic and 6.2 \mic compared to 7.7 \mic and 11.3 \mic (e.g., \citealt{draine01, draine11, maragkoudakis20}).

The PAH temperature distribution, and thus their mid-infrared emission spectrum, also depends on the hardness of the radiation field heating them (e.g., \citealt{omont86, draine21, rigopoulou21}). A harder radiation field leads to higher PAH temperatures, leading to increased emission at shorter wavelengths compared to longer ones. \citet{draine21} modeled the impact of modifying the spectral shape of the radiation field on the PAH mid-infrared emission spectrum using various stellar population models and observed interstellar radiation fields. Adjusting the radiation field from being dominated by young stars (single stellar population with an age of 3 Myr) to being dominated by 1--10 Gyr old stars (observed spectrum of M31's bulge), the mean energy of photons absorbed by the PAHs decreases from 6.73 eV to 1.22 eV. For the assumed PAH charge and size distributions, this variation results in an increase of $\sim$20\% in the 11.3 \mic/7.7 \mic band ratio, and a decrease of a factor of $\sim$3 in the 3.3 \mic/11.3 \mic ratio.

Therefore, the interpretation of PAH band ratios in the context of PAH charge and size variations must take into account the impact of the varying radiation field, especially in extragalactic observations where the radiation field is a complicated mixture of emission originating from young and old stars, and possibly from an accreting supermassive black hole in the galaxy's center. The impact of the varying radiation field poses a particular challenge when using short-to-long PAH band variations to constrain the PAH size distribution, since PAH band ratios are sensitive to both the PAH size distribution and the shape of the radiation field. Since both effects change the grain temperature distribution, they are degenerate with each other in PAH band ratio versus ratio diagrams (see discussion in \citealt{galliano08, dale23, chastenet23b, baron24, donnelly24}).

A possible way to break this degeneracy is using other observables that are sensitive to the spectral shape of the radiation field, such as atomic lines. Several studies, in particular using mid-infrared Spitzer IRS spectra, and more recently using the JWST MIRI-MRS instrument, find significant relations between PAH band ratios and the $\text{[\ion{Ne}{3}]}\lambda \, 15.6\mathrm{\mu m} / \text{[\ion{Ne}{2}]}\lambda \, 12.8\mathrm{\mu m}$ ratio (e.g., \citealt{smith07, sales10, lai22, zhang22, rigopoulou24}), though these have been typically interpreted as changes in PAH ionization as a function of the hardness of the ionizing radiation field. 

In \citealt[hereafter B24]{baron24}, we identified tight galaxy-wide correlations between the PAH band ratios 11.3 \mic/7.7 \mic and 3.3 \mic/11.3 \mic and the optical line ratios \oiiihbeta, \niihalpha, \siihalpha, \oihalpha, using high resolution multi-wavelength images of the galaxies NGC~628, NGC~1365, and NGC~7496, on 150 pc scales, obtained as part of the PHANGS-MUSE \citep{emsellem22} and PHANGS-JWST \citep{lee23} surveys. We showed that the correlations can be naturally explained in a scenario where the PAHs and the ionized gas are exposed to different parts of the same radiation field that varies spatially across the galaxies. In this scenario, most of the observed variation of the PAH 11.3 \mic/7.7 \mic band ratio and a large fraction of the observed PAH 3.3 \mic/11.3 \mic variation across nearby galaxies are in fact due to the varying radiation field, rather than driven by varying PAH size and charge distributions. Once accounting for the effect of the varying radiation field, we found a secondary variation of the PAH 3.3 \mic/11.3 \mic band ratio which we attributed to a modest variation in the PAH size distribution in different galaxies.

 \floattable
\begin{deluxetable}{ccCCCl Crr Crr}
\tablecaption{Galaxy sample\label{tab:galaxy_properties}}
\tablecolumns{10}
\tablenum{1}
\tablewidth{0pt}
\tablehead{
\colhead{(1)} & \colhead{(2)} & \colhead{(3)} & \colhead{(4)} & \colhead{(5)} & \colhead{(6)} & \colhead{(7)} & \colhead{(8)} & \colhead{(9)} & \colhead{(10)} & \colhead{(11)} \\
\colhead{Galaxy} & \colhead{D} & \colhead{i} & \colhead{SFR} & \colhead{$\log M_{*}$} & \colhead{AGN?} & \colhead{$C_{\mathrm{opt}}$} & \colhead{scale ($C_{\mathrm{opt}}$)} & \colhead{N$_{\mathrm{pix}}(C_{\mathrm{opt}})$} & \colhead{FWHM (150 pc)} & \colhead{N$_{\mathrm{pix}}$(150 pc)}  \\
\colhead{} & \colhead{[Mpc]} & \colhead{[deg]} & \colhead{[$\mathrm{M_{\odot}\,yr^{-1}}$]} & \colhead{[$\log \mathrm{M_{\odot}}$]} & \colhead{} & \colhead{[arcsec]} & \colhead{[pc]} & \colhead{} & \colhead{[arcsec]} & \colhead{}
}
\startdata
IC~5332 & 9 & 26.9 & 0.4 & 9.7 &  & 0.87 & 38 & 25\,606 (0.36) & 3.43 & 1\,446 (0.70)\\
NGC~0628 & 9.8 & 8.9 & 1.7 & 10.3 &  & 0.92 & 44 & 55\,716 (0.70) & 3.15 & 4\,088 (0.76)\\
NGC~1087 & 15.9 & 42.9 & 1.3 & 9.9 &  & 0.92 & 71 & 21\,429 (0.97) & 1.94 & 4\,279 (0.99)\\
NGC~1300 & 19 & 31.8 & 1.2 & 10.6 &  & 0.89 & 82 & 53\,724 (0.52) & 1.62 & 13\,528 (0.76)\\
NGC~1365 & 19.6 & 55.4 & 17 & 11 & yes & 1.15 & 109 & 34\,591 (0.77) & 1.57 & 17\,739 (0.81)\\
NGC~1385 & 17.2 & 44 & 2.1 & 10 &  & 0.77 & 64 & 45\,062 (0.62) & 1.79 & 6\,408 (0.68)\\
NGC~1433 & 18.6 & 28.9 & 1.1 & 10.9 &  & 0.91 & 82 & 51\,748 (0.48) & 1.66 & 13\,061 (0.76)\\
NGC~1512 & 18.8 & 42.5 & 1.3 & 10.7 &  & 1.25 & 114 & 17\,844 (0.58) & 1.64 & 10\,022 (0.77)\\
NGC~1566 & 17.7 & 29.5 & 4.6 & 10.8 & yes & 0.8 & 69 & 43\,724 (0.82) & 1.74 & 11\,030 (0.87)\\
NGC~1672 & 19.4 & 43.6 & 7.6 & 10.7 & yes & 0.96 & 90 & 46\,145 (0.70) & 1.59 & 15\,118 (0.72)\\
NGC~2835 & 12.2 & 41.3 & 1.3 & 10 &  & 1.15 & 68 & 19\,166 (0.76) & 2.53 & 3\,405 (0.87)\\
NGC~3351 & 10 & 45.1 & 1.3 & 10.4 &  & 1.05 & 51 & 25\,607 (0.83) & 3.09 & 2\,845 (0.99)\\
NGC~3627 & 11.3 & 57.3 & 3.9 & 10.8 &  & 1.05 & 58 & 25\,759 (0.91) & 2.73 & 3\,834 (0.93)\\
NGC~4254 & 13.1 & 34.4 & 3.1 & 10.4 &  & 0.89 & 57 & 49\,844 (0.86) & 2.36 & 6\,686 (0.87)\\
NGC~4303 & 17 & 23.5 & 5.4 & 10.5 & yes & 0.78 & 64 & 41\,265 (0.94) & 1.81 & 4\,662 (0.98)\\
NGC~4321 & 15.2 & 38.5 & 3.5 & 10.8 &  & 1.16 & 85 & 17\,598 (0.87) & 2.03 & 4\,459 (0.88)\\
NGC~4535 & 15.8 & 44.7 & 2.2 & 10.5 &  & 0.56 & 43 & 102\,341 (0.80) & 1.95 & 5\,070 (0.96)\\
NGC~5068 & 5.2 & 35.7 & 0.3 & 9.4 &  & 1.04 & 26 & 32\,884 (0.60) & 5.94 & 1\,026 (0.74)\\
NGC~7496 & 18.7 & 35.9 & 2.2 & 10 & yes & 0.89 & 81 & 19\,580 (0.63) & 1.65 & 4\,926 (0.71)\\
\enddata

\tablecomments{(1)-(5) Galaxy properties from \citet{lee23}: name, distance \citep{anand21a,anand21b,kourkchi17,shaya17}, inclination \citep{lang20,leroy21a}, star formation rate \citep{leroy21a}, and stellar mass \citep{leroy21a}. (6) Indicator of AGN presence from the \citet{veron_cetty10} catalog (only Seyfert nuclei are considered AGN). (7) Angular resolution of the MUSE data products. (8) Physical scale probed by $C_{\mathrm{opt}}$. (9) Number of independent pixels in the standardized multi-wavelength images at the $C_{\mathrm{opt}}$ resolution, and fraction of pixels with both $\log$PAH(11.3/7.7) band and optical \siihalpha line ratios. (10) Adopted angular resolution for the 150 pc scale products. (11) Number of independent pixels in the standardized multi-wavelength images at a 150 pc resolution, and fraction of pixels with both $\log$PAH(11.3/7.7) band and optical \siihalpha line ratios.
}
\end{deluxetable}
%\vspace{5mm}

In this work, we perform a systematic exploration of the relation between the PAH 11.3 \mic/7.7 \mic band ratio and the optical line ratios across nearby star-forming spiral galaxies on scales of 40--150 pc\footnote{Here we do not study the PAH 3.3 \mic/11.3 \mic band ratio as the 3.3 \mic PAH feature requires an accurate subtraction of the stellar continuum, which is still in progress (see Section \ref{sec:JWST}).}. We use the nineteen PHANGS galaxies with available high resolution infrared images from JWST (\citealt{lee23}) and spatially resolved optical spectroscopy from VLT-MUSE (\citealt{emsellem22}). These galaxies have also been mapped with ALMA (\citealt{leroy21a}) and HST (\citealt{lee22}). The combination of high resolution imaging probing scales of $\sim$100 pc with the wealth of multi-wavelength information makes this sample ideal for establishing a census of the relation between PAH heating and gas ionization throughout the ISM of local star-forming galaxies. Our goals are: (i) to test whether the PAH band--optical line ratios relations identified by \citetalias{baron24} are universal across nearby galaxies, (ii) to check whether the relations are scale-dependent over the range 40--150 pc, (iii) to identify special regions where the correlations break down and use the multi-wavelength information to interpret these anomalous regions, and (iv) to assess the impact of active galactic nuclei (AGN) on these correlations.

The paper is organized as follows. We describe the data we use in Section \ref{sec:data}, and present our results in section \ref{sec:results}. In section \ref{sec:discussion} we describe the emerging picture of PAH heating--gas ionization connection across nearby galaxies, and we conclude in section \ref{sec:summary}.

\section{Data}\label{sec:data}

To study the relation between PAHs and the ionized gas, we use multiwavelength observations from VLT-MUSE, JWST NIRCam and MIRI, HST, and ALMA, of the 19 PHANGS-MUSE galaxies (\citealt{leroy21a, emsellem22, lee22, lee23}). In Table \ref{tab:galaxy_properties} we summarize their main properties. They are star-forming galaxies with stellar masses $\log (M_{*}/M_{\odot}) = 9.5-11$ and metallicities $12 + \log(\mathrm{O/H}) = 8.4$--8.7, located close to the star-forming main sequence in the star formation rate versus stellar mass diagram. They are all nearby ($D < 20$ Mpc) and have modest inclinations ($i < 60^{\circ}$).

Between the VLT-MUSE, JWST, and HST observations, which are the primary products used in this paper, the limiting angular resolution of VLT-MUSE ($C_{\mathrm{opt}} \sim 0\farcs6-1\farcs2$), which differs from galaxy to galaxy, translates to a spatial resolution of $\sim$40--120 pc. The analysis in this work is carried out on standardized multi-wavelength images at both the $C_{\mathrm{opt}}$ resolution that differs from galaxy to galaxy, and at a uniform 150 pc resolution for all the galaxies. In Table \ref{tab:galaxy_properties} we list the effective angular full width at half maximum (FWHM) that corresponds to a spatial resolution of 150 pc for each of the galaxies.

\begin{figure*}
	\centering
\includegraphics[width=1\textwidth]{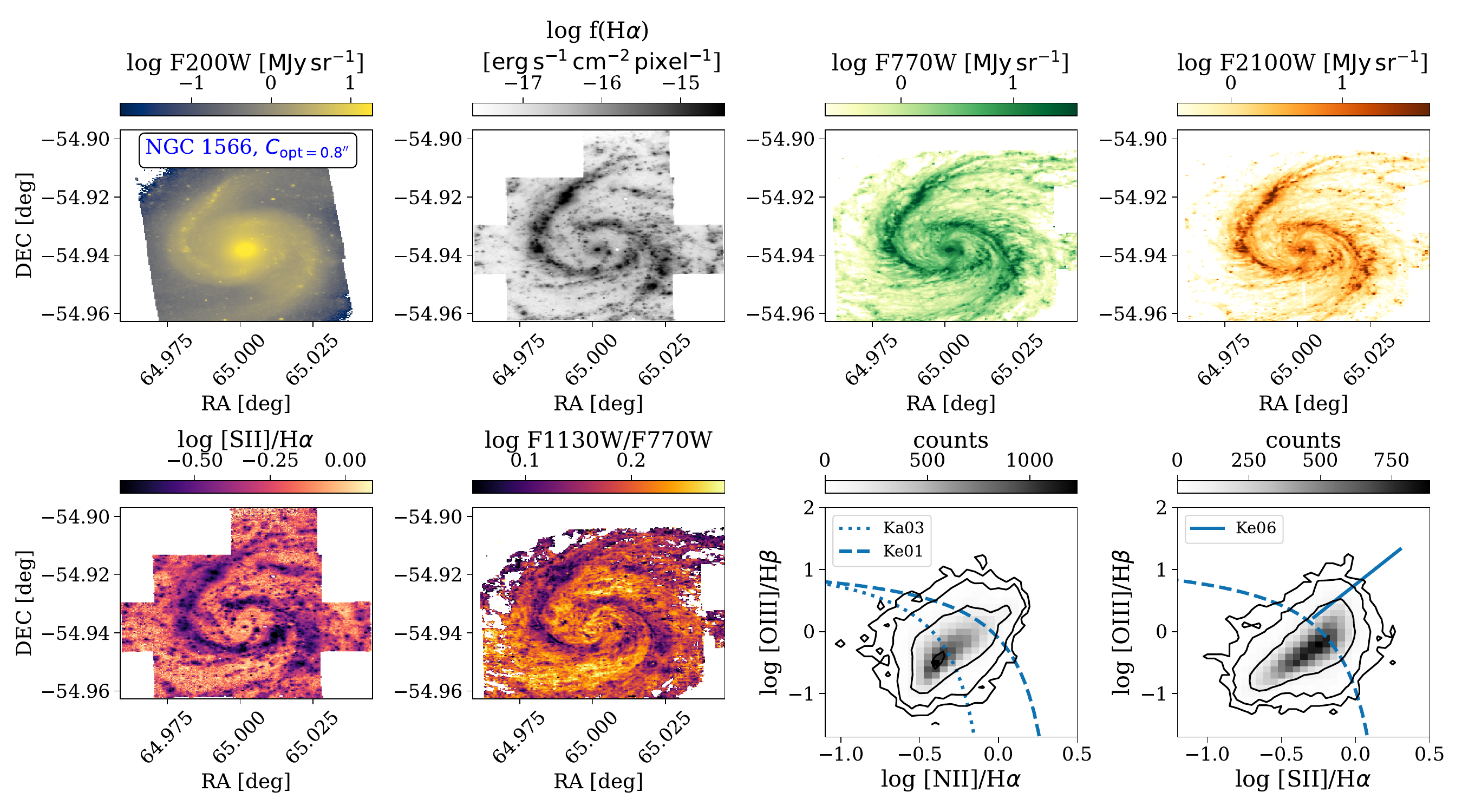}
\caption{\textbf{An example of the main observables used in this work for NGC~1566.} The first row, from left to right: surface brightness maps in the F200W JWST-NIRCam filter tracing starlight, \halpha emission tracing warm ionized gas, F770W JWST-MIRI filter tracing 7.7 \mic PAH emission, and F2100W JWST-MIRI filter tracing hot dust grains. The second row, from left to right, shows the \siihalpha ratio, the JWST-MIRI $\log$F1130W/F770W band ratio, and the 2D distribution of all the pixels ($ \mathrm{N_{pix}}=36018$) in the two line diagnostic diagrams \oiiihbeta vs. \niihalpha and \oiiihbeta vs. \siihalpha, which are used to identify the main source of ionizing radiation. The separating criteria by \citet[Ke01; dashed]{kewley01} and \citet[Ka03; dotted]{kauff03a} are used to separate ionization by young massive stars from AGN, and the criterion by \citet[Ke06; solid]{kewley06} is used to separate LINERS from Seyferts. All the maps are presented at the $C_{\mathrm{opt}}$ resolution and pixel size of 0\farcs8.}
\label{f:NGC1566_feature_display}
\end{figure*}

In Figure \ref{f:NGC1566_feature_display}, we show the main images used in our analysis for NGC~1566 as an example to showcase the data quality and completeness. The convolved and resampled images to the $C_{\mathrm{opt}}$ resolution (and to a greater extent, the 150 pc resolution) have high signal-to-noise ratios (SNRs) and are highly complete\footnote{We consider a measurement of a line or band ratio significant when both the numerator and denominator are detected at SNR $> 3$.}, even in the diffuse parts of the galaxies (see completeness fractions in Table \ref{tab:galaxy_properties}). We describe the various PHANGS data products used in this work in section \ref{sec:phangs_surveys}, and the steps we take to standardize the images from the different instruments to have common resolution and grid in section \ref{sec:conv_and_resamp}.

\subsection{PHANGS data products}\label{sec:phangs_surveys}

\subsubsection{MUSE}\label{sec:MUSE}

The PHANGS-MUSE survey mapped the 19 galaxies with the integral field spectrograph MUSE in the wavelength range 4750--9350 $\mathrm{\AA}$ with a spectral resolution of $\sim$2.5 \AA\xspace \citep{emsellem22}. The data reduction and analysis pipeline includes homogenization of individual MUSE pointings to a single point spread functions (PSFs), with FWHM labeled as $C_{\mathrm{opt}}$; fitting the stellar continuum and extraction of stellar population properties; and fitting the optical emission lines to estimate their kinematics and fluxes. These have been done at both the $C_{\mathrm{opt}}$ resolution and at a uniform 150 pc spatial resolution.  

We use the survey-derived (DR2.2) surface brightness maps of the following emission lines: the Balmer lines \halpha and \hbeta, and the collisionally-excited lines \oiiifull, \oifull, \niifull, and \siifull (\oiii, \oi, \nii, and \sii hereafter). We estimate the dust reddening of the line-emitting gas using the observed \halpha and \hbeta surface brightness maps via $\mathrm{E}(B-V) = \mathrm{2.33 \times log\, [(H\alpha/H\beta)_{obs}/2.86] \, \mathrm{mag} }$. This assumes case-B recombination with the \citet{cardelli89} extinction curve. We use maps of the light-weighted age of the stellar population derived from stellar population synthesis modeling of the continuum \citep{pessa23}. 

Throughout the paper, optical line ratios are presented as $\lambda f_{\lambda}$ ratios (line-integrated fluxes).

\subsubsection{JWST}\label{sec:JWST}

The PHANGS-JWST Cycle 1 Treasury program imaged the 19 galaxies using eight filters covering the wavelength range 2 to 21 \mic and probing scales of $\sim$5--50 pc \citep{lee23}. The filters include the four NIRCam bands F200W, F300M, F335M, and F360M, tracing emission at 2, 3, 3.35, and 3.6 \mic, and the four MIRI bands F770W, F1000W, F1130W, and F2100W, centered on the wavelengths 7.7, 10, 11.3, and 21 \mic. We use the images from the first PHANGS-JWST public data release described in \citet{williams24}. 

We perform $3 \sigma$ SNR cuts to the images in all four MIRI bands of the galaxies. \citet{sutter24} identified large-enough empty sky regions in the galaxies NGC~1087, NGC~1385, NGC~1433, NGC~1512, NGC~1566, and NGC~7496. They determined the $1 \sigma$ noise level from the standard deviation of these empty sky regions. At their common Gaussian resolution of 0\farcs9 for the different bands, the $3 \sigma$ limits are 0.09, 0.11, 0.13, and 0.29 $\mathrm{MJy\, sr^{-1}}$, for the F770W, F1000W, F1130W, and F2100W bands respectively\footnote{\citet{sutter24} estimated the $3 \sigma$ limits for the F770W, F1130W, and F2100W bands. We applied the same method to estimate the limit for the F1000W band.}. Since these $3 \sigma$ thresholds were derived at an angular resolution of 0\farcs9, for each galaxy in the sample with images at either the angular resolution $C_{\mathrm{opt}}$ or the spatial resolution 150 pc, we scale the $3 \sigma$ thresholds according to the square root of the ratio of kernel areas. At these relatively large angular scales, we expect the noise to be approximately correlated with the same angular response as the (convolved) PSF. At smaller scales, the spatial correlation of the noise can differ from the PSF (e.g., \citealt{williams24}).

We use the F770W and F1130W broad-band filters to trace the emission of the 7.7 and 11.3 \mic PAH features. Different letters published in the PHANGS-JWST Cycle 1 Focus Issue\footnote{\url{https://iopscience.iop.org/collections/2041-8205_PHANGS-JWST-First-Results}} suggest that these bands are generally dominated by PAH emission, with a contamination of about about 10--30\% from the hot dust continuum (e.g., \citealt{belfiore23, chastenet23b, dale23, leroy23, sandstrom23a, sandstrom23b}), with the exact fraction depending on the definition of continuum (see e.g., \citealt{whitcomb23}). The same works suggest that the F1000W band is probably dominated by PAH emission rather than by hot dust continuum, showing stronger correlations with the F770W and F1130W band fluxes than with the F2100W band flux. Silicate 9.7 \mic absorption can in principle affect the F1130W band. However, we expect the Silicate 9.7 \mic absorption to be negligible in the large majority of the pixels, as 99\% of them show $\mathrm{A}_{V} < 3$ mag (see e.g., \citealt{smith07}).

We use the prescription by \citet{sutter24} to subtract the starlight contribution from the F770W filter and define the 7.7 \mic PAH flux to be F770W$_{\mathrm{PAH}}$ = F770W - 0.13$\times$F200W. The prescription is based on an extensive set of {\sc cigale} \citep{boquien19} spectral energy distribution (SED) models that have varying stellar ages, star formation histories, and assumptions about the dust properties affecting its mid-infrared emission (see Section 3.1 there). In the dust-rich ISM, the correction is very small, with F770W$_{\mathrm{PAH}}$/F770W of 99\%. In the diffuse gas, F770W$_{\mathrm{PAH}}$/F770W ranges between 67\% and 90\% in most pixels, and reaches a minimum of $\sim$25\% in pixels that correspond to star formation deserts in a few of the galaxies, where the stellar radiation to PAH ratio is at its maximum. In Section \ref{sec:results:anomalous_PAHs} we study these extreme regions, and using HST+JWST SEDs, we verify the validity of the prescription and the detection of PAHs even in the most extreme environments. Regions with F770W$_{\mathrm{PAH}}$/F770W $\lessapprox 20$\% are not detected in the images, presumably due to the depth of the MIRI F770W images and our $3 \sigma$ SNR cuts.

A similar starlight subtraction can in principle be applied to the F1130W filter, with F1130W$_{\mathrm{PAH}}$ = F1130W - 0.067$\times$F200W. In practice, we find that in 95\% of the pixels of all galaxies combined, F1130W$_{\mathrm{PAH}}$/F1130W $>$ 93\%. In the small minority of pixels that trace extreme environments, F1130W$_{\mathrm{PAH}}$/F1130W can reach $\sim$70\%, and there, we subtract the expected starlight contribution (Section \ref{sec:results:anomalous_PAHs}). However, as we discuss in Section \ref{sec:results:anomalous_PAHs}, in these pixels, the main contaminant to the F1130W$_{\mathrm{PAH}}$ is the hot dust continuum rather than the starlight.

We define the PAH 11.3 \mic to 7.7 \mic flux ratio, hereafter $\log$PAH(11.3/7.7), to be $\log (\mathrm{ F1130W/F770W_{PAH}})$. Following several recent studies of the PAH-to-total dust mass fraction in nearby galaxies (e.g., \citealt{chastenet23a, egorov23, sutter24}), we also define $\mathrm{R_{PAH} = (F770W_{PAH} + F1130W)/F2100W}$. In addition, we use the following two band ratios to trace the stars-to-dust emission: F200W/F770W and F200W/F2100W, where the former is more closely related to the stars-to-PAH emission ratio, while the latter is more closely related to the stars-to-hot-dust emission ratio.

Distinct from to \citetalias{baron24}, here we do not use the PAH 3.3 \mic/11.3 \mic band ratio since the 3.3 \mic PAH feature requires accurate modeling and subtraction of the stellar continuum emission, which is still a work in progress. In the first three PHANGS galaxies studied in \citetalias{baron24}, we estimated the 3.3 \mic PAH feature using the prescription by \citet{sandstrom23a}. However, a follow-up work on the full sample of 19 PHANGS-JWST Cycle 1 galaxies suggests more diverse shapes of the stellar continuum at 3.3 \mic, as well as non-negligible starlight extinction in a fraction of the pixels, which require modifications to the \citet{sandstrom23a} prescription (Koziol et al. in prep.). The analysis of the variation of the PAH 3.3 \mic/11.3 \mic band ratio with the optical line ratios will therefore be presented in a future study.

Throughout the paper, JWST band ratios are presented as $f_{\nu}$ ratios. This is different conversion from the $\lambda f_{\lambda}$ ratios used for the optical line ratios (Section \ref{sec:MUSE}).

\subsubsection{HST}\label{sec:HST}

The PHANGS-HST survey mapped 41 galaxies, including our 19 targets, using high resolution ($\sim$0\farcs08) imaging in ultraviolet and optical wavelengths \citep{lee22}. The galaxies have been observed with five broad-band filters: F275W, F336W, F438W, F555W, and F814W\footnote{The galaxies NGC~628, NGC~1300, and NGC~1672 were observed with the F435W filter instead of the F438W.}. At this stage, the reduction pipeline does not include a flux anchoring step to to ensure accurate background levels in the images. As a result, the image reduction pipeline, which subtracts a global background taken within the field, may result in negative values for pixels with fluxes that approach the background level. In this work, we only use the HST data to produce HST+JWST stacked SEDs in specific regions that show high stellar-to-dust emission ratios using the filters F336W, F438W, F555W, and F814W (Section \ref{sec:results:anomalous_PAHs}). These regions show high SNR fluxes in the filters F438W, F555W, and F814W, with no negative values. As for the F336W filter, about 10-20\% of the pixels in these regions show negative or undetected flux values. However, including them in the stacks is not expected to have a significant impact on our conclusions, as discussed in Section \ref{sec:results:anomalous_PAHs}. We do not include fluxes measured with the F275W filter as it shows a large number of negative (or undetected) fluxes in the regions of interest.

\subsubsection{ALMA}\label{sec:ALMA}

The 19 galaxies are part of the PHANGS-ALMA survey that uses ALMA to map the CO $J = 2 \rightarrow 1$ line emission at a resolution of $\sim$1\arcsec\xspace across nearby galaxies (\citealt{leroy21a, leroy21b}). Out of the different products available in the survey (using ``strict'', ``broad'', and ``flat'' masks; see \citealt{leroy21b, leroy23}), we use the ``flat'' moment 0 maps described in \citet{leroy23} where a single fixed velocity window centered around the velocity at that position in the galaxy is used when estimating the integrated CO flux from every sightline. 

In our work, we use the CO flux maps obtained for resolutions $\sim$1\arcsec\ and 150 pc only in Section \ref{sec:results:anomalous_PAHs} to interpret a small subset of the pixels, identified in $\sim$4 of the galaxies, representing extreme environments with very high stellar-to-PAH emission ratios and anomalous $\log$PAH(11.3/7.7) ratios. Our conclusions remain unchanged when using other products instead (``broad'' masks).

\subsubsection{Environmental masks}\label{sec:env_masks}

To compare between the PAH band--optical line ratios relations across different environments, we use (i) all the identified nebulae presented in the PHANGS-MUSE nebular catalogue (\citealt{santoro22, groves23}), and (ii) the environmental maps by \citet{querejeta21}, that distinguish between different large-scale features, such as centers, bars, spiral arms, and more. 

\subsection{Convolution and resampling}\label{sec:conv_and_resamp}

The different data products used in this work are obtained from different instruments, each with a different PSF shape and size. In this section we describe our approach to standardize the images to have the same grid and angular resolution. 

For the JWST surface brightness maps, we generate kernels to convolve from the JWST images taken in one band (with the NIRCam/MIRI PSFs) to Gaussian PSFs at a specified resolution using the code {\sc jwst\_kernels}\footnote{\url{https://github.com/francbelf/jwst_kernels}}, implementing the approach described in \citet{aniano11}, using the WebbPSF library\footnote{\url{https://stsci.app.box.com/v/jwst-simulated-psf-library}}. In all but a single case, the FWHM values of the Gaussian PSFs are larger than the thresholds defined as ``safe'' by \citet{aniano11} for the NIRCam and MIRI bands (see Table 2 in \citealt{williams24} for the FWHMs corresponding to the ``safe'' and ``very safe'' thresholds for the different bands). The exception is NGC~4535, with $C_{\mathrm{opt}}$=0\farcs56, which is at a higher resolution than the F2100W band. For this galaxy only, we convolve the F2100W filter to a resolution of 0\farcs777 instead of 0\farcs56. Importantly, F2100W is used in a supporting capacity at the $C_{\mathrm{opt}}$ resolution. For the three primary JWST bands used in the work, F200W, F770W, and F1130W, the FWHMs are below the ``very safe'' threshold, for all of the galaxies in the sample. 

For the HST images, we first convert the flux units to $\mathrm{MJy\,sr^{-1}}$. We then convolve the HST images assuming that the original HST images have a Gaussian PSF with an angular resolution of $\sim$0\farcs08 (depends slightly on the specific band). Since we convolve images with angular resolutions of $\sim$0\farcs08 to $\sim$0\farcs9 ($C_{\mathrm{opt}}$) or $\sim$2\arcsec\xspace (150 pc), the simplifying assumption of a Gaussian PSF for the HST images is sufficiently accurate for our purposes. 

The ALMA moment 0 maps are available both at their native resolution ($\sim$1\arcsec\xspace) and at a 150 pc spatial resolution (see \citealt{leroy21b}), which we use as is. As for the environmental maps, the nebular catalogues were derived from the same MUSE cubes, and thus match in angular resolution, and the centers/bars masks represent large-scale features derived from a coarser resolution Spitzer map, and thus do not require a convolution.

\floattable
\begin{deluxetable}{m{5cm} m{7cm} c c c c}
\tablecaption{Featured used in unsupervised exploration with {\sc pca} and their correlation coefficients with the first four eigenvectors \label{tab:features}}
\tablecolumns{6}
\tablenum{2}
\tablewidth{0.95\linewidth}
\tablehead{
\colhead{Feature} & \colhead{Description} & \colhead{$\rho$(PCA-1)} & \colhead{$\rho$(PCA-2)} & \colhead{$\rho$(PCA-3)} & \colhead{$\rho$(PCA-4)}
}
\startdata
$\log\, \mathrm{Age} \, [yr]$ & light-weighted age of the stellar population & 0.38 & 0.28 & -0.23 & -0.04 \\
\oiiihbeta & optical line diagnostic of the shape of the radiation field, gas temperature, metallicity, and more & 0.74 & 0.02 & 0.22 & 0.23 \\
\niihalpha & '' & 0.89 & -0.05 & 0.16 & 0.10 \\
\siihalpha & '' & 0.83 & -0.35 & 0.15 & 0.01 \\
\oihalpha & '' & 0.72 & -0.42 & 0.23 & 0.13 \\
$\mathrm{E}(B-V)$ [mag] & reddening towards the line-emitting gas & -0.28 & 0.17 & 0.36 & 0.73 \\
$\log(\mathrm{H\alpha/f(7.7\, \mu m \, PAH))}$ $^{*}$ & H$\alpha$ to 7.7 \mic PAH emission & -0.64 & 0.57 & -0.40 & 0.25 \\
$\log(\mathrm{H\alpha/F2100W})$ $^{*}$ & H$\alpha$ to hot dust continuum emission & -0.77 & 0.28 & -0.43 & 0.30 \\
$\log(\mathrm{F2100W/F770W})$ & 21 to 7 \mic mid-infrared color & 0.04 & 0.71 & 0.40 & -0.21 \\
$\log(\mathrm{F2100W/f(7.7\, \mu m \, PAH)})$ & hot dust continuum to PAH emission & 0.38 & 0.76 & 0.15 & -0.23 \\
$\log(\mathrm{f(11.3\, \mu m \, PAH)/f(7.7\, \mu m \, PAH)})$ & ratio sensitive to PAH ionization, size, and temperature & 0.85 & 0.03 & -0.10 & 0.09 \\
$\log(\mathrm{F200W/F770W})$ & stellar to PAH emission ratio & 0.81 & 0.21 & -0.39 & -0.15 \\
$\log(\mathrm{F200W/F2100W})$ & stellar to hot dust continuum emission ratio & -0.77 & -0.03 & 0.51 & 0.09 \\
$\log(\mathrm{F1130W/F1000W})$ & PAH to hot dust (or silicate) emission ratio & -0.32 & -0.61 & 0.11 & -0.07 \\
$\log R_{\mathrm{PAH}}$ & PAH to small grain dust mass fraction & -0.09 & -0.79 & -0.17 & 0.30 \\
\enddata
\tablecomments{The listed correlation coefficients are the Spearman's rank correlation coefficients.\\
F770W represents the total surface brightness of the JWST-MIRI F770W filter, and $\mathrm{f(7.7\, \mu m \, PAH)}$ represents the F770W filter after subtraction of the expected starlight contribution at 7.7 \mic. \\
$^{*}$: The unit of these ratios is the ratio of the MUSE surface brightness given in the PHANGS-MUSE products to the JWST surface brightness given in the PHANGS-JWST products: $\log[(\mathrm{erg\,s^{-1}\,cm^{-2}\,pixel{^{-1}}) / (MJy\, sr^{-1}})]$.}
\end{deluxetable}

Having the convolved images, we project them onto the world coordinate system (WCS) of the MUSE observations using {\sc reproject.exact} by {\sc astropy} \citep{reproject_2020, astropy22}. To obtain independent pixels, 
we downsample the grid to have one pixel per resolution element of $C_{\mathrm{opt}}$ or 150 pc in all the images. For example, for a $C_{\mathrm{opt}}$ resolution of 0\farcs89, and the MUSE pixel size of 0\farcs2, every $4 \times 4$ pixel element matrix will be downsampled to a single pixel. For the different surface brightness maps, the downsampling is performed by taking the average (while excluding null values when present) over all the pixels. For the environmental maps, where the pixel value is a number representing a class, a majority vote is performed\footnote{A pixel can be associated with several different classes in the masks, e.g., center and bar. When combining different pixels into a single pixel at the angular resolution, in cases with equal numbers of `0' and `1', we adopt a value of `1' as the combined pixel value.}.

In Table \ref{tab:galaxy_properties} we list the number of independent pixels available for each galaxy using the $C_{\mathrm{opt}}$ or 150 pc products, that have both MUSE and JWST observations (the FOVs are only partially overlapping). We also list the fraction of the pixels that have both $\log$PAH(11.3/7.7) band and optical \siihalpha line ratios measured. Since the MUSE maps are highly complete, the fraction closely represents the 7.7 \mic and 11.3 \mic PAH detection fractions, which are the result of our adopted $3 \sigma$ masking\footnote{The two bands have comparable sensitivities, with F1130W resulting in slightly larger fraction ($\sim$10\%) of masked-out pixels.}.

\section{Results}\label{sec:results}

Our goal is to establish whether the PAH band--optical line ratios correlations are universal across a nearby star-forming galaxies, and identify regions in parameter space where they may break.  We start with an unsupervised exploration of the full PHANGS dataset using the dimensionality reduction algorithm {\sc pca} (Section \ref{sec:results:PCA}). This visualization allows for a quick and efficient exploration of the feature space, where we can examine how different features relate one to another, and importantly, quickly identify groups that show anomalous behavior. We then examine the PAH band--optical line ratios correlations in the 19 galaxies individually on 40--150 pc scales in Section \ref{sec:results:correlations}. In Section \ref{sec:results:anomalous_PAHs} we study a group of pixels that show anomalously high $\log$PAH(11.3/7.7) ratios primarily in four of the PHANGS galaxies. These pixels do not show the PAH band--optical line ratios correlations seen across galaxies. Finally, in Section \ref{sec:results:AGN_hosts}, we focus on the AGN hosts within the PHANGS sample, that show somewhat steeper PAH band--optical line ratios correlations, and try to identify the reason for this increased steepness.

Throughout this section, we use the $C_{\mathrm{opt}}$ products when studying individual galaxies separately and the 150 pc products when combining different galaxies within the same analysis.

\subsection{Unsupervised exploration with PCA}\label{sec:results:PCA}

Principal Component Analysis ({\sc pca}; \citealt{jolliffe02}) is a linear decomposition of a dataset into orthogonal components, also called the principal components. They are constructed such that each successive component accounts for the maximum possible variance in the data that remains after accounting for the variance explained by the preceding components. After the decomposition, each object in the dataset can be represented as a point in the space defined by the orthogonal principal components. {\sc pca} can be used to perform a dimensionality reduction of a complex dataset by selecting only a handful of leading components to represent each object. In particular, selecting only the two leading components and representing objects as points in this two-dimensional space, the dataset can be easily visualized while retaining most of the possible variance observed in the features. {\sc pca} is a simple and powerful technique, and its formulation through orthogonal components and their explained variance makes its output less challenging to interpret. In \citetalias{baron24} we used the non-linear dimensionality reduction algorithm {\sc umap}, and found that the data can, to first order, be represented as a one-dimensional sequence (the identified groups formed a continuous sequence in most of the features). We therefore use the simpler {\sc pca} technique in this work.

We apply {\sc pca} to a set of 15 features from 108\,403 spatially independent pixels. Since our goal is to study the PAH band--optical line ratios correlation, we consider features we believe may show connection either to the heating of PAHs or the ionization of the ionized gas. We list these features in Table \ref{tab:features}. The set of features included in the analysis is by no means exhaustive. We do not include ALMA CO observations since the maps have comparatively lower sensitivity, so including the CO flux would require us to exclude a significant number of pixels from the diffuse regions of the galaxies from the analysis. For a similar reason, we do not include the PAH 3.3 \mic feature (see Section 5.2 in \citetalias{baron24}). We do not normalize or rescale the features as they all have comparable dynamical ranges ($\sim$1 dex).

\begin{figure*}
	\centering
\includegraphics[width=1\textwidth]{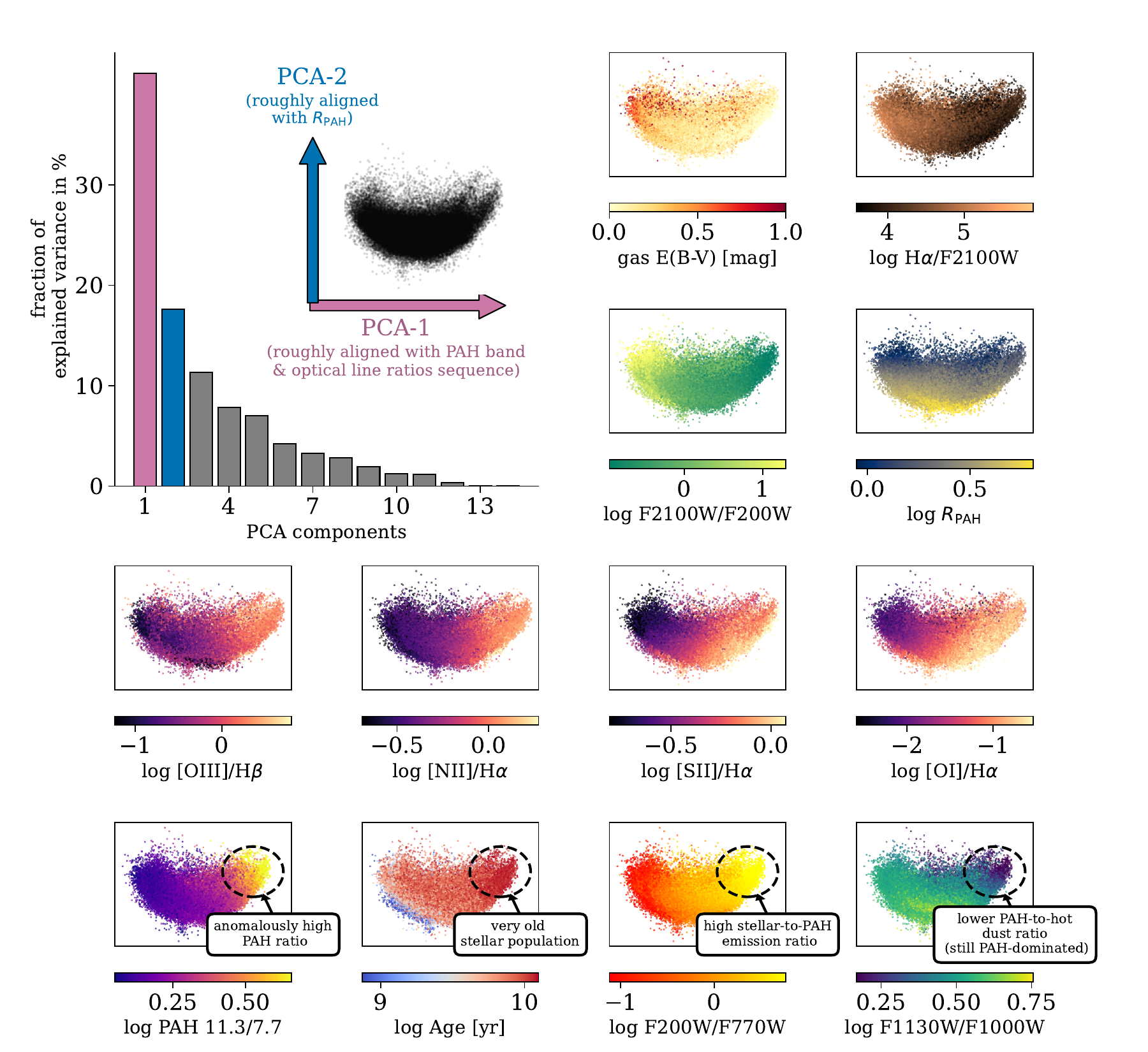}
\caption{\textbf{Two-dimensional visualization by {\sc pca} of the optical-infrared feature space spanned by $\sim$100\,000 150 pc-scale pixels from the 19 PHANGS galaxies.}
The top left panel shows the result of the {\sc pca} decomposition applied to 108\,403 spatially independent pixels that trace different optical and infrared features measured over a 150 pc scale. The bars represent the fraction of explained variance by each of the principal components, with the first component accounting for 41\% of the full variance, and the second for 17\%. Together, they account for 58\% of the total variance in the data. The inset in the top left panel shows the location of the pixels in the two-dimensional plane spanned by these first two orthogonal principal components. In the rest of the panels, the distribution of the pixels in this 2D plane is color-coded by different features of interest, where it can be seen that the first principal component (x-axis) aligns roughly with the $\log$PAH(11.3/7.7) versus optical line sequence, and the second component (y-axis) aligns roughly with the PAH-to-total dust mass fraction, $R_{\mathrm{PAH}}$. In the bottom row, we mark a group of pixels that show anomalously high $\log$PAH(11.3/7.7) ratios of $\sim$0.7 dex (typical $\log$PAH(11.3/7.7) ratios show a maximum of $\sim$0.4 dex, see Section \ref{sec:results:correlations}). The same group of pixels originates from regions dominated by old stellar populations, very high stellar-to-mid infrared emission ratio (suggesting old and bright populations), and relatively low PAH-to-hot dust ratio. We study this anomalous group in Section \ref{sec:results:anomalous_PAHs}.}
\label{f:feature_display}
\end{figure*}

The top left panel of Figure \ref{f:feature_display} shows the fraction of explained variance of the 15 principal components, which by construction, decreases from one component to the next. The first component accounts for 41\% of the total variance, while the second for 17\%, the third for 1.1\%, and the fourth for 0.78\%. In Table \ref{tab:features} we list the Spearman's rank correlation coefficients between our adopted features and the four leading principal components. The inset in the panel displays the dataset in the space defined by the first two principal components.

In the rest of the panels of Figure \ref{f:feature_display}, we color-code this 2D distribution by various features of interest. Inspection of the correlation coefficients in Table \ref{tab:features} and the color gradients in the different panels reveals the following. (1) The first principal component, PCA-1, is roughly aligned with the PAH-ionized gas correlation sequence, with both $\log$PAH(11.3/7.7) and the optical line ratios changing primarily along its direction (x-axis). This is not surprising, given that the correlation was known prior to the application of {\sc pca} (\citetalias{baron24}), and given that we use 5 individual features ($\log$PAH(11.3/7.7), \oiiihbeta, \niihalpha, \siihalpha, and \oihalpha) that encode a single one-dimensional sequence. This increases the relative weight of this sequence when constructing the principal components. (2) The second principal component, PCA-2, is roughly aligned with the PAH-to-total dust mass fraction, traced by $R_{\mathrm{PAH}}$ (y-axis), though it shows some continuous variation with PCA-3 as well (see \ref{app:pca_3_and_4} for details). This is also not surprising given that this information is encoded in several features considered in the analysis.

The bottom row of Figure \ref{f:feature_display} reveals a group of pixels with unusually high $\log$PAH(11.3/7.7) ratios, occupying the upper right corner of the 2D distribution. In the large majority of pixels, the $\log$PAH(11.3/7.7) ratio reaches a maximum of $\sim$0.4 dex (see Section \ref{sec:results:correlations} for additional details), while this group shows ratios of $\sim$0.6--0.7 dex. This group does not seem to follow the typical PAH band--optical line ratios correlations, showing high, but not unusually high, optical line ratios. Their low \halpha/F2100W ratios suggest that they originate in diffuse regions. Inspection of the bottom panels of Figure \ref{f:feature_display} shows that these pixels correspond to regions with old stellar populations (light-weighted mean age of $\sim$10 Gyr); high stellar-to-mid infrared emission ratio, suggesting bright stellar emission with respect to dust; and with relatively lower PAH-to-hot dust ratio. These properties resemble those observed the bulge of M31 (e.g., \citealt{groves12, draine14}). We study these regions in detail in Section \ref{sec:results:anomalous_PAHs}. 

Inspection of the color gradients in Figure \ref{f:feature_display} further reveals a diagonal stripe in the bottom left corner of the distribution that does not follow the same color gradient trends seen in the rest of the distribution, particularly in the optical line ratios \siihalpha and \oihalpha. It is characterized by a younger stellar population ($\log$ Age/yr $\sim 9$). In \ref{app:pca_3_and_4} we compare the properties of this group to a control group with the same stellar population age, and find that this group stands out in its high dust mid-infrared continuum emission (traced by F2100W) and exceptionally bright PAH emission (traced by F770W) with respect to the stellar emission (traced by F200W). Inspection of optical line diagnostic diagrams shows that the group has comparable optical line ratios to those of the control group. Since this group follows the typical PAH-ionized gas correlations seen in the rest of the population, we do not study it further here.

In \ref{app:pca_3_and_4} we present the 2D distribution of the pixels in the space defined by the third and fourth {\sc pca} components. These components approximately align with the F2100W/F200W ratio, tracing hot dust to stellar emission, and $\mathrm{E}(B-V)$, the dust reddening of the line-emitting gas.

\subsection{PAH-ionized gas correlations in individual galaxies}\label{sec:results:correlations}

In this section we study the correlation between the $\log$PAH(11.3/7.7) band ratio and the \siihalpha optical line ratio across the individual galaxies. All 19 galaxies show significant correlations between $\log$PAH(11.3/7.7) and \niihalpha, \siihalpha, and \oihalpha, and in cases where the \oiii line is significantly detected throughout the field of view, also with the \oiiihbeta ratio. The collisionally-excited \oiii, \nii, \sii, and \oi lines trace different ionization potentials, and they peak in different regions within the ionized (and once reaching the ionization front, the neutral) clouds. In this work, we focus on the relation with the \siihalpha. The \siihalpha is sensitive to several properties of the ionized gas in galaxies such as the metallicity, ionization parameter, and hardness of the radiation field. Given the observed range in metallicity in the PHANGS galaxies (e.g., \citealt{kreckel20, williams22}; Brazzini et al. submitted), we believe that this ratio is mostly tied to the separation of HII regions and diffuse ionized gas (see \citealt{belfiore22}), meaning it is primarily sensitive to the radiation field hardness. However, as we show in Section \ref{sec:results:AGN_hosts}, its diagnostic power in separating gas ionized by starlight and by an AGN is limited compared other line ratios, such as the \oiiihbeta. In a future study, we plan to study the relations with all the optical line ratios simultaneously, and use photoionization models to aid with the interpretation of these relations. 

\begin{figure*}
	\centering
\includegraphics[width=0.87\textwidth]{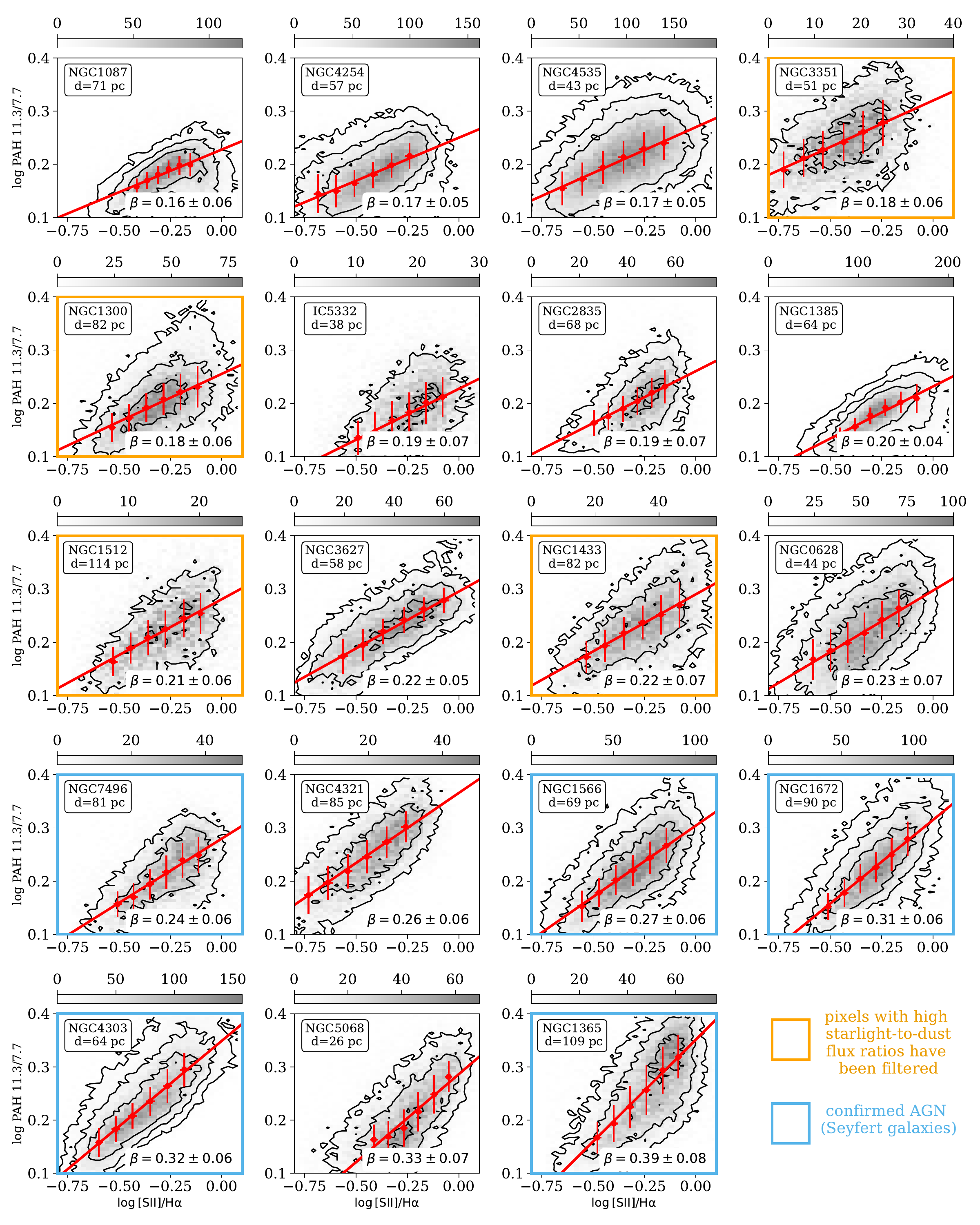}
\caption{\textbf{The $\mathrm{\log}$PAH(11.3/7.7) versus \siihalpha relations across individual PHANGS galaxies on scales of 40--120 pc.} Each panel shows the 2D distribution of the $\log$PAH(11.3/7.7) band ratio versus the \siihalpha optical line ratio across a single PHANGS galaxy. The gray color-coding represent the number of pixels with the corresponding PAH and optical line ratios, where the total number of pixels ranges from 9\,000 to 80\,000, with most galaxies having around 20\,000 spatially independent pixels. The black contours encompass the regions within which the counts are 5, 20, and 50. The relations are obtained using the maps at the $C_{\mathrm{opt}}$ resolution, and each panel notes the spatial scale in parsec probed for the galaxy. The red error bars represent 6 bins in \siihalpha and their medians and median absolute deviations of the $\log$PAH(11.3/7.7) ratio in the bin. The red solid lines represent the best-fitting linear relations of the bins. The galaxies are ordered by their best-fitting slope, from the shallowest (NGC~1087) to the steepest (NGC~1365). Galaxies with pixels belonging to the anomalous group identified in Section \ref{sec:results:PCA} and studied separately in Section \ref{sec:results:anomalous_PAHs} are marked with orange edges. These pixels are filtered out and excluded from 2D histograms, contours, and best fits. Galaxies with known Seyfert nuclei are marked with blue edges.}
\label{f:correlations_sep_ordered_by_slope}
\end{figure*} 

\begin{figure*}
	\centering
\includegraphics[width=0.87\textwidth]{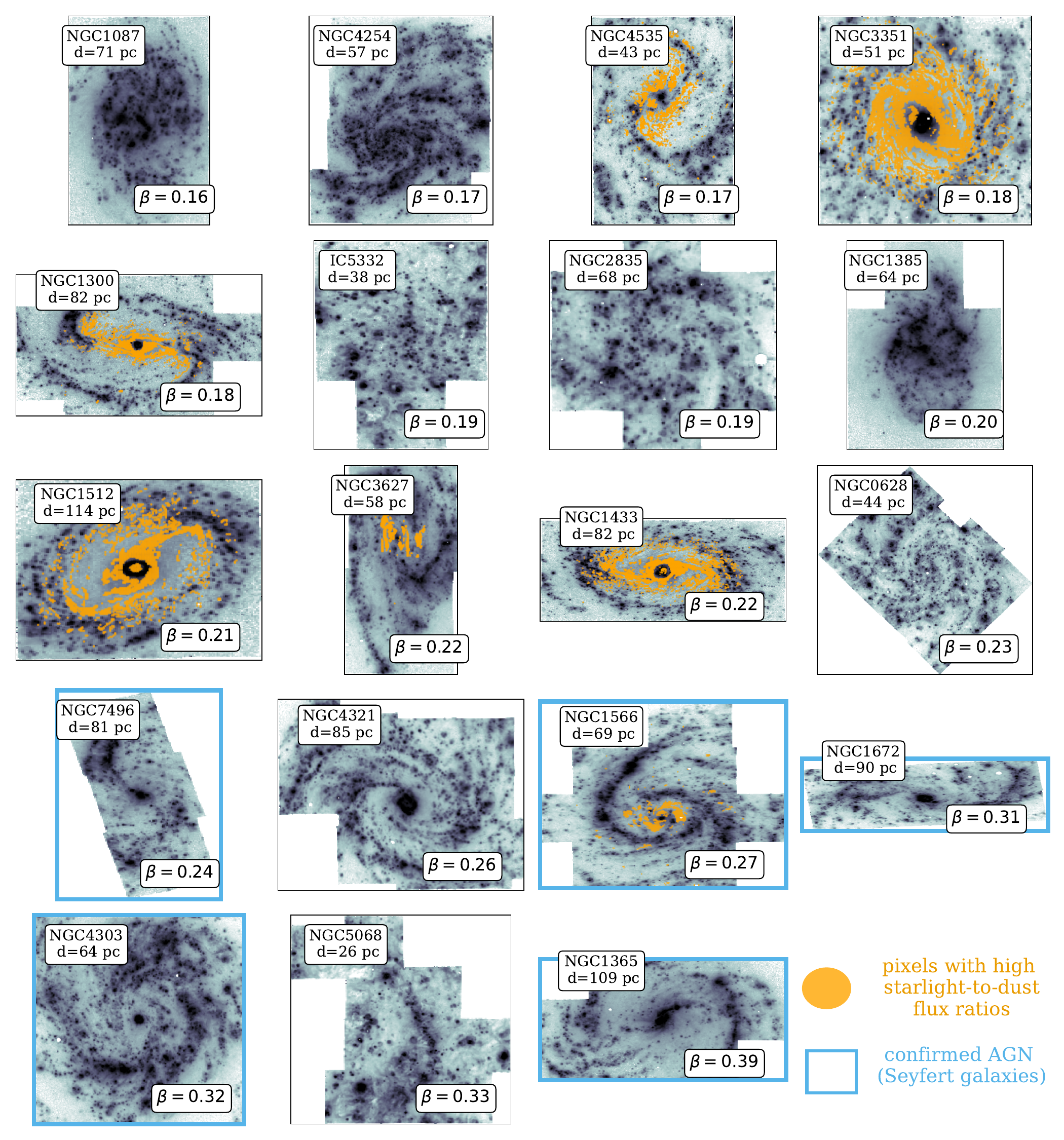}
\caption{\textbf{\halpha maps of the individual PHANGS galaxies ordered by the best-fitting slope from Figure \ref{f:correlations_sep_ordered_by_slope}.} Each panel shows the \halpha surface brightness of a single PHANGS galaxy. The maps are at the $C_{\mathrm{opt}}$ resolution, and each panel shows the spatial scale in parsec probed for the galaxy. The galaxies are ordered according to the best-fitting slope of the $\log$PAH(11.3/7.7) versus \siihalpha relation from Figure \ref{f:correlations_sep_ordered_by_slope}. Pixels corresponding to regions with high starlight-to-dust flux ratios are marked with orange. For the four galaxies that have a significant fraction of such pixels (10--25\%; NGC~3351, NGC~1300, NGC~1512, and NGC~1433), they are excluded from the analysis in Figure \ref{f:correlations_sep_ordered_by_slope} and of this section. For the three galaxies that have lower fractions of such pixels (2--5\%; NGC~4535, NGC~3627, and NGC~1566), we mark the pixels in this Figure, but include them in the analysis of this section. Known AGN-hosts are marked with blue edges.}
\label{f:galaxies_sorted_by_slope}
\end{figure*} 

In Section \ref{sec:results:PCA} above, we identified a group of pixels with anomalous PAH ratios that do not follow the PAH band--optical line ratios correlation. They are primarily seen in four galaxies (NGC~1300, NGC~1433, NGC~1512, and NGC~3351), and their fraction out of all the pixels of each respective galaxy is between 10\% and 25\%. The same pixels show high stellar-to-mid infrared emission ratios, generally above $\log$(F200W/F770W) $> 0.4\,\mathrm{dex}$. We use this threshold to filter out these pixels and exclude them from the analysis in this section. We study the properties of these regions separately in Section \ref{sec:results:anomalous_PAHs} below. Such pixels are also present in several additional galaxies (e.g., NGC~1566, NGC~3627, and NGC~4535), but their fraction is lower, $\sim$2--5\%, and thus including them does not affect the derived slopes. 

In Figure \ref{f:correlations_sep_ordered_by_slope} we show the $\log$PAH(11.3/7.7) versus \siihalpha relations across the individual PHANGS galaxies. The relation is clearly detected across each of the 19 galaxies, on scales as small as 40 pc, markedly below the scale of 150 pc probed in our \citetalias{baron24} work\footnote{The only major difference between the relation presented in Figure 13 in \citetalias{baron24} and our analysis here is that here we subtract the starlight contribution from the F770W filter before deriving the $\log$PAH(11.3/7.7) ratio. This makes the range in $\log$PAH(11.3/7.7) ratios somewhat larger here compared to the range seen in \citetalias{baron24}}. We fit the measurements with a linear relation for each galaxy as described below. We divide the \siihalpha measurements into 6 bins, and estimate the median and median absolute deviation (MAD) of \siihalpha and $\log$PAH(11.3/7.7) within each bin. We then fit the medians with a linear relation, considering the MADs as uncertainties. We estimate the uncertainty of the best-fitting slope and intercept by bootstrapping the samples to produce the bins for 1000 times, refitting the relation in every iteration. The best-fitting slopes are consistent, within the uncertainties, for a number of bins ranging between 6 and 12. The galaxies in Figure \ref{f:correlations_sep_ordered_by_slope} are sorted according to their best-fitting slope, from the shallowest (NGC~1087; $\beta = 0.16 \pm 0.06$) to the steepest (NGC~1365; $\beta = 0.38 \pm 0.08$). In Figure \ref{f:galaxies_sorted_by_slope} we show the \halpha surface brightness maps of the galaxies, ordered by the best-fitting slope from Figure \ref{f:correlations_sep_ordered_by_slope}.

\begin{figure}
\includegraphics[width=\columnwidth]{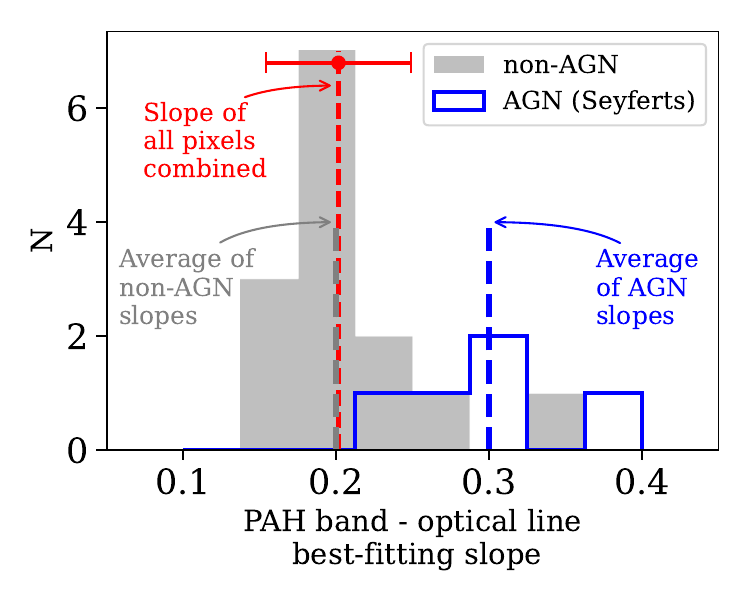}
\caption{\textbf{Best-fitting slopes of the $\log$PAH(11.3/7.7) versus \siihalpha relations in individual PHANGS galaxies.} The best-fitting slopes are obtained for the $C_{\mathrm{opt}}$ resolution maps. The red dashed line represents the best-fitting slope obtained when combining all the pixels from all the galaxies, dividing into 6 bins, and fitting the median values. The red horizontal error bar represents the uncertainty of the derived slope, obtained with bootstrapping the samples. The subset of the PHANGS galaxies with identified Seyfert nuclei (NGC~1365, NGC~1566, NGC~1672, NGC~4303, and NGC~7496) are marked with an empty blue histogram. The rest are marked with a gray histogram. AGN-hosts have on average steeper slopes in their PAH-ionized gas relation than non-AGN hosts. Putting the AGN hosts aside, the figure shows that there exist a universal relation between $\log$PAH(11.3/7.7) and \siihalpha across nearby galaxies, with a slope of about 0.2.} 
\label{f:best_fitting_slope_distribution}
\end{figure} 

We use the \citet{veron_cetty10} catalog and the spatially resolved PHANGS-MUSE optical line ratios to identify PHANGS galaxies with AGN. Since LINER-like optical line ratios can represent ionization by either an AGN or by hot and evolved stellar populations (see \citealt{belfiore22} and references therein), here we use the more stringent AGN definition of Seyfert-like line ratios in the \siihalpha versus \oiiihbeta line diagnostic diagram, using the \citet{kewley06} criterion. Since AGN activity can be manifested as LINER-like optical line ratios, our choice to focus only on Seyfert-like line ratios has implications to the shape of the ionizing radiation field and/or the ionization parameter (see e.g., photoionization models by \citealt{baron19b}). Among the 19 galaxies, 5 are identified as Seyfert AGN: with NGC~1365, NGC~1566, NGC~1672, NGC~4303, and NGC~7496. Although NGC~3627 has been considered by various studies as a Seyfert galaxy, its classification as a Seyfert is based on the study by \citet{ho97}, which employs a different Seyfert-LINER separation scheme. Using the \citet{kewley06} criterion, NGC~3627 is classified as a LINER, and we therefore do not consider it as an AGN in our study. In Figures \ref{f:correlations_sep_ordered_by_slope} and \ref{f:galaxies_sorted_by_slope} we mark the AGN hosts with blue edges. Of the seven PHANGS galaxies with the steepest PAH-ionized gas slopes, five are AGN hosts. Figure \ref{f:best_fitting_slope_distribution} shows the distribution of the best-fitting slopes for the PHANGS galaxies, where AGN hosts show on average steeper slopes. In Section \ref{sec:results:AGN_hosts} below we focus on the AGN hosts and attempt to identify the reason for the increased steepness.

 \floattable
\begin{deluxetable}{c CCCC CCCC}
\tablecaption{$\log$PAH(11.3/7.7) versus \siihalpha best-fitting relations\label{tab:separate_slopes}}
\tablecolumns{10}
\tablenum{3}
\tablewidth{0pt}
\tablehead{
\colhead{(1)} & \colhead{(2)} & \colhead{(3)} & \colhead{(4)} & \colhead{(5)} & \colhead{(6)} & \colhead{(7)} & \colhead{(8)} & \colhead{(9)} \\
\colhead{Galaxy} & \colhead{$\alpha$} & \colhead{$\beta$} & \colhead{$y(x=-0.5)$} & \colhead{$\sigma_{y}$} & \colhead{$\alpha$} & \colhead{$\beta$} & \colhead{$y(x=-0.5)$} & \colhead{$\sigma_{y}$} \\ 
\colhead{} & \colhead{($C_{\mathrm{opt}}$)} & \colhead{($C_{\mathrm{opt}}$)} & \colhead{($C_{\mathrm{opt}}$)} & \colhead{($C_{\mathrm{opt}}$)} & \colhead{(150 pc)} & \colhead{(150 pc)} & \colhead{(150 pc)} & \colhead{(150 pc)}
} 
\startdata
IC~5332 & 0.23 \pm 0.02 & 0.20 \pm 0.07 & 0.13 & 0.037 & 0.21 \pm 0.03 & 0.19 \pm 0.08 & 0.12 & 0.035 \\
NGC~0628 & 0.30 \pm 0.03 & 0.23 \pm 0.07 & 0.18 & 0.038 & 0.31 \pm 0.03 & 0.26 \pm 0.06 & 0.18 & 0.033 \\
NGC~1087 & 0.23 \pm 0.02 & 0.16 \pm 0.06 & 0.15 & 0.017 & 0.24 \pm 0.02 & 0.19 \pm 0.05 & 0.14 & 0.015 \\
NGC~1300 & 0.26 \pm 0.02 & 0.18 \pm 0.06 & 0.16 & 0.031 & 0.27 \pm 0.02 & 0.23 \pm 0.06 & 0.16 & 0.025 \\
NGC~1365 & 0.35 \pm 0.03 & 0.38 \pm 0.08 & 0.16 & 0.039 & 0.36 \pm 0.02 & 0.40 \pm 0.07 & 0.16 & 0.036 \\
NGC~1385 & 0.23 \pm 0.02 & 0.19 \pm 0.05 & 0.13 & 0.019 & 0.23 \pm 0.01 & 0.19 \pm 0.04 & 0.14 & 0.016 \\
NGC~1433 & 0.29 \pm 0.02 & 0.22 \pm 0.06 & 0.18 & 0.034 & 0.31 \pm 0.02 & 0.26 \pm 0.06 & 0.18 & 0.029 \\
NGC~1512 & 0.28 \pm 0.02 & 0.21 \pm 0.06 & 0.18 & 0.033 & 0.29 \pm 0.02 & 0.23 \pm 0.06 & 0.17 & 0.030 \\
NGC~1566 & 0.30 \pm 0.02 & 0.26 \pm 0.06 & 0.17 & 0.030 & 0.31 \pm 0.02 & 0.28 \pm 0.05 & 0.17 & 0.027 \\
NGC~1672 & 0.32 \pm 0.02 & 0.32 \pm 0.07 & 0.15 & 0.029 & 0.32 \pm 0.02 & 0.32 \pm 0.06 & 0.16 & 0.026 \\
NGC~2835 & 0.26 \pm 0.02 & 0.20 \pm 0.06 & 0.16 & 0.027 & 0.26 \pm 0.02 & 0.19 \pm 0.06 & 0.17 & 0.023 \\
NGC~3351 & 0.32 \pm 0.04 & 0.17 \pm 0.07 & 0.23 & 0.038 & 0.33 \pm 0.04 & 0.19 \pm 0.07 & 0.23 & 0.031 \\
NGC~3627 & 0.30 \pm 0.02 & 0.21 \pm 0.05 & 0.19 & 0.026 & 0.30 \pm 0.01 & 0.23 \pm 0.04 & 0.19 & 0.023 \\
NGC~4254 & 0.25 \pm 0.02 & 0.16 \pm 0.05 & 0.17 & 0.027 & 0.26 \pm 0.02 & 0.18 \pm 0.05 & 0.17 & 0.023 \\
NGC~4303 & 0.35 \pm 0.03 & 0.32 \pm 0.06 & 0.19 & 0.028 & 0.36 \pm 0.02 & 0.34 \pm 0.05 & 0.19 & 0.026 \\
NGC~4321 & 0.37 \pm 0.03 & 0.26 \pm 0.05 & 0.23 & 0.032 & 0.37 \pm 0.03 & 0.26 \pm 0.05 & 0.23 & 0.028 \\
NGC~4535 & 0.27 \pm 0.02 & 0.17 \pm 0.05 & 0.18 & 0.031 & 0.29 \pm 0.02 & 0.22 \pm 0.04 & 0.18 & 0.023 \\
NGC~5068 & 0.29 \pm 0.02 & 0.34 \pm 0.07 & 0.12 & 0.033 & 0.29 \pm 0.02 & 0.31 \pm 0.06 & 0.14 & 0.028 \\
NGC~7496 & 0.28 \pm 0.02 & 0.25 \pm 0.06 & 0.16 & 0.030 & 0.29 \pm 0.02 & 0.26 \pm 0.06 & 0.16 & 0.026 \\
\enddata
\tablecomments{(1) Galaxy name. (2) and (3) Best-fitting intercept and slope of the $\log$PAH(11.3/7.7) versus \siihalpha relation. (4) $\log$PAH(11.3/7.7) value for \siihalpha=$-0.5$ of the best-fitting relation. (5) RMS scatter in  $\log$PAH(11.3/7.7) values around the best-fitting relation. (2)-(5) represent the best-fitting relation when using the MUSE $C_{\mathrm{opt}}$ resolution, and (6)-(9) when using the 150 pc resolution.
}
\end{deluxetable}
%\vspace{5mm}

We do not find a relation between the best-fitting slope value and the location of the galaxies with respect to the star-forming main sequence, or with inclination. In addition, we do not find a relation between the value of the best-fitting slope and the physical scale at which the relation is probed ($C_{\mathrm{opt}}$ in pc). In fact, in \ref{app:pah_gas_corr_150pc} we repeat the analysis using $\log$PAH(11.3/7.7) and \siihalpha measurements in individual galaxies over the uniform 150 pc scale. The best-fitting slopes for the 150 pc scale case are comparable to those derived using the $C_{\mathrm{opt}}$ resolution, though they are somewhat larger, especially for small $\beta$ values (Figure \ref{f:comparison_of_slopes_Copt_vs_150pc}). The distribution of the best-fitting slopes using the 150 pc scale maps is slightly narrower than that seen at the $C_{\mathrm{opt}}$ resolution (Figure \ref{f:beta_distribution_Copt_vs_150pc}). The lack of a strong dependence of the slope on the scale at which the relation is probed within 40--150 pc suggests that the process driving this correlation is in place on scales as small as 40 pc, and that it does not have a significant spatial variation. This is in line with our preferred interpretation of the correlation in \citetalias{baron24}, suggesting that the PAH band--optical line ratio sequence is driven by a varying shape of the interstellar radiation field. The radiation field is a weighted mixture of emission originating from young massive stars; emission from hot and evolved stars; and a hard ionizing radiation by the AGN; with the weights changing spatially.

Putting aside the AGN hosts and the group of pixels with anomalous PAH ratios, Figures \ref{f:correlations_sep_ordered_by_slope} and \ref{f:best_fitting_slope_distribution} suggest that the PAH band--optical line ratio relation is a universal feature across nearby star-forming galaxies, with a slope of about 0.2 (0.22 at a 150 pc scale). It suggests a fundamental connection between the ionized gas and neutral gas, and can be used to place constraints on intrinsic PAH properties and physics (size, charge, temperature) as well as on gas ionization. The scatter around the best-fitting relations, $\sigma[\log \mathrm{PAH(11.3/7.7)}] \sim 0.025$ dex, is quite uniform across the galaxies and surprisingly small, suggesting very limited variation in properties that may affect this ratio, such as the PAH ionization fraction. In Table \ref{tab:separate_slopes} we list the parameters of the best-fitting linear relations for the galaxies in the sample, on the $C_{\mathrm{opt}}$ and the 150 pc scales.

\begin{figure*}[h]
	\centering
\includegraphics[width=1\textwidth]{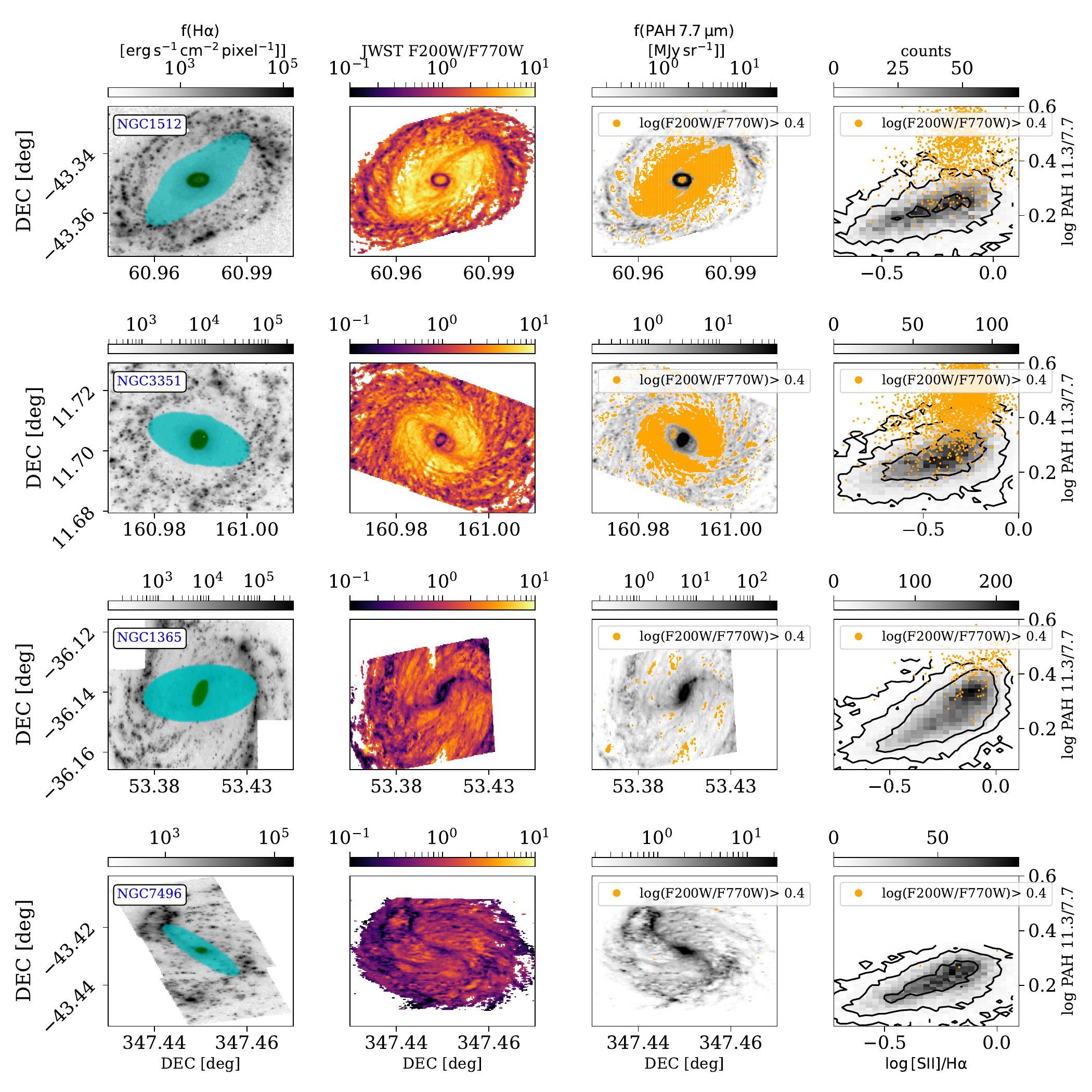}
\caption{\textbf{Identification of regions with anomalous $\log$PAH(11.3/7.7) ratios using the stellar-to-mid infrared emission ratio F200W/F770W.} Each row represents a single galaxy, where we show two examples of galaxies with a significant fraction of anomalous PAH ratios (NGC~1512 and NGC~3351) and two with far less (NGC~1365 and NGC~7496). In each row, from left to right: (i) The black-white background shows the \halpha surface brightness. The cyan (green) color represents pixels that are identified as bars (centers) in the environmental maps. (2) Image showing the  F200W/F2100W ratio throughout the galaxy. (3) The black-white background shows the surface brightness of the 7.7 \mic PAH feature (F770W$_{\mathrm{PAH}}$), and the orange points represent all pixels with $\log$(F200W/F770W)$\;>0.4$. (4) $\log$PAH(11.3/7.7) versus \siihalpha relations for pixels below the $\log$(F200W/F770W)$\;=0.4$ threshold (considered `normal'; shown in the gray-scale color-coding and the contours), and pixels above $\log$(F200W/F770W)$\;=0.4$ threshold (orange points). There is a correspondence between the presence of PAHs with anomalous ratios and very bright stellar-to-mid infrared emission ($\log$(F200W/F770W)$\;>0.4$). Galaxies having a small number of pixels with $\log$(F200W/F770W)$\;>0.4$ also do not show PAHs with anomalous ratios, even if they are barred galaxies. We therefore use the threshold $\log$(F200W/F770W)$\;>0.4$ to identify pixels that belong to the anomalous PAH group. }
\label{f:stellar_to_PAH_threshold}
\end{figure*}

Figure \ref{f:correlations_sep_ordered_by_slope} does suggest some second-order variation in the observed $\log$PAH(11.3/7.7) versus \siihalpha relation when considering individual galaxies. For example, NGC~5068 and NGC~4321 show steeper than average slopes ($0.26 \pm 0.05$ and $0.33 \pm 0.07$ respectively), with a maximum $\log$PAH(11.3/7.7) ratio of about 0.4 dex, although they have not been identified as AGN hosts. At the other side of the range, NGC~1087 and NGC~4254 show the least steep slopes of $0.16 \pm 0.06$ and $0.17 \pm 0.05$ respectively, with a maximum $\log$PAH(11.3/7.7) ratio of about 0.3 dex. NGC~4535 and NGC~0628 show larger scatter in $\log$PAH(11.3/7.7) around the best-fitting relation. Finally, looking at different environments within a given galaxy, we find deviations from the best-fitting relation obtained when considering all the pixels. In particular, pixels that are identified as centers show higher $\log$PAH(11.3/7.7) ratios on average in some of the galaxies. Some of these pixels are also classified as pixels belonging to the anomalous group described in Section \ref{sec:results:anomalous_PAHs} below. Additional analysis is required to explain the second-order variations in slopes and the dependence on the environment, which we leave to a future study.

\begin{figure*}
	\centering
\includegraphics[width=1\textwidth]{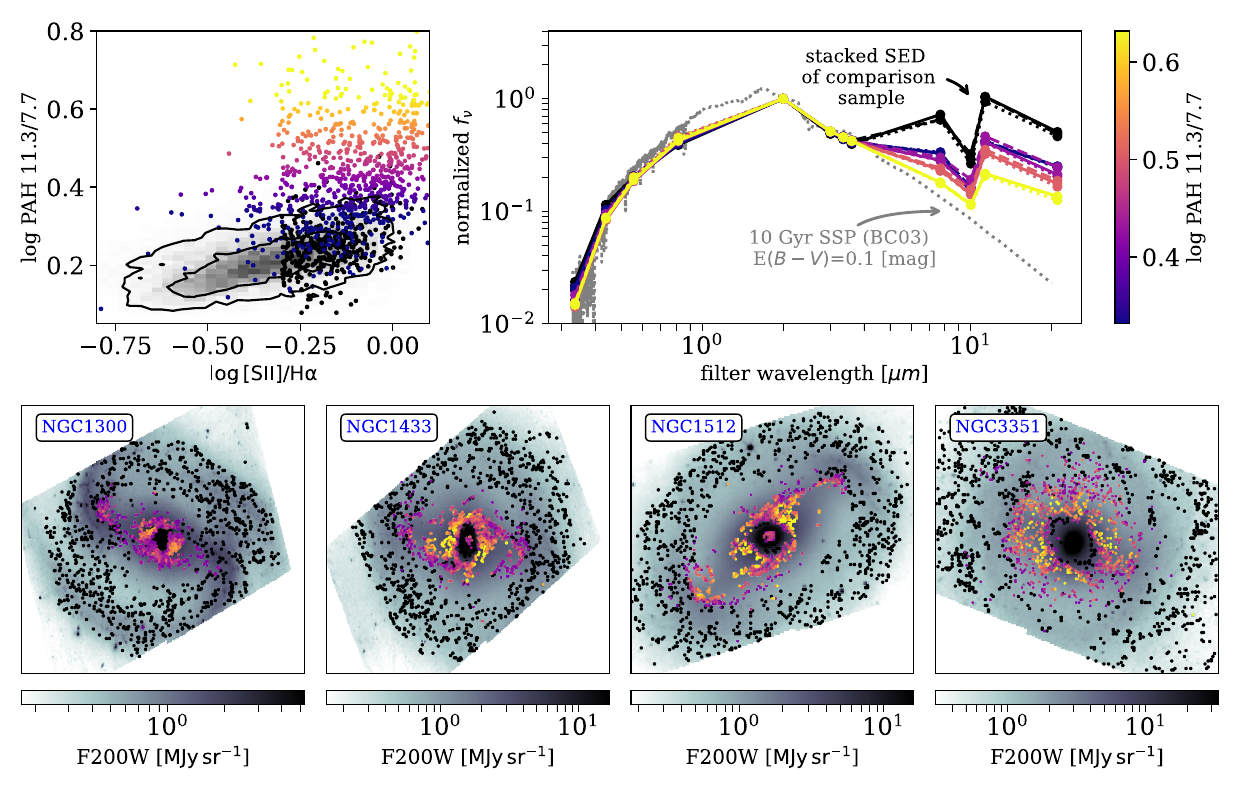}
\caption{\textbf{Stacked HST+JWST SEDs and spatial distribution of regions showing anomalous PAH ratios.}\textbf{ The top left panel} shows the distribution of a sample of pixels with anomalous PAH ratios with respect to the general observed correlation between $\log$PAH(11.3/7.7) and \siihalpha across nearby galaxies. The contours represent `normal' pixels from the four galaxies NGC~1300, NGC~1433, NGC~1512, and NGC~3351. The black points represent the comparison sample (which includes 7557 pixels, out of which a random subset of 400 is shown), and purple-yellow represent the anomalous group (which includes 6404 pixels, out of which a random subset of 800 is shown). \textbf{The top right panel} shows the HST+JWST stacked SEDs of the comparison sample (black lines) and a few groups with anomalous PAH ratios, binned by their $\log$PAH(11.3/7.7). The gray dashed line represents a 10 Gyr SSP model by \citet{bruzual03} with a reddening of $\mathrm{E}(B-V)=0.1$ mag. \textbf{The bottom row} shows the spatial distribution of pixels from the comparison sample (black points) and pixels from the anomalous PAH ratio group (purple-yellow points). The group of pixels with anomalous PAH ratios is spatially distinct from the comparison sample, concentrated primarily along the gas inflow lines to the galactic centers.}
\label{f:stacked_SEDs_weird_PAHs_fnu}
\end{figure*}

\subsection{Anomalous PAH ratios}\label{sec:results:anomalous_PAHs}

Our low-dimensional visualization using {\sc pca} in Section \ref{sec:results:PCA} reveals a group of pixels with unusually high $\log$PAH(11.3/7.7) ratios that do not seem to follow the general PAH-ionized gas correlation established in Section \ref{sec:results:correlations}. These pixels generally originate from regions with very old ($\sim 10^{10}$ yr) stellar populations and high starlight-to-dust emission as traced by F200W/F770W (see Figure \ref{f:feature_display}). In this section, we focus on this group and use multi-wavelength observations, including HST and ALMA, to interpret the observed PAH band ratios in these regions. 

In Figure \ref{f:stellar_to_PAH_threshold} we demonstrate our method to identify pixels with anomalous PAH ratios. We find a strong correspondence between the presence of PAH emission with anomalous band ratios and bright stellar-to-mid infrared emission. Galaxies that show pixels with $\log$(F200W/F770W)$>0.4\,\mathrm{dex}$ also show PAH ratios with unusually high values that are above the general PAH band--optical line ratio correlations seen across galaxies. Galaxies that have a small number of pixels with $\log$(F200W/F770W)$>0.4\,\mathrm{dex}$ do not show PAHs with anomalous ratios, even if they are barred galaxies. We therefore use the stellar-to-mid infrared emission ratio to define our sample of pixels with anomalous PAH ratios, using the following threshold: $\log$(F200W/F770W)$>0.4\,\mathrm{dex}$\footnote{As described later in the section, there is a continuous variation of increasing $\log$PAH(11.3/7.7) with $\log$(F200W/F770W), making the selected threshold of $\log$(F200W/F770W)$>0.4\,\mathrm{dex}$ somewhat arbitrary. We find qualitatively similar results when using thresholds of 0.3 and 0.5 $\mathrm{dex}$ instead.}. Among the 19 PHANGS galaxies in our sample, there are four sources that show a large fraction of pixels above this threshold: NGC~1300 (10\%), NGC~1433 (21\%), NGC~1512 (21\%), and NGC~3351 (25\%), and we focus on these in this section. In all four cases, the $\log$PAH(11.3/7.7) are unusually high and above the typical correlation with \siihalpha. In Figure \ref{f:stellar_to_PAH_threshold_outlying_gals_worst} in \ref{app:anomalous_pahs} we show them in the same format as in Figure \ref{f:stellar_to_PAH_threshold}.

There are three additional galaxies that show a more modest fraction of pixels above this threshold: NGC~1566 (2.5\%), NGC~3627 (2.6\%), and NGC~4535 (5\%). Due to the lower fraction of pixels with anomalous ratios, these are not included in the analysis of this section. In Figure \ref{f:stellar_to_PAH_threshold_outlying_gals_notworst} in \ref{app:anomalous_pahs} we show them in the same format as in Figure \ref{f:stellar_to_PAH_threshold}. For the rest of the galaxies, the fraction of pixels above the threshold is lower than 5\%, and in most cases, below 2\%, and they do not show pixels that are significantly offset from the PAH band--optical line ratio correlation.

\begin{figure*}
	\centering
\includegraphics[width=1\textwidth]{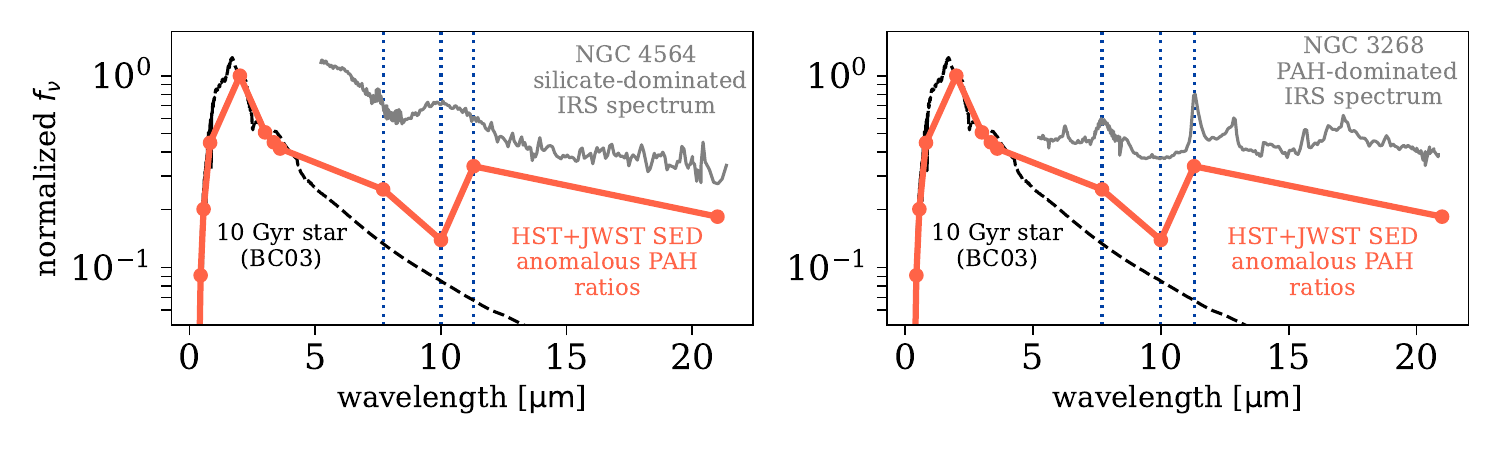}
\caption{\textbf{Comparison of the HST+JWST SED of the anomalous PAH group with Spitzer spectra of elliptical galaxies.} The orange points and line represents a single HST+JWST stacked spectrum of pixels showing anomalous PAH ratios. The black dashed line shows the SED of a 10 Gyr SSP model from the \citet{bruzual03} library. The gray lines show Spitzer IRS spectra of two elliptical galaxies: NGC~4564, which is dominated by silicate emission at 9.7 \mic, and NGC~3268, which is dominated by PAH features with an unusual PAH 11.3/7.7 ratio. The blue dotted vertical lines correspond to wavelengths 7.7, 10, and 11.3 \mic. Our observed SEDs are inconsistent with a silicate emission spectrum, since they show F1000W$<$F1130W, and for silicate-dominated spectra, we expect F1000W$>$F1130W.}
\label{f:anomalous_PAH_ratios_silicates_vs_PAHs}
\end{figure*}

We start by examining the SEDs and spatial distribution of the pixels in the anomalous group. To aid with the interpretation, we define a comparison sample that includes all the pixels from the same four galaxies that are \emph{below} the $\log$(F200W/F770W)$=0.4\,\mathrm{dex}$ threshold, and are thus considered `normal', but with \siihalpha ratios in the range observed in the anomalous group (-0.3 to 0.1). The latter is done to control for variations in the shape of the ionizing radiation seen in the different regions. We then produce stacked HST+JWST SEDs for the comparison and the anomalous groups, where we divide the anomalous group into a 4 bins according to their $\log$PAH(11.3/7.7). 

In Figure \ref{f:stacked_SEDs_weird_PAHs_fnu}, we compare between the two groups in the $\log$PAH(11.3/7.7) versus \siihalpha plane (top left panel), and compare their optical-infrared SEDs (top right panel). The stacked SEDs of both the comparison and the anomalous group are quite similar to each other in optical-to-near infrared wavelengths, and both resemble the SED of an old stellar population. None of the SEDs show detectable PAH 3.3 \mic emission. Above 3 \mic, the SEDs show differences of a factor of a few, with the F770W filter decreasing dramatically from the comparison sample to the anomalous group, and throughout different bins of $\log$PAH(11.3/7.7) in the anomalous group. As a reference, we show the SED of a 10 Gyr single stellar population (SSP) model from the \citet{bruzual03} library, to which we apply dust reddening of $\mathrm{E}(B-V)=0.1$ mag to match the observed optical-to-near infrared SEDs. The reddening is quite modest, and does not affect wavelengths longer than $\sim$2 \mic. The comparison to the 10 Gyr SSP model suggests that the $\log$PAH(11.3/7.7) ratios observed in the anomalous group are not due to our starlight subtraction prescription from the F770W filter -- the observed F770W filter in all the stacks is clearly elevated with respect to the expected emission at 7.7 \mic from old stellar populations. 

In the bottom row of Figure \ref{f:stacked_SEDs_weird_PAHs_fnu}, we show the spatial distribution of pixels from the comparison and anomalous groups. The groups show clear differences, where the comparison pixels are located along the spiral arms of the galaxies, while the anomalous pixels are located close to inflowing gas filaments feeding the centers of the galaxies (bar lanes hereafter). The $\log$PAH(11.3/7.7) ratios observed for the anomalous group display clear spatial structures with respect to the bar lanes.  They show lower values of $\log$PAH(11.3/7.7)$\sim$0.4--0.5 dex closer to the bar lanes, and higher values of $\log$PAH(11.3/7.7)$\sim$0.6 dex farther from the lanes.

\begin{figure*}
	\centering
\includegraphics[width=1\textwidth]{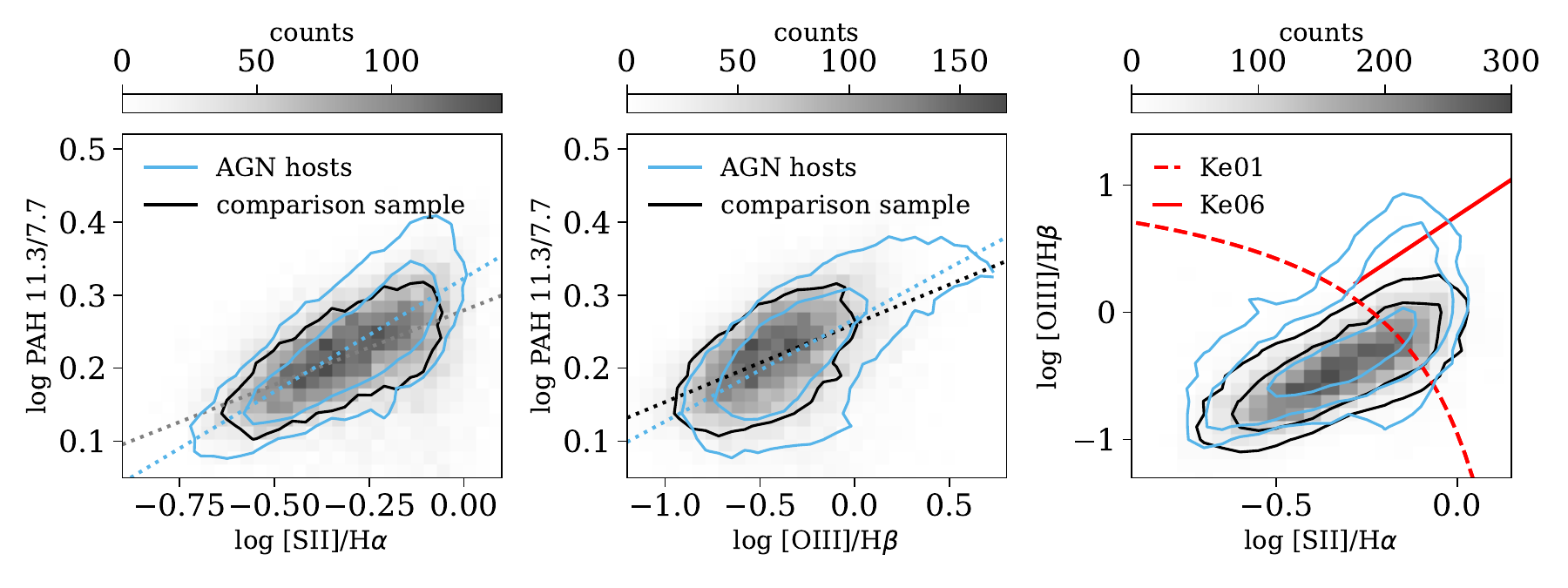}
\caption{\textbf{Comparison of the PAH and ionized gas properties in AGN and non-AGN hosts.} The left panel shows $\log$PAH(11.3/7.7) versus \siihalpha for the AGN (blue contours) and non-AGN hosts (gray contours; grayscale background). The dashed lines represent the best-fitting relations, where the AGN show a steeper slope. The steeper slope is caused by a small increase in $\log$PAH(11.3/7.7) with respect to \siihalpha, for large \siihalpha values of $\sim$-0.1. The middle panel shows $\log$PAH(11.3/7.7) versus \oiiihbeta, with the latter being a more sensitive tracer of AGN ionization than the \siihalpha. In this panel, the increase in $\log$PAH(11.3/7.7) corresponds to increased \oiiihbeta. The right panel shows the \oiiihbeta versus \siihalpha diagnostic diagram. The two separating criteria are used to separate between SF and AGN \citep[Ke01; dashed]{kewley01}, and between LINERs and Seyferts \citep[Ke06; solid]{kewley06}. The excess $\log$PAH(11.3/7.7) from the left panel, which corresponds to increased \oiiihbeta in the middle panel, is consistent with Seyfert-like line ratios in the right panel. All together suggest that the radiation originating from the AGN alters the \oiiihbeta and $\log$PAH(11.3/7.7) ratios.}
\label{f:PAH_ionized_gas_relation_AGN_non_AGN_comparison_SII_and_OIII}
\end{figure*}

In \ref{app:anomalous_pahs} we examine the spatial distribution of the anomalous pixels with respect to the \halpha surface brightness and CO flux (figures \ref{f:Halpha_of_anomalous_PAH_ratios} and \ref{f:CO_of_anomalous_PAH_ratios}). The pixels are generally located in regions with little \halpha emission, in the so-called star formation deserts of the galaxies. Pixels with the highest values of $\log$PAH(11.3/7.7)$\sim$0.6 dex are typically located outside the bar lanes, where the CO emission is weaker. Although the pixels in the anomalous group show a variation in $\log$PAH(11.3/7.7) with varying $I(\mathrm{CO})$, the distributions of $I(\mathrm{CO})$ and $I(\mathrm{CO})$/f(\halpha) are similar between the comparison and the anomalous groups. In both cases, $I(\mathrm{CO})$ and $I(\mathrm{CO})$/f(\halpha) are consistent with values seen in the diffuse gas, and cannot be used to differentiate between pixels that follow and those that do not follow the PAH band--optical line ratios correlation.

Next, we attempt to identify the process that drives the range of observed $\log$PAH(11.3/7.7) ratios in the anomalous group. Within the anomalous group, $\log$PAH(11.3/7.7) does not vary significantly with the \siihalpha ratio. We do not find significant relations between $\log$PAH(11.3/7.7) and the \halpha surface brightness, CO surface brightness, or $I(\mathrm{CO})$/f(\halpha). Unfortunately, the CO flux in these regions is not bright enough to deduce reliable kinematic information, so it is not clear whether the range in anomalous $\log$PAH(11.3/7.7) ratios is related to the cold molecular gas velocity or velocity dispersion. Among the different multi-wavelength features available in these regions, the only significant correlation we find is with the stellar-to-PAH/dust emission ratios, defined using F200W/F770W and F200W/F2100W. The former, F200W/F770W, inversely depends on F770W, so a correlation with $\log$PAH(11.3/7.7) is expected. The latter is a ratio of the stellar emission to hot (small grain) dust continuum, and may represent the ratio of available photons that heat the dust (F200W) to the dust column density (F2100W)\footnote{We find weaker correlations when considering F200W or F2100W separately.}. This is similar to observations in the bulge of M31 (e.g., \citealt{groves12, draine14}). In \ref{app:anomalous_pahs} we examine how the correlation depends on our assumptions about the hot dust continuum emission between 7.7 \mic and 21 \mic. 

The anomalous pixels trace regions with old and bright (relative to the dust) stellar populations, resembling the stellar populations in elliptical galaxies. Spitzer observations of elliptical galaxies revealed two main types of mid-infrared spectra: (1) spectra that show no PAH emission, with a strong emission feature from silicate at 9.7 \mic, and (2) spectra with PAH emission, often with unusually strong 11.3 \mic PAH band compared to the 7.7 \mic band (e.g., \citealt{kaneda05, bressan06, kaneda08, rampazzo13}). In Figure \ref{f:anomalous_PAH_ratios_silicates_vs_PAHs} we compare the stacked HST+JWST SED of the pixels with anomalous PAH ratios to two Spitzer spectra of elliptical galaxies, one dominated by silicate emission, and the other by PAH emission. The Spitzer spectra are extracted from {\sc CASSIS}\footnote{\url{https://cassis.sirtf.com/atlas/query.shtml}}, most recently described in \citet{lebouteiller11}. The stacked SEDs of the anomalous PAH groups are inconsistent with spectra that are dominated by silicate emission, since they show F1000W$<$F1130W, which is inconsistent with a broad emission that peaks at $\sim$10 \mic. Therefore, our adopted interpretation is that the regions showing anomalous $\log$PAH(11.3/7.7) ratios are indeed dominated by PAHs, rather than by silicate or hot dust continuum emission, and they show unusually high 11.3 \mic PAH feature compared to the 7.7 \mic. In Section \ref{sec:discussion} we propose different explanations for the increased PAH band ratio.

\begin{figure*}
	\centering
\includegraphics[width=1\textwidth]{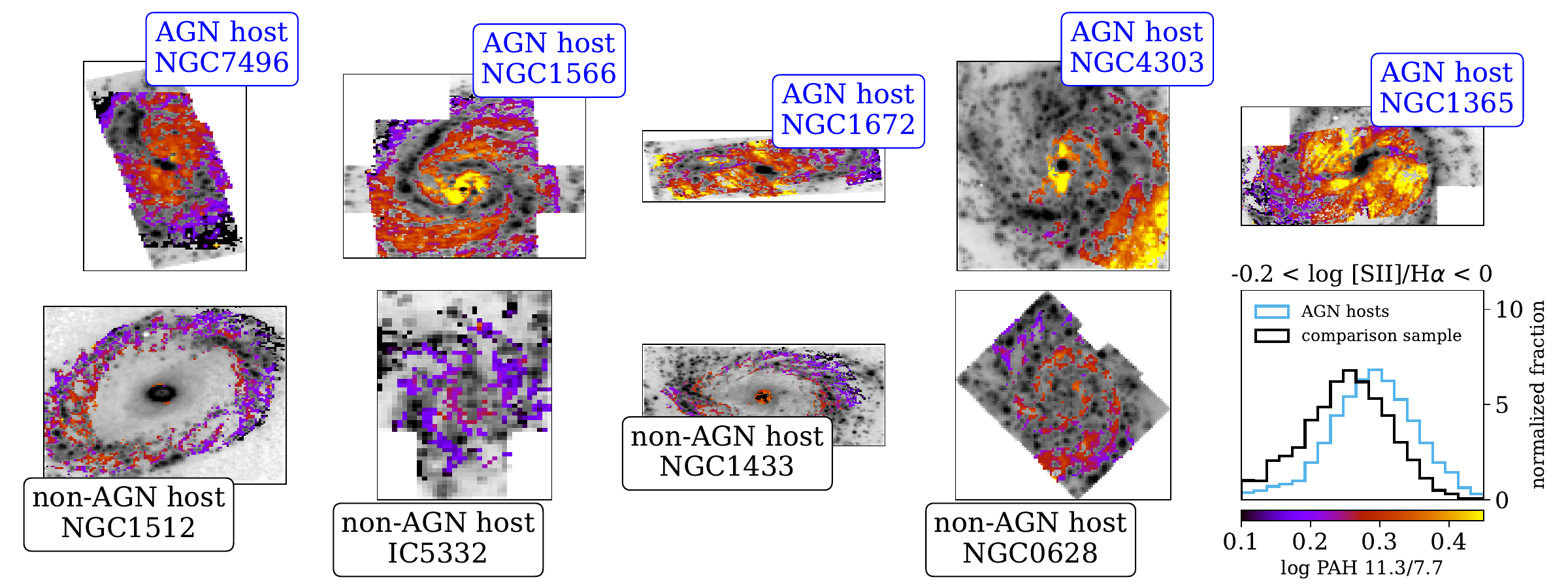}
\caption{\textbf{Distribution of PAH ratios in AGN and non-AGN hosts for high \siihalpha values.} The top row shows the five galaxies hosting AGN, and the bottom row shows the four galaxies selected as our comparison sample. The grayscale background represents the surface brightness of the \halpha line. The bottom right-most panel compares the distribution of $\log$PAH(11.3/7.7) in AGN and non-AGN hosts, for \siihalpha in the range -0.2 to 0. This is the range where a difference in the PAH ratios has been observed in Figure \ref{f:PAH_ionized_gas_relation_AGN_non_AGN_comparison_SII_and_OIII}. In addition, in non-AGN galaxies, this range represents diffuse ionized gas. This panel shows that $\log$PAH(11.3/7.7) is about 0.05 dex larger in AGN hosts than in non-AGN. The purple-to-yellow color-coding represents the $\log$PAH(11.3/7.7) ratios in the $-0.2 <$ \siihalpha $< 0$ range. The higher $\log$PAH(11.3/7.7) ratios observed in AGN hosts are located on kpc scales, and in some cases (NGC~1365 and NGC~7496), may originate from AGN ionization cones. See Figure \ref{f:log_OIII_hbeta_distribution_AGN_and_non_AGN} for the distribution of \oiiihbeta ratios.} 
\label{f:PAH_ratio_distribution_AGN_and_non_AGN}
\end{figure*}

\begin{figure*}
	\centering
\includegraphics[width=1\textwidth]{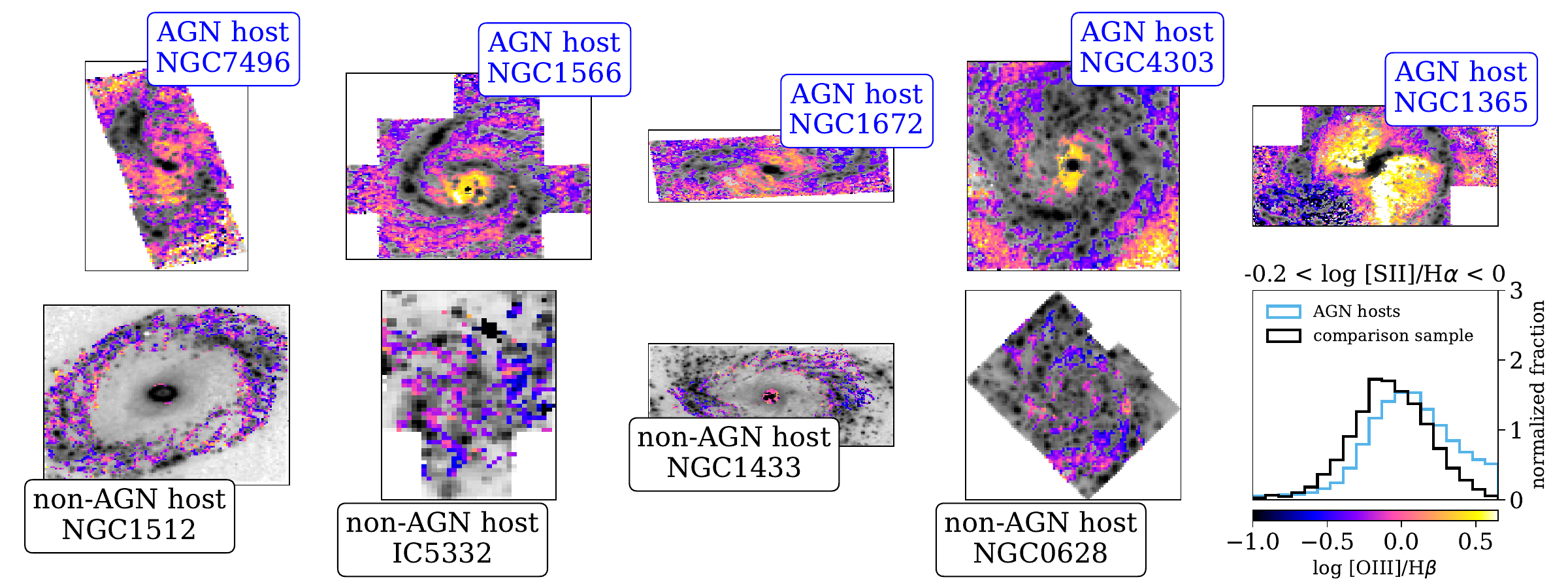}
\caption{\textbf{Distribution of \oiiihbeta ratios in AGN and non-AGN hosts for high \siihalpha values.} Similar to Figure \ref{f:PAH_ratio_distribution_AGN_and_non_AGN} but showing the \oiiihbeta ratio instead of $\log$PAH(11.3/7.7).  
Higher $\log$PAH(11.3/7.7) ratios observed in AGN hosts correspond to higher \oiiihbeta ratios, which indicates an increased contribution of the hard ionizing radiation by the AGN.}
\label{f:log_OIII_hbeta_distribution_AGN_and_non_AGN}
\end{figure*}

\subsection{Small $\log$PAH(11.3/7.7) enhancement in AGN hosts}\label{sec:results:AGN_hosts}

Out of the 19 PHANGS galaxies considered in this study, 5 host Seyfert nuclei in their center: NGC~1365, NGC~1566, NGC~1672, NGC~4303, and NGC~7496. In Section \ref{sec:results:correlations} we find that they show on average steeper slopes in their $\log$PAH(11.3/7.7) versus \siihalpha relations compared to non-AGN host galaxies. In this section, we compare the PAH and ionized gas properties of the AGN hosts to those of a comparison sample, with the goal of identifying the reason for the increased steepness. We focus on two main questions: (1) Are the observed slopes steeper due to a change in the PAH band ratios or the \siihalpha\ line ratios? (2) Is the AGN radiation responsible for the change in PAH-to-optical line ratios, and thus for the steeper slopes?

Our AGN sample includes all the pixels from the 5 AGN hosts, regardless of their main source of ionizing radiation. We build a comparison sample that includes that four galaxies NGC~0628, NGC~1433, NGC~1512, and IC~5332, that show relatively average slopes ($\beta \sim 0.22$). In both cases, we exclude pixels above the threshold $\log$(F200W/F770W)$>0.4$. In addition to the $\log$PAH(11.3/7.7) and \siihalpha ratios, we also use the \oiiihbeta ratio, as it is a more sensitive tracer of the hard ionizing spectrum originating from an AGN. 

As a first test, we use the environmental masks and exclude all pixels identified as bars and/or centers. We find no difference in the $\log$PAH(11.3/7.7) versus \siihalpha relations before and after excluding the bars and centers, suggesting that the difference in slopes between AGN and non-AGN hosts is not related to the presence of bars or due to the inclusion of the galaxies' centers.

In Figure \ref{f:PAH_ionized_gas_relation_AGN_non_AGN_comparison_SII_and_OIII} we show $\log$PAH(11.3/7.7) versus \siihalpha for the AGN and comparison samples, where the AGN group indeed shows a steeper slope. The steeper slope is due to a $\sim 0.05$ dex increase in $\log$PAH(11.3/7.7) at a fixed \siihalpha ratio of about $\sim$-0.1. This suggests that the increased slope is due to a slight increase in the PAH band ratio, rather than the ionized line ratio \siihalpha. In the middle panel, we show  $\log$PAH(11.3/7.7) versus \oiiihbeta for the AGN and comparison samples. Due to the higher ionization potential of $\mathrm{O^{++}}$, the \oiiihbeta ratio is a more sensitive tracer of the hard ionizing radiation originating from AGN. The panel shows that the increase in $\log$PAH(11.3/7.7) corresponds to increased \oiiihbeta. The distributions in these panels strongly resemble those we found for the first three PHANGS galaxies (Figures 12 and 13 in \citetalias{baron24}). In the right panel, we compare the AGN and non-AGN groups in the line diagnostic diagram \oiiihbeta-\siihalpha (\citealt{baldwin81, veilleux87, kewley01, kauff03a}). The excess in $\log$PAH(11.3/7.7) that causes steeper slopes in AGN hosts, which corresponds to increased \oiiihbeta ratios in the middle panel, corresponds to optical line ratios consistent with Seyfert ionization. It suggests that the increase in \oiiihbeta and, probably, in $\log$PAH(11.3/7.7) is due to an increasing contribution of the AGN to the total radiation field affecting the ionized gas and PAHs. 

In Figure \ref{f:PAH_ratio_distribution_AGN_and_non_AGN} we study the spatial distribution of regions showing elevated $\log$PAH(11.3/7.7) ratios in the AGN hosts, and compare them to those observed in non-AGN hosts in the diffuse ionized gas. We focus on regions where $-0.2 <$ \siihalpha $< 0$, which is the range where we find a difference between AGN and non-AGN in Figure \ref{f:PAH_ionized_gas_relation_AGN_non_AGN_comparison_SII_and_OIII}. This range corresponds to diffuse ionized gas regions in the non-AGN hosts. First, comparing the $\log$PAH(11.3/7.7) distributions in AGN and non-AGN hosts in the range $-0.2 <$ \siihalpha $< 0$, we find a small $\log$PAH(11.3/7.7) enhancement in the AGN hosts, of about $\sim$0.05 dex, corresponding to $\sim$10\%. Comparing the spatial distribution of the $\log$PAH(11.3/7.7) ratios in AGN and non-AGN hosts, we find: (1) The enhanced $\log$PAH(11.3/7.7) ratios are seen on kpc scales, and (2) In some cases (e.g., NGC~1365 and NGC~7496), the enhanced ratios may originate from regions along the AGN ionization cones, as can be seen in the \oiiihbeta maps in Figure \ref{f:log_OIII_hbeta_distribution_AGN_and_non_AGN}. This suggests a global impact of the AGN on the PAH band and optical line ratios, on kpc scales, and not only in the galaxies' centers. The effect is very small, of the order of $\sim$10\% in PAH(11.3/7.7), suggesting that the old stellar population in the bulges is the dominant heating source of the PAHs.

In \ref{app:agn_SEDs} we compare the HST+JWST stacked SEDs of AGN and non-AGN hosts, and find that their ultraviolet-optical stellar continua are comparable, suggesting that the difference in $\log$PAH(11.3/7.7) is not driven due to a variation in the stellar population properties. We also confirm that the variation is not driven by an increased contribution of hot dust continuum emission to the F1130W filter.

Finally, we emphasize that the analysis in this section and our conclusions apply to PAHs and ionized gas in diffuse regions on kpc scales, as these dominate in number of pixels considered. On very small scales close to the accreting black holes, the AGN radiation field may have additional impacts on the PAH band ratios (see e.g., \citealt{jensen17, lai22, donnelly24}). We leave the analysis of the galaxy centers to a future work.

\section{Discussion}\label{sec:discussion}

\begin{figure*}
	\centering
\includegraphics[width=1\textwidth]{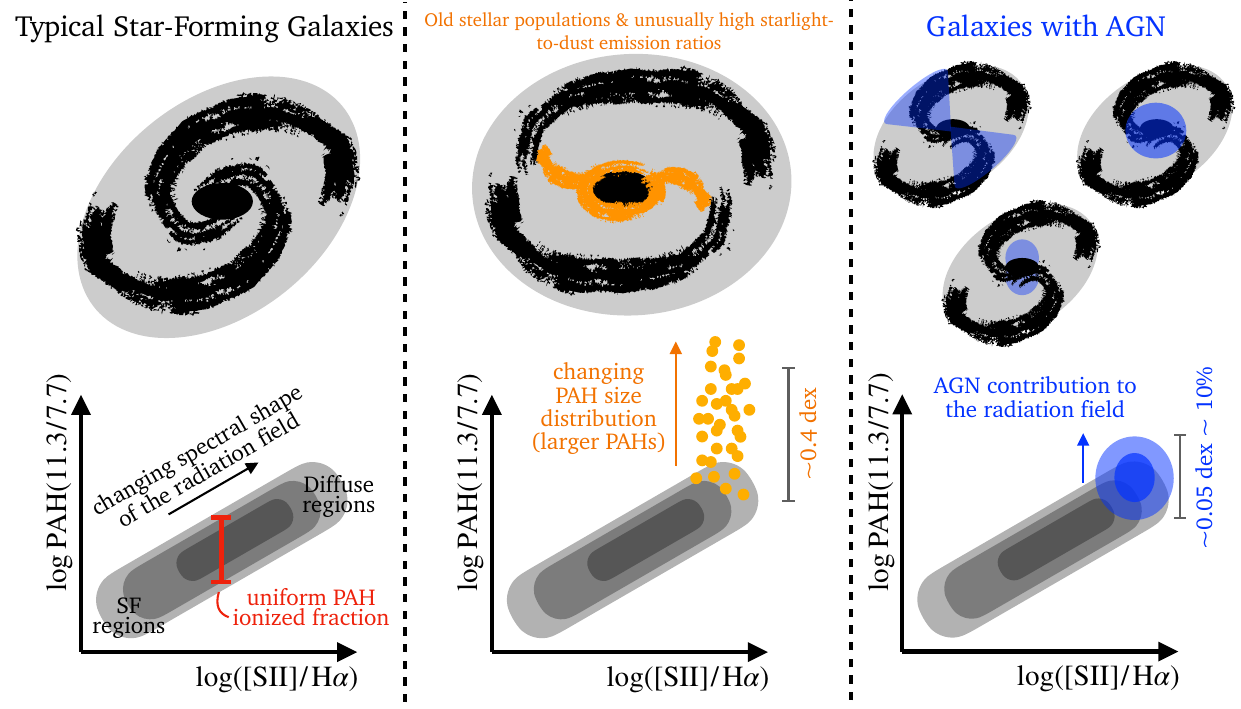}
\caption{\textbf{Emerging picture of PAH band--optical line ratios correlations across nearby star-forming galaxies.} \textbf{The left panel} depicts a typical nearby star-forming galaxy, where there is a tight correlation between $\log$PAH(11.3/7.7) and \siihalpha ratios on 40--150 pc scales. Similarly tight correlations are observed with other optical line ratios: \oiiihbeta, \niihalpha, and \oihalpha. The bottom left part of the relation is dominated by star-forming regions, where young massive stars ionize the gas. The top right part of the relation corresponds to diffuse regions on kpc scales, where a combination of radiation leaking from HII regions and from hot and evolved stars ionizes the gas, giving rise to LINER/LIER-like optical line ratios. The slope of the relation is $\beta \sim 0.2$, roughly matching the expected relation for a varying radiation field that heats the PAHs and ionizes the gas (see \citetalias{baron24}). The small scatter in the relation, $\sim$ 0.025 dex, suggests uniform PAH ionization fraction across different environments and galaxies. \textbf{The middle panel} represents a small group of pixels observed in a handful of galaxies, showing anomalously-high $\log$PAH(11.3/7.7) ratios for a constant \siihalpha ratio. Such PAH ratios are observed in regions with old stellar populations and unusually high starlight-to-dust emission ratio, $\log$F200W/F770W$>$ 0.4 dex. The anomalous ratios are consistent with PAH populations with larger grain sizes, which can be either the result of processes that change the typical ISM grain population as it streams into the center (destruction via shocks or under-production in grain-grain collisions), or the result of a varying mixing between two grain populations: those of the typical dusty ISM and those produced in AGB star atmospheres. \textbf{The right panel} shows a star-forming galaxy with a low luminosity AGN in its center. The AGN contributes to the total radiation field affecting the PAHs and ionized gas on kpc scales, resulting in a small enhancement of $\log$PAH(11.3/7.7)$\sim$0.05 dex$\sim$10\%. The AGN has little impact on the \siihalpha ratio, but a stronger impact on the \oiiihbeta ratio, with the two optical line ratios consistent with Seyfert-like ionization in standard line diagnostic diagrams.}
\label{f:emerging_picture_PAH_ionized_gas}
\end{figure*}

In Figure \ref{f:emerging_picture_PAH_ionized_gas} we outline the emerging picture of PAH band--optical line ratios correlations across nearby star-forming galaxies. Typical nearby galaxies show tight correlations between $\log$PAH(11.3/7.7) and the optical line ratios \oiiihbeta, \niihalpha, \siihalpha, and \oihalpha. These correlations are based on observations of $\sim$700\,000 spatially independent regions in 19 galaxies probing scales of 40--150 pc. The correlation is a sequence in ionized gas properties (see \citetalias{baron24}), where small $\log$PAH(11.3/7.7) and optical line ratios are observed in star-forming regions, with optical line ratios consistent with ionization by young and massive stars, and large $\log$PAH(11.3/7.7) and optical line ratios originating in diffuse regions on kpc scales, where the LINER/LIER-like optical line ratios are consistent with
the ionizing radiation being a combination of radiation leaking from HII regions and radiation of hot and evolved stars.

The correlation is universal across our sample, with a best-fitting slope around $\beta \sim 0.2$ for $\log$PAH(11.3/7.7) versus \siihalpha, and with a scatter of about $\sim$0.025 dex, which does not vary significantly from galaxy to galaxy. Such a tight and global correlation may be surprising a-priori, as the PAH mid-infrared and optical line spectra depend on multiple properties that are expected to vary within the ISM of galaxies. In particular, the PAH 11.3/7.7 ratio strongly depends on the PAH ionized fraction, and shows a secondary dependence on the PAH size distribution. Both of these properties are expected to be sensitive to the intensity of the interstellar radiation field and the ISM characteristics (e.g., \citealt{draine01, hony01, ohalloran06, draine07, gordon08, boersma16, croiset16, peeters17, boersma18, maragkoudakis20, draine21, knight21, maragkoudakis22}; see reviews \citealt{tielens08, li20}). More recently, the PAH ratio was also shown to depend on the spectral shape of the radiation field, which can alter the PAH temperature distribution and thus the relative emission in the different bands (\citealt{draine21, rigopoulou21}). The optical line ratios show a strong dependence on the gas-phase metallicity, in particular the nitrogen and oxygen abundances, the shape of the ionizing radiation field, and the ionization parameter (e.g., \citealt{pilyugin04, liang06, perez-montero09, marino13, steidel14, blanc15, steidel16, vincenzo16, byler17} and review by \citealt{kewley19}). 

The universal and tight relation between the PAH band and optical line ratios suggests a one-dimensional sequence driven by a single varying property. The small scatter and minimal galaxy-to-galaxy variations observed in the relation suggest that other properties are either relatively uniform, or that they do not impact the observed ratios as much as the leading varying property. In \citetalias{baron24} we suggested that the correlation is driven by the changing shape of the radiation field, which is a mixture of radiation of massive and young stars as well as hot and evolved stars. We combined PAH models with a large set of model stellar SEDs to show that the observed slope of the PAH band--optical line ratios relation roughly matches the expected value when varying the radiation field from being dominated by young to being dominated by old stellar populations. In this picture, as we increase the contribution of the old stellar population with respect to the young, the ionizing part of the radiation field becomes harder, leading to increased electron temperatures, which results in increasing optical line ratios. At the same time, the non-ionizing ultraviolet radiation that heats the PAHs becomes softer, leading to colder PAHs and increasing $\log$PAH(11.3/7.7) ratios. The total PAH and optical line fluxes decrease as expected since younger stellar populations are much brighter than older ones. On the scales probed here, the strength of the interstellar radiation field is not expected to impact the $\log$PAH(11.3/7.7) ratio (see figure 19 in \citealt{draine21}).

An alternative interpretation is that the $\log$PAH(11.3/7.7) ratio varies primarily due to a change in the PAH charge distribution, with more ionized PAHs close to star-forming regions and more neutral PAHs in the diffuse parts of the galaxies. We prefer the interpretation of the varying shape of the radiation field as it is unavoidable -- the radiation field is observed to vary in HST UV-optical images from young to old stellar populations, with the optical line ratios confirming the observed variations, and this variation alone can already account for the observed $\log$PAH(11.3/7.7) ratio variation. Therefore, a major implication of our adopted interpretation is that the PAH charge distribution is surprisingly uniform on scales of 40--150 pc, as it can vary only within the observed $\sim$0.025 dex scatter observed in the $\log$PAH(11.3/7.7) versus \siihalpha relation. For comparison, the three possible PAH ionization fractions assumed by \citet[Figure 9b]{draine21} lead to a much larger variation in $\log$PAH(11.3/7.7), of $\sim 0.2\,\mathrm{dex}$. 

Visualizing the optical-infrared feature space with {\sc pca}, we identify a group of pixels where the PAH band--optical line ratios relation breaks down (depicted in the middle panel of Figure \ref{f:emerging_picture_PAH_ionized_gas}). We identify these pixels primarily in 4 out of the 19 galaxies, where they constitute 10--25\% of the spatially independent pixels within each galaxy. In the other galaxies, their fraction ranges between 5\% to less than 2\%. These regions show unusually high $\log$PAH(11.3/7.7) ratios, with up to $\sim 0.4$ dex increase with respect to the expected ratio from the universal PAH band--optical line ratios correlation. The anomalous PAH ratios are found in regions with old ($\sim 10^{10}$ Gyr) stellar populations and high starlight-to-dust mid-infrared emission ratios of $\log$(F200W/F770W)$\;>\;$0.4 dex. They originate in the so-called `star-formation deserts' of the galaxies, which show very little \halpha emission. The unusual PAHs are detected close to the CO-traced bar lanes that are likely associated with strong radial streaming motions, which may suggest that shocks produced by the inflowing gas may play a role. However, the highest $\log$PAH(11.3/7.7) values are farther away from the bar lanes, in regions with fainter CO emission.The PAH band ratios show a clear dependence on the stellar-to-dust emission ratio F200W/F2100W, which may trace the ratio of stellar photons that heat the dust to the dust column density. Our stacked HST+JWST SEDs in these regions suggest that they are regions with peculiar PAH spectra rather than spectra that are dominated by silicate 9.7 \mic emission or by hot dust continuum emission. 

The elevated $\log$PAH(11.3/7.7) ratios may be caused by a change in the PAH size distribution, where compared to the size distribution in the typical ISM, the smallest PAHs are missing (e.g., \citealt{smith07, diamond_stanic10, draine21, rigopoulou21}). Smaller PAHs can be absent either because they have been destroyed by shocks, which destroy smaller grains more efficiently than larger ones (e.g., \citealt{micelotta10}), or because shattering of larger grains, which is believed to be a formation channel of the small grains, is occurring at a slower rate (e.g., \citealt{hirashita09, hirashita10}). Another possible explanation is that the pixels corresponding to the anomalous group trace two PAH populations projected onto the same line of sight: (i) typical PAHs seen throughout the ISM, flowing along the bar lanes onto the centers, and (ii) PAHs with larger average sizes that are located in the bulges of these galaxies, which are dominated by the old stellar population. Indeed, mid-infrared observations in elliptical galaxies with old stellar populations reveal unusually high PAH 11.3/7.7 \mic band ratios (see e.g., \citealt{kaneda05, bressan06, kaneda08, vega10, rampazzo13}). In this case, the high starlight-to-dust emission ratio may suggest an increased contribution of the second population to the PAH bands, thus increasing the $\log$PAH(11.3/7.7) ratio.

An alternative interpretation is that the elevated $\log$PAH(11.3/7.7) ratios correspond to regions with more neutral PAHs, but we find this unlikely as we do not find any difference in CO/\halpha ratios in this group compared to those observed in typical diffuse regions. In addition, the regions showing the highest $\log$PAH(11.3/7.7) ratios correspond to regions with the lowest gas and dust surface density, where less shielding is taking place. Distinguishing between the different scenarios requires spatially resolved mid-infrared spectra that include the various PAH transitions, shock-tracing emission lines such as the rotational $\mathrm{H_{2}}$, and lines tracing the kinematics of the gas as it transitions from the bar lanes to the gas and dust disc. 

The unusual PAH 11.3/7.7 ratio may be connected to two presumably separate observational peculiarities of PAHs from the Spitzer era: unusually high PAH 11.3/7.7 ratios in low-luminosity AGN and in elliptical galaxies (e.g., \citealt{kaneda05, smith07, galliano08, kaneda08, diamond_stanic10, vega10, zhang22}). Various explanations have been put forward to explain these peculiarities, such as (i) a hard AGN spectrum directly modifying the PAH grain size distribution and even serving as the main excitation source for PAH emission, (ii) destruction of the smaller PAH grains in shocks, and (iii) an increase in the fraction of neutral-to-charged PAHs. The common property in all three cases -- anomalous PAH ratios in the PHANGS galaxies; in low luminosity AGN hosts; and in elliptical galaxies -- is the presence of old stellar populations and unusually high starlight-to-dust emission ratios. As discussed in Section \ref{sec:results:AGN_hosts}, the presence of a low luminosity AGN has a small impact on the observed PAH 11.3/7.7 ratio in our case, altering $\log$PAH(11.3/7.7) by $\sim 0.05$ dex ($\sim$10\%). This is compared to a factor of 2--10 difference observed in Spitzer IRS spectra by \citet{smith07} and in the anomalous PAH ratio group in this work. We therefore suggest that the peculiar PAH 11.3/7.7 ratios observed in AGN in the Spitzer era may be connected to the stellar population in the bulges of these hosts, rather than the AGN itself, grouping all these peculiarities into a single group. Testing this hypothesis requires re-analysis of the Spitzer IRS spectra and the inclusion of new JWST MIRI-MRS spectra, where the presence of a low luminosity AGN and of a bright and old stellar population can be disentangled for a large statistical sample.

Five of the nineteen PHANGS galaxies host an AGN in their center. These five galaxies show steeper slopes in their $\log$PAH(11.3/7.7) versus \siihalpha relations, which is the result of a slight ($\sim$0.05 dex, corresponding to $\sim$10\%) enhancement of the $\log$PAH(11.3/7.7) ratio for $-0.2 < $\siihalpha$< 0$ (depicted in the right panel of Figure \ref{f:emerging_picture_PAH_ionized_gas}). The regions showing the enhancement also show Seyfert-like \oiiihbeta versus \siihalpha line ratios in standard line diagnostic diagrams. They are distributed on kpc scales, and in some cases, may originate in the AGN ionization cones. This suggests that the AGN drives the slight $\log$PAH(11.3/7.7) enhancement. The enhancement can be due to the contribution of the AGN to the total radiation field on kpc scales, which can potentially alter the PAH temperature distribution and thus the relative emission in the different bands (e.g., \citealt{smith07, donnelly24}); or a slight change in the PAH size distribution due to interaction of AGN-driven winds with diffuse gas clouds (e.g., \citealt{diamond_stanic10, zhang22}). Regardless of the physical process responsible for the change, we highlight that it has a minor impact on the observed PAH ratios, of the order of 10\%, which is in a stark contrast to the large impact observed in some other AGN hosts by Spitzer (e.g., \citealt{smith07, diamond_stanic10, zhang22}). It is the combination of spatially resolved optical spectroscopy and mid-infrared imaging over unprecedentedly small scales of 40--150 pc that allows us to isolate such a minor effect for the first time.

The near-infrared coverage of JWST allows to measure the 3.3 \mic PAH feature throughout the galaxies, which can be used to place additional constraints on the PAH size variation and temperature distribution. The PAH 3.3 \mic/11.3 \mic band ratio and its relation to the optical line ratios can be used to (i) confirm our interpretation that the PAH band--optical line ratio relation is driven by the varying radiation field (left panel of Figure \ref{f:emerging_picture_PAH_ionized_gas}), and (ii) refine our interpretation about the slight increase in slopes seen in AGN hosts (right panel of Figure \ref{f:emerging_picture_PAH_ionized_gas}). At this stage, it is unclear whether the 3.3 \mic PAH feature is detected in pixels showing unusual PAH 11.3/7.7 ratios (Koziol et al. in prep.). An analysis of the variation of the PAH 3.3 \mic/11.3 \mic band ratio with varying optical line ratios will be presented in a future study.

\section{Summary}\label{sec:summary}

In this paper, we use the PHANGS survey to establish a census of the relation between PAH heating and gas ionization across nearby star-forming galaxies. We particularly focus on the PAH band ratio 11.3 \mic/7.7 \mic versus \siihalpha relation, though strong correlations are seen with other optical line ratios as well. We use the 19 PHANGS galaxies with high-resolution JWST and MUSE-VLT observations, and at the limiting angular resolution of MUSE-VLT (0.56--1.2\arcsec), we have $\sim$700\,000 spatially independent pixels probing PAH and optical line emission on scales of 40--150 pc. Our results and their implications are summarized below. They are also depicted in Figure \ref{f:emerging_picture_PAH_ionized_gas}.\vspace{0.3cm}

\noindent \textbf{(I) There is a universal relation between PAH 11.3 \mic/7.7 \mic and \siihalpha across star-forming galaxies on 40--150 pc scales (Section \ref{sec:results:correlations} and Figures \ref{f:correlations_sep_ordered_by_slope} and \ref{f:best_fitting_slope_distribution}).} For each galaxy, we use images convolved to the MUSE optimal angular resolution that translates to different scales in the range 40--120 pc per source, as well as images convolved to a uniform spatial resolution of 150 pc. We fit linear relations between $\log$PAH(11.3/7.7) and \siihalpha to all the spatially independent pixels of each galaxy separately. The best-fitting slopes are 0.16--0.38 for the MUSE optimal resolution, and 0.19--0.40 for the 150 pc resolution. There is no significant dependence of the slope on the spatial scale probed, and the correlation is well-detected on scales as small as 40 pc.  Among the seven galaxies with the steepest slopes, five are systems with identified Seyfert nuclei. Putting aside anomalous regions with unusually high starlight-to-dust emission ratios (see below) and AGN hosts, nearby star-forming galaxies show a universal relation with a slope of $\beta \sim 0.2$ and a scatter of $\sim 0.025$ dex around the best-fitting relation. 

PAH band and optical line ratios depend on multiple properties that are expected to vary in the ISM of star-forming galaxies (PAH charge and size, gas-phase abundances, ionization parameter, spectral shape of the interstellar radiation field). The universal relation identified between $\log$PAH(11.3/7.7) and \siihalpha suggests a one-dimensional sequence that is driven by the variation of a single leading property. The small and uniform scatter suggests that other properties are either relatively uniform on these scales, or that they have a smaller impact on the observed PAH band and optical line ratios. Our interpretation is that the relation is driven by the changing interstellar radiation field, which is a mixture of radiation originating from young massive stars and radiation of hot and evolved stars. The expected slope in this scenario roughly matches the observed $\beta \sim 0.2$ slope. This interpretation implies that the PAH ionization fraction is surprisingly uniform on 40--150 pc scales, accounting for a variation in $\log$PAH(11.3/7.7) of  $\sim 0.025$ dex. For comparison, the three examples of PAH ionization fractions examined by \citet{draine21}, where they note that it would not be surprising if even larger variations are observed, result in a much larger variation of $\sim 0.2$ dex.\vspace{0.3cm}

\noindent \textbf{(II) Regions with old stellar populations and high starlight-to-dust emission ratios do not follow the universal relation, showing unusually high PAH 11.3 \mic/7.7 \mic ratios resembling those observed in elliptical galaxies (Section \ref{sec:results:anomalous_PAHs} and Figure \ref{f:stacked_SEDs_weird_PAHs_fnu}).} We use {\sc pca} to visualize the optical-infrared feature space, and identify a group of anomalous pixels showing unusually high PAH ratios ($\sim$$\times 2$) that do not follow the universal PAH band--optical line ratios correlation. These pixels constitute 10--25\% of the pixels in four of the PHANGS galaxies, and a smaller fraction of 5\% to $< 2$\% in the rest. They trace regions with old stellar populations ($\sim 10$ Gyr) and high stellar-to-mid infrared dust emission ($\log$F200W/F770W$>$0.4 dex), in the so-called `star-formation deserts'. The PAH emission is detected close to or along CO-traced bar lanes that feed the centers of these galaxies, that are associated with strong radial gas motions. The unusually high PAH ratios show a clear spatial variation with respect of where the filaments meet the gas and dust discs, which may suggest that shocks play a role in shaping the properties of the mid-infrared-emitting population. The ratios also show a strong dependence on the stellar-to-small grain continuum emission ratio F200W/F2100W. The HST+JWST stacked SEDs of these regions are inconsistent with the emission being powered by silicate emission or purely by small grain continuum emission. Instead, the SEDs resemble mid-infrared spectra of PAHs with peculiar ratios observed in elliptical galaxies. A possible explanation for the unusually high PAH ratios is a change in the observed PAH size distribution, where the typical grain population in the dusty ISM is altered due to processes such as shocks or grain shattering. Alternatively, an increased contribution to the mid-infrared emission by a second population of (larger) grains that is associated with evolved stellar populations, typically seen in early-type galaxies, may account for the observed ratios. \vspace{0.3cm}

\noindent \textbf{(III) AGN hosts in our sample show slightly steeper PAH 11.3 \mic/7.7 \mic versus \siihalpha relations that are the result of a small ($\sim$10\%) increase in the PAH 11.3 \mic/7.7 \mic ratio in regions exposed to AGN radiation on kpc scales (Section \ref{sec:results:AGN_hosts} and Figure \ref{f:PAH_ionized_gas_relation_AGN_non_AGN_comparison_SII_and_OIII}).} Out of the 19 galaxies in our sample, 5 host AGN with identified Seyfert nuclei. These sources show steeper PAH band--optical line ratios relations with $\beta$ of 0.25--0.38, which are the result of a small increase in the PAH 11.3 \mic/7.7 \mic ratio for high \siihalpha ratios in the range -0.2 to 0. The regions showing elevated PAH ratios show \oiiihbeta versus \siihalpha ratios consistent with Seyfert in standard line diagnostic diagrams, suggesting that the AGN contributes to the total radiation field that ionizes the gas, and possibly to PAH heating. The regions are identified on kpc scales, and in some cases, display a cone-like morphology resembling AGN ionization cones. The effect of the AGN on the PAH ratio is very small, of the order of 10\%, in contrast to other AGN hosts where PAH ratios are observed to vary significantly. It is the combination of high angular resolution imaging with observations across the electromagnetic spectrum that allows us to identify such a relatively minor effect.\vspace{0.3cm}

The emerging picture is that there is a strong connection between PAH heating and gas ionization across different environments in the ISM of star-forming galaxies (Figure \ref{f:emerging_picture_PAH_ionized_gas}). The tight relation between $\log$PAH(11.3/7.7) and \siihalpha is driven by the varying radiation field, while other properties, such as the PAH ionized fraction, remain relatively uniform on 40--150 pc scales. This implies that the $\log$PAH(11.3/7.7) versus optical line ratios diagrams can potentially be used to probe the interstellar radiation field in a unique way -- while optical line ratios diagrams are only sensitive to the spectral shape of the ionizing radiation field, PAHs are sensitive to the shape of the non-ionizing ultraviolet radiation field. Combining the 11.3 \mic/7.7 \mic PAH ratio with optical line ratios may therefore place novel constraints on the ionizing+non-ionizing spectra of young and old stars. We plan to explore this possibility in a future work. 

\acknowledgments{

We are grateful to our anonymous referee for suggestions that helped improve the paper structure and presentation. 

D. Baron is grateful to L. Armus, G. Donnelly, B. Draine, B. Hensley, T. Lai, and J. D. Smith, for discussions that significantly shaped the interpretations and conclusions presented in this paper.

During the work on this project, D. Baron was supported by the Carnegie-Princeton fellowship. JS, KS acknowledge funding from JWST-GO- 2107.006-A. MB gratefully acknowledges support from the ANID BASAL project FB210003 and from the FONDECYT regular grant 1211000. This work was supported by the French government through the France 2030 investment plan managed by the National Research Agency (ANR), as part of the Initiative of Excellence of Université Côte d'Azur under reference number ANR-15-IDEX-01. JPe acknowledges support by the French Agence Nationale de la Recherche through the DAOISM grant ANR-21-CE31-0010. JC acknowledges funding from the Belgian Science Policy Office (BELSPO) through the PRODEX project ``JWST/MIRI Science exploitation'' (C4000142239). MC gratefully acknowledges funding from the DFG through an Emmy Noether Research Group (grant number CH2137/1-1). COOL Research DAO is a Decentralized Autonomous Organization supporting research in astrophysics aimed at uncovering our cosmic origins. HH and ER acknowledge the support of the Canadian Space Agency (23JWGO2A07), and the support of the Natural Sciences and Engineering Research Council of Canada (NSERC), funding reference number RGPIN-2022-03499. RSK acknowledges financial support from the European Research Council via the ERC Synergy Grant ``ECOGAL'' (project ID 855130),  from the Heidelberg Cluster of Excellence (EXC 2181 - 390900948) ``STRUCTURES'', funded by the German Excellence Strategy, and from the German Ministry for Economic Affairs and Climate Action in project ``MAINN'' (funding ID 50OO2206). RSK is grateful to the Harvard Radcliffe Institute for Advanced Studies and Harvard-Smithsonian Center for Astrophysics for their hospitality and support during his sabbatical. The team in Heidelberg also thanks for computing resources provided by the Ministry of Science, Research and the Arts (MWK) of {\em The L\"{a}nd} through bwHPC and the German Science Foundation (DFG) through grant INST 35/1134-1 FUGG and 35/1597-1 FUGG, and also for data storage at SDS@hd funded through grants INST 35/1314-1 FUGG and INST 35/1503-1 FUGG.}

\acknowledgments{
Some of the data presented in this paper were obtained from the Mikulski Archive for Space Telescopes (MAST) at the Space Telescope Science Institute. The specific observations analyzed can be accessed via \dataset[https://doi.org/10.17909/ew88-jt15]{https://doi.org/10.17909/ew88-jt15}. STScI is operated by the Association of Universities for Research in Astronomy, Inc., under NASA contract NAS5–26555. Support to MAST for these data is provided by the NASA Office of Space Science via grant NAG5–7584 and by other grants and contracts. 

This paper makes use of the following ALMA data: ADS/JAO.ALMA\#2012.1.00650.S, 2013.1.01161.S, 2015.1.00925.S, 2017.1.00392.S, and 2017.1.00886.L.
ALMA is a partnership of ESO (representing its member states), NSF (USA), and NINS (Japan), together with NRC (Canada), NSC and ASIAA (Taiwan), and KASI (Republic of Korea), in cooperation with the Republic of Chile. The Joint ALMA Observatory is operated by ESO, AUI/NRAO, and NAOJ. The National Radio Astronomy Observatory is a facility of the National Science Foundation operated under cooperative agreement by Associated Universities, Inc.

Some of the data used in this work is based on observations collected at the European Southern Observatory under ESO programmes 094.C-0623 (PI: Kreckel), 095.C-0473,  098.C-0484 (PI: Blanc), 1100.B-0651 (PHANGS-MUSE; PI: Schinnerer), as well as 094.B-0321 (MAGNUM; PI: Marconi), 099.B-0242, 0100.B-0116, 098.B-0551 (MAD; PI: Carollo) and 097.B-0640 (TIMER; PI: Gadotti).
}%\vspace{1.8cm }

\software{Astropy \citep{astropy13, astropy18, astropy22},
		  IPython \citep{perez07},
          scikit-learn \citep{pedregosa11}, 
          SciPy \citep{scipy01},
		  matplotlib \citep{hunter07},
		  reproject \citep{reproject_2020}}\vspace{2cm}
         
\newpage

\bibliography{ref.bib}

\begin{thebibliography}{}
\expandafter\ifx\csname natexlab\endcsname\relax\def\natexlab#1{#1}\fi
\providecommand{\url}[1]{\href{#1}{#1}}
\providecommand{\dodoi}[1]{doi:~\href{http://doi.org/#1}{\nolinkurl{#1}}}
\providecommand{\doeprint}[1]{\href{http://ascl.net/#1}{\nolinkurl{http://ascl.net/#1}}}
\providecommand{\doarXiv}[1]{\href{https://arxiv.org/abs/#1}{\nolinkurl{https://arxiv.org/abs/#1}}}

\bibitem[{{Allamandola} {et~al.}(1999){Allamandola}, {Hudgins}, \&
  {Sandford}}]{allamandola99}
{Allamandola}, L.~J., {Hudgins}, D.~M., \& {Sandford}, S.~A. 1999, \apjl, 511,
  L115, \dodoi{10.1086/311843}

\bibitem[{{Allamandola} {et~al.}(1985){Allamandola}, {Tielens}, \&
  {Barker}}]{allamandola85}
{Allamandola}, L.~J., {Tielens}, A.~G.~G.~M., \& {Barker}, J.~R. 1985, \apjl,
  290, L25, \dodoi{10.1086/184435}

\bibitem[{{Anand} {et~al.}(2021{\natexlab{a}}){Anand}, {Lee}, {Van Dyk},
  {Leroy}, {Rosolowsky}, {Schinnerer}, {Larson}, {Kourkchi}, {Kreckel},
  {Scheuermann}, {Rizzi}, {Thilker}, {Tully}, {Bigiel}, {Blanc}, {Boquien},
  {Chandar}, {Dale}, {Emsellem}, {Deger}, {Glover}, {Grasha}, {Groves}, {S.
  Klessen}, {Kruijssen}, {Querejeta}, {S{\'a}nchez-Bl{\'a}zquez}, {Schruba},
  {Turner}, {Ubeda}, {Williams}, \& {Whitmore}}]{anand21a}
{Anand}, G.~S., {Lee}, J.~C., {Van Dyk}, S.~D., {et~al.} 2021{\natexlab{a}},
  \mnras, 501, 3621, \dodoi{10.1093/mnras/staa3668}

\bibitem[{{Anand} {et~al.}(2021{\natexlab{b}}){Anand}, {Rizzi}, {Tully},
  {Shaya}, {Karachentsev}, {Makarov}, {Makarova}, {Wu}, {Dolphin}, \&
  {Kourkchi}}]{anand21b}
{Anand}, G.~S., {Rizzi}, L., {Tully}, R.~B., {et~al.} 2021{\natexlab{b}}, \aj,
  162, 80, \dodoi{10.3847/1538-3881/ac0440}

\bibitem[{{Aniano} {et~al.}(2011){Aniano}, {Draine}, {Gordon}, \&
  {Sandstrom}}]{aniano11}
{Aniano}, G., {Draine}, B.~T., {Gordon}, K.~D., \& {Sandstrom}, K. 2011, \pasp,
  123, 1218, \dodoi{10.1086/662219}

\bibitem[{{Astropy Collaboration} {et~al.}(2013){Astropy Collaboration},
  {Robitaille}, {Tollerud}, {Greenfield}, {Droettboom}, {Bray}, {Aldcroft},
  {Davis}, {Ginsburg}, {Price-Whelan}, {Kerzendorf}, {Conley}, {Crighton},
  {Barbary}, {Muna}, {Ferguson}, {Grollier}, {Parikh}, {Nair}, {Unther},
  {Deil}, {Woillez}, {Conseil}, {Kramer}, {Turner}, {Singer}, {Fox}, {Weaver},
  {Zabalza}, {Edwards}, {Azalee Bostroem}, {Burke}, {Casey}, {Crawford},
  {Dencheva}, {Ely}, {Jenness}, {Labrie}, {Lim}, {Pierfederici}, {Pontzen},
  {Ptak}, {Refsdal}, {Servillat}, \& {Streicher}}]{astropy13}
{Astropy Collaboration}, {Robitaille}, T.~P., {Tollerud}, E.~J., {et~al.} 2013,
  \aap, 558, A33, \dodoi{10.1051/0004-6361/201322068}

\bibitem[{{Astropy Collaboration} {et~al.}(2018){Astropy Collaboration},
  {Price-Whelan}, {Sip{\H{o}}cz}, {G{\"u}nther}, {Lim}, {Crawford}, {Conseil},
  {Shupe}, {Craig}, {Dencheva}, {Ginsburg}, {VanderPlas}, {Bradley},
  {P{\'e}rez-Su{\'a}rez}, {de Val-Borro}, {Aldcroft}, {Cruz}, {Robitaille},
  {Tollerud}, {Ardelean}, {Babej}, {Bach}, {Bachetti}, {Bakanov}, {Bamford},
  {Barentsen}, {Barmby}, {Baumbach}, {Berry}, {Biscani}, {Boquien}, {Bostroem},
  {Bouma}, {Brammer}, {Bray}, {Breytenbach}, {Buddelmeijer}, {Burke},
  {Calderone}, {Cano Rodr{\'\i}guez}, {Cara}, {Cardoso}, {Cheedella}, {Copin},
  {Corrales}, {Crichton}, {D'Avella}, {Deil}, {Depagne}, {Dietrich}, {Donath},
  {Droettboom}, {Earl}, {Erben}, {Fabbro}, {Ferreira}, {Finethy}, {Fox},
  {Garrison}, {Gibbons}, {Goldstein}, {Gommers}, {Greco}, {Greenfield},
  {Groener}, {Grollier}, {Hagen}, {Hirst}, {Homeier}, {Horton}, {Hosseinzadeh},
  {Hu}, {Hunkeler}, {Ivezi{\'c}}, {Jain}, {Jenness}, {Kanarek}, {Kendrew},
  {Kern}, {Kerzendorf}, {Khvalko}, {King}, {Kirkby}, {Kulkarni}, {Kumar},
  {Lee}, {Lenz}, {Littlefair}, {Ma}, {Macleod}, {Mastropietro}, {McCully},
  {Montagnac}, {Morris}, {Mueller}, {Mumford}, {Muna}, {Murphy}, {Nelson},
  {Nguyen}, {Ninan}, {N{\"o}the}, {Ogaz}, {Oh}, {Parejko}, {Parley}, {Pascual},
  {Patil}, {Patil}, {Plunkett}, {Prochaska}, {Rastogi}, {Reddy Janga},
  {Sabater}, {Sakurikar}, {Seifert}, {Sherbert}, {Sherwood-Taylor}, {Shih},
  {Sick}, {Silbiger}, {Singanamalla}, {Singer}, {Sladen}, {Sooley},
  {Sornarajah}, {Streicher}, {Teuben}, {Thomas}, {Tremblay}, {Turner},
  {Terr{\'o}n}, {van Kerkwijk}, {de la Vega}, {Watkins}, {Weaver}, {Whitmore},
  {Woillez}, {Zabalza}, \& {Astropy Contributors}}]{astropy18}
{Astropy Collaboration}, {Price-Whelan}, A.~M., {Sip{\H{o}}cz}, B.~M., {et~al.}
  2018, \aj, 156, 123, \dodoi{10.3847/1538-3881/aabc4f}

\bibitem[{{Astropy Collaboration} {et~al.}(2022){Astropy Collaboration},
  {Price-Whelan}, {Lim}, {Earl}, {Starkman}, {Bradley}, {Shupe}, {Patil},
  {Corrales}, {Brasseur}, {N{\"o}the}, {Donath}, {Tollerud}, {Morris},
  {Ginsburg}, {Vaher}, {Weaver}, {Tocknell}, {Jamieson}, {van Kerkwijk},
  {Robitaille}, {Merry}, {Bachetti}, {G{\"u}nther}, {Aldcroft},
  {Alvarado-Montes}, {Archibald}, {B{\'o}di}, {Bapat}, {Barentsen},
  {Baz{\'a}n}, {Biswas}, {Boquien}, {Burke}, {Cara}, {Cara}, {Conroy},
  {Conseil}, {Craig}, {Cross}, {Cruz}, {D'Eugenio}, {Dencheva}, {Devillepoix},
  {Dietrich}, {Eigenbrot}, {Erben}, {Ferreira}, {Foreman-Mackey}, {Fox},
  {Freij}, {Garg}, {Geda}, {Glattly}, {Gondhalekar}, {Gordon}, {Grant},
  {Greenfield}, {Groener}, {Guest}, {Gurovich}, {Handberg}, {Hart},
  {Hatfield-Dodds}, {Homeier}, {Hosseinzadeh}, {Jenness}, {Jones}, {Joseph},
  {Kalmbach}, {Karamehmetoglu}, {Ka{\l}uszy{\'n}ski}, {Kelley}, {Kern},
  {Kerzendorf}, {Koch}, {Kulumani}, {Lee}, {Ly}, {Ma}, {MacBride}, {Maljaars},
  {Muna}, {Murphy}, {Norman}, {O'Steen}, {Oman}, {Pacifici}, {Pascual},
  {Pascual-Granado}, {Patil}, {Perren}, {Pickering}, {Rastogi}, {Roulston},
  {Ryan}, {Rykoff}, {Sabater}, {Sakurikar}, {Salgado}, {Sanghi}, {Saunders},
  {Savchenko}, {Schwardt}, {Seifert-Eckert}, {Shih}, {Jain}, {Shukla}, {Sick},
  {Simpson}, {Singanamalla}, {Singer}, {Singhal}, {Sinha}, {Sip{\H{o}}cz},
  {Spitler}, {Stansby}, {Streicher}, {{\v{S}}umak}, {Swinbank}, {Taranu},
  {Tewary}, {Tremblay}, {de Val-Borro}, {Van Kooten}, {Vasovi{\'c}}, {Verma},
  {de Miranda Cardoso}, {Williams}, {Wilson}, {Winkel}, {Wood-Vasey}, {Xue},
  {Yoachim}, {Zhang}, {Zonca}, \& {Astropy Project Contributors}}]{astropy22}
{Astropy Collaboration}, {Price-Whelan}, A.~M., {Lim}, P.~L., {et~al.} 2022,
  \apj, 935, 167, \dodoi{10.3847/1538-4357/ac7c74}

\bibitem[{{Bakes} \& {Tielens}(1994)}]{bakes94}
{Bakes}, E.~L.~O., \& {Tielens}, A.~G.~G.~M. 1994, \apj, 427, 822,
  \dodoi{10.1086/174188}

\bibitem[{{Baldwin} {et~al.}(1981){Baldwin}, {Phillips}, \&
  {Terlevich}}]{baldwin81}
{Baldwin}, J.~A., {Phillips}, M.~M., \& {Terlevich}, R. 1981, \pasp, 93, 5,
  \dodoi{10.1086/130766}

\bibitem[{{Baron} \& {Netzer}(2019)}]{baron19b}
{Baron}, D., \& {Netzer}, H. 2019, \mnras, 486, 4290,
  \dodoi{10.1093/mnras/stz1070}

\bibitem[{{Baron} {et~al.}(2024){Baron}, {Sandstrom}, {Rosolowsky}, {Egorov},
  {Klessen}, {Leroy}, {Boquien}, {Schinnerer}, {Belfiore}, {Groves},
  {Chastenet}, {Dale}, {Blanc}, {M{\'e}ndez-Delgado}, {Koch}, {Grasha},
  {Chevance}, {Thilker}, {Colombo}, {Williams}, {Pathak}, {Sutter}, {Brown},
  {Wu}, {Peek}, {Emsellem}, {Larson}, \& {Neumann}}]{baron24}
{Baron}, D., {Sandstrom}, K.~M., {Rosolowsky}, E., {et~al.} 2024, \apj, 968,
  24, \dodoi{10.3847/1538-4357/ad39e5}

\bibitem[{{Belfiore} {et~al.}(2022){Belfiore}, {Santoro}, {Groves},
  {Schinnerer}, {Kreckel}, {Glover}, {Klessen}, {Emsellem}, {Blanc}, {Congiu},
  {Barnes}, {Boquien}, {Chevance}, {Dale}, {Kruijssen}, {Leroy}, {Pan},
  {Pessa}, {Schruba}, \& {Williams}}]{belfiore22}
{Belfiore}, F., {Santoro}, F., {Groves}, B., {et~al.} 2022, \aap, 659, A26,
  \dodoi{10.1051/0004-6361/202141859}

\bibitem[{{Belfiore} {et~al.}(2023){Belfiore}, {Leroy}, {Williams}, {Barnes},
  {Bigiel}, {Boquien}, {Cao}, {Chastenet}, {Congiu}, {Dale}, {Egorov},
  {Eibensteiner}, {Emsellem}, {Glover}, {Groves}, {Hassani}, {Klessen},
  {Kreckel}, {Neumann}, {Neumann}, {Querejeta}, {Rosolowsky},
  {Sanchez-Blazquez}, {Sandstrom}, {Schinnerer}, {Sun}, {Sutter}, \&
  {Watkins}}]{belfiore23}
{Belfiore}, F., {Leroy}, A.~K., {Williams}, T.~G., {et~al.} 2023, \aap, 678,
  A129, \dodoi{10.1051/0004-6361/202347175}

\bibitem[{{Blanc} {et~al.}(2015){Blanc}, {Kewley}, {Vogt}, \&
  {Dopita}}]{blanc15}
{Blanc}, G.~A., {Kewley}, L., {Vogt}, F. P.~A., \& {Dopita}, M.~A. 2015, \apj,
  798, 99, \dodoi{10.1088/0004-637X/798/2/99}

\bibitem[{{Boersma} {et~al.}(2016){Boersma}, {Bregman}, \&
  {Allamandola}}]{boersma16}
{Boersma}, C., {Bregman}, J., \& {Allamandola}, L.~J. 2016, \apj, 832, 51,
  \dodoi{10.3847/0004-637X/832/1/51}

\bibitem[{{Boersma} {et~al.}(2018){Boersma}, {Bregman}, \&
  {Allamandola}}]{boersma18}
---. 2018, \apj, 858, 67, \dodoi{10.3847/1538-4357/aabcbe}

\bibitem[{{Boquien} {et~al.}(2019){Boquien}, {Burgarella}, {Roehlly}, {Buat},
  {Ciesla}, {Corre}, {Inoue}, \& {Salas}}]{boquien19}
{Boquien}, M., {Burgarella}, D., {Roehlly}, Y., {et~al.} 2019, \aap, 622, A103,
  \dodoi{10.1051/0004-6361/201834156}

\bibitem[{{Bressan} {et~al.}(2006){Bressan}, {Panuzzo}, {Buson}, {Clemens},
  {Granato}, {Rampazzo}, {Silva}, {Valdes}, {Vega}, \& {Danese}}]{bressan06}
{Bressan}, A., {Panuzzo}, P., {Buson}, L., {et~al.} 2006, \apjl, 639, L55,
  \dodoi{10.1086/502970}

\bibitem[{{Bruzual} \& {Charlot}(2003)}]{bruzual03}
{Bruzual}, G., \& {Charlot}, S. 2003, \mnras, 344, 1000,
  \dodoi{10.1046/j.1365-8711.2003.06897.x}

\bibitem[{{Byler} {et~al.}(2017){Byler}, {Dalcanton}, {Conroy}, \&
  {Johnson}}]{byler17}
{Byler}, N., {Dalcanton}, J.~J., {Conroy}, C., \& {Johnson}, B.~D. 2017, \apj,
  840, 44, \dodoi{10.3847/1538-4357/aa6c66}

\bibitem[{{Cardelli} {et~al.}(1989){Cardelli}, {Clayton}, \&
  {Mathis}}]{cardelli89}
{Cardelli}, J.~A., {Clayton}, G.~C., \& {Mathis}, J.~S. 1989, \apj, 345, 245,
  \dodoi{10.1086/167900}

\bibitem[{{Cecchi-Pestellini} {et~al.}(2008){Cecchi-Pestellini}, {Malloci},
  {Mulas}, {Joblin}, \& {Williams}}]{cecchi_pestellini08}
{Cecchi-Pestellini}, C., {Malloci}, G., {Mulas}, G., {Joblin}, C., \&
  {Williams}, D.~A. 2008, \aap, 486, L25, \dodoi{10.1051/0004-6361:200810015}

\bibitem[{{Chastenet} {et~al.}(2023{\natexlab{a}}){Chastenet}, {Sutter},
  {Sandstrom}, {Belfiore}, {Egorov}, {Larson}, {Leroy}, {Liu}, {Rosolowsky},
  {Thilker}, {Watkins}, {Williams}, {Barnes}, {Bigiel}, {Boquien}, {Chevance},
  {Dale}, {Kruijssen}, {Emsellem}, {Grasha}, {Groves}, {Hassani}, {Hughes},
  {Kreckel}, {Meidt}, {Pan}, {Querejeta}, {Schinnerer}, \&
  {Whitcomb}}]{chastenet23b}
{Chastenet}, J., {Sutter}, J., {Sandstrom}, K., {et~al.} 2023{\natexlab{a}},
  \apjl, 944, L12, \dodoi{10.3847/2041-8213/acac94}

\bibitem[{{Chastenet} {et~al.}(2023{\natexlab{b}}){Chastenet}, {Sutter},
  {Sandstrom}, {Belfiore}, {Egorov}, {Larson}, {Leroy}, {Liu}, {Rosolowsky},
  {Thilker}, {Watkins}, {Williams}, {Barnes}, {Bigiel}, {Boquien}, {Chevance},
  {Chiang}, {Dale}, {Kruijssen}, {Emsellem}, {Grasha}, {Groves}, {Hassani},
  {Hughes}, {Kreckel}, {Meidt}, {Rickards Vaught}, {Sardone}, \&
  {Schinnerer}}]{chastenet23a}
---. 2023{\natexlab{b}}, \apjl, 944, L11, \dodoi{10.3847/2041-8213/acadd7}

\bibitem[{{Croiset} {et~al.}(2016){Croiset}, {Candian}, {Bern{\'e}}, \&
  {Tielens}}]{croiset16}
{Croiset}, B.~A., {Candian}, A., {Bern{\'e}}, O., \& {Tielens}, A.~G.~G.~M.
  2016, \aap, 590, A26, \dodoi{10.1051/0004-6361/201527714}

\bibitem[{{Dale} {et~al.}(2023){Dale}, {Boquien}, {Barnes}, {Belfiore},
  {Bigiel}, {Cao}, {Chandar}, {Chastenet}, {Chevance}, {Deger}, {Egorov},
  {Grasha}, {Groves}, {Hassani}, {Henny}, {Klessen}, {Kreckel}, {Kruijssen},
  {Larson}, {Lee}, {Leroy}, {Liu}, {Murphy}, {Rosolowsky}, {Sandstrom},
  {Schinnerer}, {Sutter}, {Thilker}, {Watkins}, {Whitmore}, \&
  {Williams}}]{dale23}
{Dale}, D.~A., {Boquien}, M., {Barnes}, A.~T., {et~al.} 2023, \apjl, 944, L23,
  \dodoi{10.3847/2041-8213/aca769}

\bibitem[{{DeFrees} {et~al.}(1993){DeFrees}, {Miller}, {Talbi}, {Pauzat}, \&
  {Ellinger}}]{defrees93}
{DeFrees}, D.~J., {Miller}, M.~D., {Talbi}, D., {Pauzat}, F., \& {Ellinger}, Y.
  1993, \apj, 408, 530, \dodoi{10.1086/172610}

\bibitem[{{Diamond-Stanic} \& {Rieke}(2010)}]{diamond_stanic10}
{Diamond-Stanic}, A.~M., \& {Rieke}, G.~H. 2010, \apj, 724, 140,
  \dodoi{10.1088/0004-637X/724/1/140}

\bibitem[{{Donnelly} {et~al.}(2024){Donnelly}, {Smith}, {Draine}, {Togi},
  {Lai}, {Armus}, {Dale}, \& {Charmandaris}}]{donnelly24}
{Donnelly}, G.~P., {Smith}, J.~D.~T., {Draine}, B.~T., {et~al.} 2024, \apj,
  965, 75, \dodoi{10.3847/1538-4357/ad2169}

\bibitem[{{Draine}(2011)}]{draine11}
{Draine}, B.~T. 2011, {Physics of the Interstellar and Intergalactic Medium}
  (Princeton University Press)

\bibitem[{{Draine} \& {Li}(2001)}]{draine01}
{Draine}, B.~T., \& {Li}, A. 2001, \apj, 551, 807, \dodoi{10.1086/320227}

\bibitem[{{Draine} \& {Li}(2007)}]{draine07}
---. 2007, \apj, 657, 810, \dodoi{10.1086/511055}

\bibitem[{{Draine} {et~al.}(2021){Draine}, {Li}, {Hensley}, {Hunt},
  {Sandstrom}, \& {Smith}}]{draine21}
{Draine}, B.~T., {Li}, A., {Hensley}, B.~S., {et~al.} 2021, \apj, 917, 3,
  \dodoi{10.3847/1538-4357/abff51}

\bibitem[{{Draine} {et~al.}(2014){Draine}, {Aniano}, {Krause}, {Groves},
  {Sandstrom}, {Braun}, {Leroy}, {Klaas}, {Linz}, {Rix}, {Schinnerer},
  {Schmiedeke}, \& {Walter}}]{draine14}
{Draine}, B.~T., {Aniano}, G., {Krause}, O., {et~al.} 2014, \apj, 780, 172,
  \dodoi{10.1088/0004-637X/780/2/172}

\bibitem[{{Egorov} {et~al.}(2023){Egorov}, {Kreckel}, {Sandstrom}, {Leroy},
  {Glover}, {Groves}, {Kruijssen}, {Barnes}, {Belfiore}, {Bigiel}, {Blanc},
  {Boquien}, {Cao}, {Chastenet}, {Chevance}, {Congiu}, {Dale}, {Emsellem},
  {Grasha}, {Klessen}, {Larson}, {Liu}, {Murphy}, {Pan}, {Pessa}, {Pety},
  {Rosolowsky}, {Scheuermann}, {Schinnerer}, {Sutter}, {Thilker}, {Watkins}, \&
  {Williams}}]{egorov23}
{Egorov}, O.~V., {Kreckel}, K., {Sandstrom}, K.~M., {et~al.} 2023, \apjl, 944,
  L16, \dodoi{10.3847/2041-8213/acac92}

\bibitem[{{Emsellem} {et~al.}(2022){Emsellem}, {Schinnerer}, {Santoro},
  {Belfiore}, {Pessa}, {McElroy}, {Blanc}, {Congiu}, {Groves}, {Ho}, {Kreckel},
  {Razza}, {Sanchez-Blazquez}, {Egorov}, {Faesi}, {Klessen}, {Leroy}, {Meidt},
  {Querejeta}, {Rosolowsky}, {Scheuermann}, {Anand}, {Barnes},
  {Be{\v{s}}li{\'c}}, {Bigiel}, {Boquien}, {Cao}, {Chevance}, {Dale},
  {Eibensteiner}, {Glover}, {Grasha}, {Henshaw}, {Hughes}, {Koch}, {Kruijssen},
  {Lee}, {Liu}, {Pan}, {Pety}, {Saito}, {Sandstrom}, {Schruba}, {Sun},
  {Thilker}, {Usero}, {Watkins}, \& {Williams}}]{emsellem22}
{Emsellem}, E., {Schinnerer}, E., {Santoro}, F., {et~al.} 2022, \aap, 659,
  A191, \dodoi{10.1051/0004-6361/202141727}

\bibitem[{{Flagey} {et~al.}(2006){Flagey}, {Boulanger}, {Verstraete}, {Miville
  Desch{\^e}nes}, {Noriega Crespo}, \& {Reach}}]{flagey06}
{Flagey}, N., {Boulanger}, F., {Verstraete}, L., {et~al.} 2006, \aap, 453, 969,
  \dodoi{10.1051/0004-6361:20053949}

\bibitem[{{Galliano} {et~al.}(2008){Galliano}, {Madden}, {Tielens}, {Peeters},
  \& {Jones}}]{galliano08}
{Galliano}, F., {Madden}, S.~C., {Tielens}, A. G.~G.~M., {Peeters}, E., \&
  {Jones}, A.~P. 2008, \apj, 679, 310, \dodoi{10.1086/587051}

\bibitem[{{Genzel} {et~al.}(1998){Genzel}, {Lutz}, {Sturm}, {Egami}, {Kunze},
  {Moorwood}, {Rigopoulou}, {Spoon}, {Sternberg}, {Tacconi-Garman}, {Tacconi},
  \& {Thatte}}]{genzel98}
{Genzel}, R., {Lutz}, D., {Sturm}, E., {et~al.} 1998, \apj, 498, 579,
  \dodoi{10.1086/305576}

\bibitem[{{Gordon} {et~al.}(2008){Gordon}, {Engelbracht}, {Rieke}, {Misselt},
  {Smith}, \& {Kennicutt}}]{gordon08}
{Gordon}, K.~D., {Engelbracht}, C.~W., {Rieke}, G.~H., {et~al.} 2008, \apj,
  682, 336, \dodoi{10.1086/589567}

\bibitem[{{Gordon} {et~al.}(2024){Gordon}, {Fitzpatrick}, {Massa}, {Bohlin},
  {Chastenet}, {Murray}, {Clayton}, {Lennon}, {Misselt}, \&
  {Sandstrom}}]{gordon24}
{Gordon}, K.~D., {Fitzpatrick}, E.~L., {Massa}, D., {et~al.} 2024, \apj, 970,
  51, \dodoi{10.3847/1538-4357/ad4be1}

\bibitem[{{Gregg} {et~al.}(2024){Gregg}, {Calzetti}, {Adamo}, {Bajaj}, {Ryon},
  {Linden}, {Correnti}, {Cignoni}, {Messa}, {Sabbi}, {Gallagher}, {Grasha},
  {Pedrini}, {Gutermuth}, {Melinder}, {Kotulla}, {P{\'e}rez}, {Krumholz},
  {Bik}, {{\"O}stlin}, {Johnson}, {Bortolini}, {Smith}, {Tosi}, {Maji}, \&
  {Faustino Vieira}}]{gregg24}
{Gregg}, B., {Calzetti}, D., {Adamo}, A., {et~al.} 2024, arXiv e-prints,
  arXiv:2405.09667, \dodoi{10.48550/arXiv.2405.09667}

\bibitem[{{Groves} {et~al.}(2012){Groves}, {Krause}, {Sandstrom}, {Schmiedeke},
  {Leroy}, {Linz}, {Kapala}, {Rix}, {Schinnerer}, {Tabatabaei}, {Walter}, \&
  {da Cunha}}]{groves12}
{Groves}, B., {Krause}, O., {Sandstrom}, K., {et~al.} 2012, \mnras, 426, 892,
  \dodoi{10.1111/j.1365-2966.2012.21696.x}

\bibitem[{{Groves} {et~al.}(2023){Groves}, {Kreckel}, {Santoro}, {Belfiore},
  {Zavodnik}, {Congiu}, {Egorov}, {Emsellem}, {Grasha}, {Leroy}, {Scheuermann},
  {Schinnerer}, {Watkins}, {Barnes}, {Bigiel}, {Dale}, {Glover}, {Pessa},
  {Sanchez-Blazquez}, \& {Williams}}]{groves23}
{Groves}, B., {Kreckel}, K., {Santoro}, F., {et~al.} 2023, \mnras, 520, 4902,
  \dodoi{10.1093/mnras/stad114}

\bibitem[{{Hirashita}(2010)}]{hirashita10}
{Hirashita}, H. 2010, \mnras, 407, L49,
  \dodoi{10.1111/j.1745-3933.2010.00902.x}

\bibitem[{{Hirashita} \& {Yan}(2009)}]{hirashita09}
{Hirashita}, H., \& {Yan}, H. 2009, \mnras, 394, 1061,
  \dodoi{10.1111/j.1365-2966.2009.14405.x}

\bibitem[{{Ho} {et~al.}(1997){Ho}, {Filippenko}, \& {Sargent}}]{ho97}
{Ho}, L.~C., {Filippenko}, A.~V., \& {Sargent}, W.~L.~M. 1997, \apjs, 112, 315

\bibitem[{{Hony} {et~al.}(2001){Hony}, {Van Kerckhoven}, {Peeters}, {Tielens},
  {Hudgins}, \& {Allamandola}}]{hony01}
{Hony}, S., {Van Kerckhoven}, C., {Peeters}, E., {et~al.} 2001, \aap, 370,
  1030, \dodoi{10.1051/0004-6361:20010242}

\bibitem[{Hunter(2007)}]{hunter07}
Hunter, J.~D. 2007, Computing In Science \& Engineering, 9, 90,
  \dodoi{10.1109/MCSE.2007.55}

\bibitem[{{Jensen} {et~al.}(2017){Jensen}, {H{\"o}nig}, {Rakshit},
  {Alonso-Herrero}, {Asmus}, {Gandhi}, {Kishimoto}, {Smette}, \&
  {Tristram}}]{jensen17}
{Jensen}, J.~J., {H{\"o}nig}, S.~F., {Rakshit}, S., {et~al.} 2017, \mnras, 470,
  3071, \dodoi{10.1093/mnras/stx1447}

\bibitem[{{Joblin} {et~al.}(1992){Joblin}, {Leger}, \& {Martin}}]{joblin92}
{Joblin}, C., {Leger}, A., \& {Martin}, P. 1992, \apjl, 393, L79,
  \dodoi{10.1086/186456}

\bibitem[{Jolliffe(2002)}]{jolliffe02}
Jolliffe, I.~T. 2002, Principal Component Analysis, Springer Series in
  Statistics (New York: Springer-Verlag), \dodoi{10.1007/b98835}.
\newblock
  \url{http://www.springer.com/statistics/statistical+theory+and+methods/book/978-0-387-95442-4}

\bibitem[{Jones {et~al.}(2001--)Jones, Oliphant, Peterson, {et~al.}}]{scipy01}
Jones, E., Oliphant, T., Peterson, P., {et~al.} 2001--, {SciPy}: Open source
  scientific tools for {Python}.
\newblock \url{http://www.scipy.org/}

\bibitem[{{Kaneda} {et~al.}(2005){Kaneda}, {Onaka}, \& {Sakon}}]{kaneda05}
{Kaneda}, H., {Onaka}, T., \& {Sakon}, I. 2005, \apjl, 632, L83,
  \dodoi{10.1086/497913}

\bibitem[{{Kaneda} {et~al.}(2008){Kaneda}, {Onaka}, {Sakon}, {Kitayama},
  {Okada}, \& {Suzuki}}]{kaneda08}
{Kaneda}, H., {Onaka}, T., {Sakon}, I., {et~al.} 2008, \apj, 684, 270,
  \dodoi{10.1086/590243}

\bibitem[{{Kauffmann} {et~al.}(2003){Kauffmann}, {Heckman}, {Tremonti},
  {Brinchmann}, {Charlot}, {White}, {Ridgway}, {Brinkmann}, {Fukugita}, {Hall},
  {Ivezi{\'c}}, {Richards}, \& {Schneider}}]{kauff03a}
{Kauffmann}, G., {Heckman}, T.~M., {Tremonti}, C., {et~al.} 2003, \mnras, 346,
  1055, \dodoi{10.1111/j.1365-2966.2003.07154.x}

\bibitem[{{Kewley} {et~al.}(2001){Kewley}, {Dopita}, {Sutherland}, {Heisler},
  \& {Trevena}}]{kewley01}
{Kewley}, L.~J., {Dopita}, M.~A., {Sutherland}, R.~S., {Heisler}, C.~A., \&
  {Trevena}, J. 2001, \apj, 556, 121, \dodoi{10.1086/321545}

\bibitem[{{Kewley} {et~al.}(2006){Kewley}, {Groves}, {Kauffmann}, \&
  {Heckman}}]{kewley06}
{Kewley}, L.~J., {Groves}, B., {Kauffmann}, G., \& {Heckman}, T. 2006, \mnras,
  372, 961, \dodoi{10.1111/j.1365-2966.2006.10859.x}

\bibitem[{{Kewley} {et~al.}(2019){Kewley}, {Nicholls}, \&
  {Sutherland}}]{kewley19}
{Kewley}, L.~J., {Nicholls}, D.~C., \& {Sutherland}, R.~S. 2019, \araa, 57,
  511, \dodoi{10.1146/annurev-astro-081817-051832}

\bibitem[{{Klessen} \& {Glover}(2016)}]{klessen16}
{Klessen}, R.~S., \& {Glover}, S. C.~O. 2016, Saas-Fee Advanced Course, 43, 85,
  \dodoi{10.1007/978-3-662-47890-5_2}

\bibitem[{{Knight} {et~al.}(2021){Knight}, {Peeters}, {Stock}, {Vacca}, \&
  {Tielens}}]{knight21}
{Knight}, C., {Peeters}, E., {Stock}, D.~J., {Vacca}, W.~D., \& {Tielens},
  A.~G.~G.~M. 2021, \apj, 918, 8, \dodoi{10.3847/1538-4357/ac02c6}

\bibitem[{{Kourkchi} \& {Tully}(2017)}]{kourkchi17}
{Kourkchi}, E., \& {Tully}, R.~B. 2017, \apj, 843, 16,
  \dodoi{10.3847/1538-4357/aa76db}

\bibitem[{{Kreckel} {et~al.}(2020){Kreckel}, {Ho}, {Blanc}, {Glover}, {Groves},
  {Rosolowsky}, {Bigiel}, {Boqu{\'\i}en}, {Chevance}, {Dale}, {Deger},
  {Emsellem}, {Grasha}, {Kim}, {Klessen}, {Kruijssen}, {Lee}, {Leroy}, {Liu},
  {McElroy}, {Meidt}, {Pessa}, {Sanchez-Blazquez}, {Sandstrom}, {Santoro},
  {Scheuermann}, {Schinnerer}, {Schruba}, {Utomo}, {Watkins}, \&
  {Williams}}]{kreckel20}
{Kreckel}, K., {Ho}, I.~T., {Blanc}, G.~A., {et~al.} 2020, \mnras, 499, 193,
  \dodoi{10.1093/mnras/staa2743}

\bibitem[{{Lai} {et~al.}(2022){Lai}, {Armus}, {U}, {D{\'\i}az-Santos},
  {Larson}, {Evans}, {Malkan}, {Appleton}, {Rich}, {M{\"u}ller-S{\'a}nchez},
  {Inami}, {Bohn}, {McKinney}, {Finnerty}, {Law}, {Linden}, {Medling},
  {Privon}, {Song}, {Stierwalt}, {van der Werf}, {Barcos-Mu{\~n}oz}, {Smith},
  {Togi}, {Aalto}, {B{\"o}ker}, {Charmandaris}, {Howell}, {Iwasawa}, {Kemper},
  {Mazzarella}, {Murphy}, {Brown}, {Hayward}, {Marshall}, {Sanders}, \&
  {Surace}}]{lai22}
{Lai}, T. S.~Y., {Armus}, L., {U}, V., {et~al.} 2022, \apjl, 941, L36,
  \dodoi{10.3847/2041-8213/ac9ebf}

\bibitem[{{Lai} {et~al.}(2023){Lai}, {Armus}, {Bianchin}, {D{\'\i}az-Santos},
  {Linden}, {Privon}, {Inami}, {U}, {Bohn}, {Evans}, {Larson}, {Hensley},
  {Smith}, {Malkan}, {Song}, {Stierwalt}, {van der Werf}, {McKinney}, {Aalto},
  {Buiten}, {Rich}, {Charmandaris}, {Appleton}, {Barcos-Mu{\~n}oz},
  {B{\"o}ker}, {Finnerty}, {Kader}, {Law}, {Medling}, {Brown}, {Hayward},
  {Howell}, {Iwasawa}, {Kemper}, {Marshall}, {Mazzarella},
  {M{\"u}ller-S{\'a}nchez}, {Murphy}, {Sanders}, \& {Surace}}]{lai23}
{Lai}, T. S.~Y., {Armus}, L., {Bianchin}, M., {et~al.} 2023, \apjl, 957, L26,
  \dodoi{10.3847/2041-8213/ad0387}

\bibitem[{{Lang} {et~al.}(2020){Lang}, {Meidt}, {Rosolowsky}, {Nofech},
  {Schinnerer}, {Leroy}, {Emsellem}, {Pessa}, {Glover}, {Groves}, {Hughes},
  {Kruijssen}, {Querejeta}, {Schruba}, {Bigiel}, {Blanc}, {Chevance},
  {Colombo}, {Faesi}, {Henshaw}, {Herrera}, {Liu}, {Pety}, {Puschnig}, {Saito},
  {Sun}, \& {Usero}}]{lang20}
{Lang}, P., {Meidt}, S.~E., {Rosolowsky}, E., {et~al.} 2020, \apj, 897, 122,
  \dodoi{10.3847/1538-4357/ab9953}

\bibitem[{{Le Page} {et~al.}(2003){Le Page}, {Snow}, \& {Bierbaum}}]{lepage03}
{Le Page}, V., {Snow}, T.~P., \& {Bierbaum}, V.~M. 2003, \apj, 584, 316,
  \dodoi{10.1086/345595}

\bibitem[{{Lebouteiller} {et~al.}(2011){Lebouteiller}, {Barry}, {Spoon},
  {Bernard-Salas}, {Sloan}, {Houck}, \& {Weedman}}]{lebouteiller11}
{Lebouteiller}, V., {Barry}, D.~J., {Spoon}, H.~W.~W., {et~al.} 2011, \apjs,
  196, 8, \dodoi{10.1088/0067-0049/196/1/8}

\bibitem[{{Lee} {et~al.}(2022){Lee}, {Whitmore}, {Thilker}, {Deger}, {Larson},
  {Ubeda}, {Anand}, {Boquien}, {Chandar}, {Dale}, {Emsellem}, {Leroy},
  {Rosolowsky}, {Schinnerer}, {Schmidt}, {Lilly}, {Turner}, {Van Dyk}, {White},
  {Barnes}, {Belfiore}, {Bigiel}, {Blanc}, {Cao}, {Chevance}, {Congiu},
  {Egorov}, {Glover}, {Grasha}, {Groves}, {Henshaw}, {Hughes}, {Klessen},
  {Koch}, {Kreckel}, {Kruijssen}, {Liu}, {Lopez}, {Mayker}, {Meidt}, {Murphy},
  {Pan}, {Pety}, {Querejeta}, {Razza}, {Saito}, {S{\'a}nchez-Bl{\'a}zquez},
  {Santoro}, {Sardone}, {Scheuermann}, {Schruba}, {Sun}, {Usero}, {Watkins}, \&
  {Williams}}]{lee22}
{Lee}, J.~C., {Whitmore}, B.~C., {Thilker}, D.~A., {et~al.} 2022, \apjs, 258,
  10, \dodoi{10.3847/1538-4365/ac1fe5}

\bibitem[{{Lee} {et~al.}(2023){Lee}, {Sandstrom}, {Leroy}, {Thilker},
  {Schinnerer}, {Rosolowsky}, {Larson}, {Egorov}, {Williams}, {Schmidt},
  {Emsellem}, {Anand}, {Barnes}, {Belfiore}, {Be{\v{s}}li{\'c}}, {Bigiel},
  {Blanc}, {Bolatto}, {Boquien}, {den Brok}, {Cao}, {Chandar}, {Chastenet},
  {Chevance}, {Chiang}, {Congiu}, {Dale}, {Deger}, {Eibensteiner}, {Faesi},
  {Glover}, {Grasha}, {Groves}, {Hassani}, {Henny}, {Henshaw}, {Hoyer},
  {Hughes}, {Jeffreson}, {Jim{\'e}nez-Donaire}, {Kim}, {Kim}, {Klessen},
  {Koch}, {Kreckel}, {Kruijssen}, {Li}, {Liu}, {Lopez}, {Maschmann}, {Chen},
  {Meidt}, {Murphy}, {Neumann}, {Neumayer}, {Pan}, {Pessa}, {Pety},
  {Querejeta}, {Pinna}, {Rodr{\'\i}guez}, {Saito}, {S{\'a}nchez-Bl{\'a}zquez},
  {Santoro}, {Sardone}, {Smith}, {Sormani}, {Scheuermann}, {Stuber}, {Sutter},
  {Sun}, {Teng}, {Tre{\ss}}, {Usero}, {Watkins}, {Whitmore}, \&
  {Razza}}]{lee23}
{Lee}, J.~C., {Sandstrom}, K.~M., {Leroy}, A.~K., {et~al.} 2023, \apjl, 944,
  L17, \dodoi{10.3847/2041-8213/acaaae}

\bibitem[{{Leger} \& {Puget}(1984)}]{leger84}
{Leger}, A., \& {Puget}, J.~L. 1984, \aap, 137, L5

\bibitem[{{Leroy} {et~al.}(2021{\natexlab{a}}){Leroy}, {Schinnerer}, {Hughes},
  {Rosolowsky}, {Pety}, {Schruba}, {Usero}, {Blanc}, {Chevance}, {Emsellem},
  {Faesi}, {Herrera}, {Liu}, {Meidt}, {Querejeta}, {Saito}, {Sandstrom}, {Sun},
  {Williams}, {Anand}, {Barnes}, {Behrens}, {Belfiore}, {Benincasa},
  {Be{\v{s}}li{\'c}}, {Bigiel}, {Bolatto}, {den Brok}, {Cao}, {Chandar},
  {Chastenet}, {Chiang}, {Congiu}, {Dale}, {Deger}, {Eibensteiner}, {Egorov},
  {Garc{\'\i}a-Rodr{\'\i}guez}, {Glover}, {Grasha}, {Henshaw}, {Ho}, {Kepley},
  {Kim}, {Klessen}, {Kreckel}, {Koch}, {Kruijssen}, {Larson}, {Lee}, {Lopez},
  {Machado}, {Mayker}, {McElroy}, {Murphy}, {Ostriker}, {Pan}, {Pessa},
  {Puschnig}, {Razza}, {S{\'a}nchez-Bl{\'a}zquez}, {Santoro}, {Sardone},
  {Scheuermann}, {Sliwa}, {Sormani}, {Stuber}, {Thilker}, {Turner}, {Utomo},
  {Watkins}, \& {Whitmore}}]{leroy21a}
{Leroy}, A.~K., {Schinnerer}, E., {Hughes}, A., {et~al.} 2021{\natexlab{a}},
  \apjs, 257, 43, \dodoi{10.3847/1538-4365/ac17f3}

\bibitem[{{Leroy} {et~al.}(2021{\natexlab{b}}){Leroy}, {Hughes}, {Liu}, {Pety},
  {Rosolowsky}, {Saito}, {Schinnerer}, {Schruba}, {Usero}, {Faesi}, {Herrera},
  {Chevance}, {Hygate}, {Kepley}, {Koch}, {Querejeta}, {Sliwa}, {Will},
  {Wilson}, {Anand}, {Barnes}, {Belfiore}, {Be{\v{s}}li{\'c}}, {Bigiel},
  {Blanc}, {Bolatto}, {Boquien}, {Cao}, {Chandar}, {Chastenet}, {Chiang},
  {Congiu}, {Dale}, {Deger}, {den Brok}, {Eibensteiner}, {Emsellem},
  {Garc{\'\i}a-Rodr{\'\i}guez}, {Glover}, {Grasha}, {Groves}, {Henshaw},
  {Jim{\'e}nez Donaire}, {Kim}, {Klessen}, {Kreckel}, {Kruijssen}, {Larson},
  {Lee}, {Mayker}, {McElroy}, {Meidt}, {Mok}, {Pan}, {Puschnig}, {Razza},
  {S{\'a}nchez-Bl'azquez}, {Sandstrom}, {Santoro}, {Sardone}, {Scheuermann},
  {Sun}, {Thilker}, {Turner}, {Ubeda}, {Utomo}, {Watkins}, \&
  {Williams}}]{leroy21b}
{Leroy}, A.~K., {Hughes}, A., {Liu}, D., {et~al.} 2021{\natexlab{b}}, \apjs,
  255, 19, \dodoi{10.3847/1538-4365/abec80}

\bibitem[{{Leroy} {et~al.}(2023){Leroy}, {Sandstrom}, {Rosolowsky}, {Belfiore},
  {Bolatto}, {Cao}, {Koch}, {Schinnerer}, {Barnes}, {Be{\v{s}}li{\'c}},
  {Bigiel}, {Blanc}, {Chastenet}, {Chen}, {Chevance}, {Chown}, {Congiu},
  {Dale}, {Egorov}, {Emsellem}, {Eibensteiner}, {Faesi}, {Glover}, {Grasha},
  {Groves}, {Hassani}, {Henshaw}, {Hughes}, {Jim{\'e}nez-Donaire}, {Kim},
  {Klessen}, {Kreckel}, {Kruijssen}, {Larson}, {Lee}, {Levy}, {Liu}, {Lopez},
  {Meidt}, {Murphy}, {Neumann}, {Pessa}, {Pety}, {Saito}, {Sardone}, {Sun},
  {Thilker}, {Usero}, {Watkins}, {Whitcomb}, \& {Williams}}]{leroy23}
{Leroy}, A.~K., {Sandstrom}, K., {Rosolowsky}, E., {et~al.} 2023, \apjl, 944,
  L9, \dodoi{10.3847/2041-8213/acaf85}

\bibitem[{{Li}(2020)}]{li20}
{Li}, A. 2020, Nature Astronomy, 4, 339, \dodoi{10.1038/s41550-020-1051-1}

\bibitem[{{Li} \& {Draine}(2001)}]{li01}
{Li}, A., \& {Draine}, B.~T. 2001, \apj, 554, 778, \dodoi{10.1086/323147}

\bibitem[{{Liang} {et~al.}(2006){Liang}, {Yin}, {Hammer}, {Deng}, {Flores}, \&
  {Zhang}}]{liang06}
{Liang}, Y.~C., {Yin}, S.~Y., {Hammer}, F., {et~al.} 2006, \apj, 652, 257,
  \dodoi{10.1086/507592}

\bibitem[{{Lutz} {et~al.}(2007){Lutz}, {Sturm}, {Tacconi}, {Valiante},
  {Schweitzer}, {Netzer}, {Maiolino}, {Andreani}, {Shemmer}, \&
  {Veilleux}}]{lutz07}
{Lutz}, D., {Sturm}, E., {Tacconi}, L.~J., {et~al.} 2007, \apjl, 661, L25,
  \dodoi{10.1086/518537}

\bibitem[{{Maragkoudakis} {et~al.}(2022){Maragkoudakis}, {Boersma}, {Temi},
  {Bregman}, \& {Allamandola}}]{maragkoudakis22}
{Maragkoudakis}, A., {Boersma}, C., {Temi}, P., {Bregman}, J.~D., \&
  {Allamandola}, L.~J. 2022, \apj, 931, 38, \dodoi{10.3847/1538-4357/ac666f}

\bibitem[{{Maragkoudakis} {et~al.}(2020){Maragkoudakis}, {Peeters}, \&
  {Ricca}}]{maragkoudakis20}
{Maragkoudakis}, A., {Peeters}, E., \& {Ricca}, A. 2020, \mnras, 494, 642,
  \dodoi{10.1093/mnras/staa681}

\bibitem[{{Marino} {et~al.}(2013){Marino}, {Rosales-Ortega}, {S{\'a}nchez},
  {Gil de Paz}, {V{\'\i}lchez}, {Miralles-Caballero}, {Kehrig},
  {P{\'e}rez-Montero}, {Stanishev}, {Iglesias-P{\'a}ramo}, {D{\'\i}az},
  {Castillo-Morales}, {Kennicutt}, {L{\'o}pez-S{\'a}nchez}, {Galbany},
  {Garc{\'\i}a-Benito}, {Mast}, {Mendez-Abreu}, {Monreal-Ibero}, {Husemann},
  {Walcher}, {Garc{\'\i}a-Lorenzo}, {Masegosa}, {Del Olmo Orozco},
  {Mour{\~a}o}, {Ziegler}, {Moll{\'a}}, {Papaderos},
  {S{\'a}nchez-Bl{\'a}zquez}, {Gonz{\'a}lez Delgado}, {Falc{\'o}n-Barroso},
  {Roth}, {van de Ven}, \& {CALIFA Team}}]{marino13}
{Marino}, R.~A., {Rosales-Ortega}, F.~F., {S{\'a}nchez}, S.~F., {et~al.} 2013,
  \aap, 559, A114, \dodoi{10.1051/0004-6361/201321956}

\bibitem[{{Micelotta} {et~al.}(2010){Micelotta}, {Jones}, \&
  {Tielens}}]{micelotta10}
{Micelotta}, E.~R., {Jones}, A.~P., \& {Tielens}, A.~G.~G.~M. 2010, \aap, 510,
  A36, \dodoi{10.1051/0004-6361/200911682}

\bibitem[{{Mulas} {et~al.}(2013){Mulas}, {Zonca}, {Casu}, \&
  {Cecchi-Pestellini}}]{mulas13}
{Mulas}, G., {Zonca}, A., {Casu}, S., \& {Cecchi-Pestellini}, C. 2013, \apjs,
  207, 7, \dodoi{10.1088/0067-0049/207/1/7}

\bibitem[{{O'Halloran} {et~al.}(2006){O'Halloran}, {Satyapal}, \&
  {Dudik}}]{ohalloran06}
{O'Halloran}, B., {Satyapal}, S., \& {Dudik}, R.~P. 2006, \apj, 641, 795,
  \dodoi{10.1086/500529}

\bibitem[{{Omont}(1986)}]{omont86}
{Omont}, A. 1986, \aap, 164, 159

\bibitem[{Pedregosa {et~al.}(2011)Pedregosa, Varoquaux, Gramfort, Michel,
  Thirion, Grisel, Blondel, Prettenhofer, Weiss, Dubourg, Vanderplas, Passos,
  Cournapeau, Brucher, Perrot, \& Duchesnay}]{pedregosa11}
Pedregosa, F., Varoquaux, G., Gramfort, A., {et~al.} 2011, Journal of Machine
  Learning Research, 12, 2825

\bibitem[{{Peeters} {et~al.}(2017){Peeters}, {Bauschlicher}, {Allamandola},
  {Tielens}, {Ricca}, \& {Wolfire}}]{peeters17}
{Peeters}, E., {Bauschlicher}, Charles~W., J., {Allamandola}, L.~J., {et~al.}
  2017, \apj, 836, 198, \dodoi{10.3847/1538-4357/836/2/198}

\bibitem[{P\'erez \& Granger(2007)}]{perez07}
P\'erez, F., \& Granger, B.~E. 2007, Computing in Science and Engineering, 9,
  21, \dodoi{10.1109/MCSE.2007.53}

\bibitem[{{P{\'e}rez-Montero} \& {Contini}(2009)}]{perez-montero09}
{P{\'e}rez-Montero}, E., \& {Contini}, T. 2009, \mnras, 398, 949,
  \dodoi{10.1111/j.1365-2966.2009.15145.x}

\bibitem[{{Pessa} {et~al.}(2023){Pessa}, {Schinnerer}, {Sanchez-Blazquez},
  {Belfiore}, {Groves}, {Emsellem}, {Neumann}, {Leroy}, {Bigiel}, {Chevance},
  {Dale}, {Glover}, {Grasha}, {Klessen}, {Kreckel}, {Kruijssen}, {Pinna},
  {Querejeta}, {Rosolowsky}, \& {Williams}}]{pessa23}
{Pessa}, I., {Schinnerer}, E., {Sanchez-Blazquez}, P., {et~al.} 2023, \aap,
  673, A147, \dodoi{10.1051/0004-6361/202245673}

\bibitem[{{Pilyugin} {et~al.}(2004){Pilyugin}, {V{\'\i}lchez}, \&
  {Contini}}]{pilyugin04}
{Pilyugin}, L.~S., {V{\'\i}lchez}, J.~M., \& {Contini}, T. 2004, \aap, 425,
  849, \dodoi{10.1051/0004-6361:20034522}

\bibitem[{{Pope} {et~al.}(2008){Pope}, {Chary}, {Alexander}, {Armus},
  {Dickinson}, {Elbaz}, {Frayer}, {Scott}, \& {Teplitz}}]{pope08}
{Pope}, A., {Chary}, R.-R., {Alexander}, D.~M., {et~al.} 2008, \apj, 675, 1171,
  \dodoi{10.1086/527030}

\bibitem[{{Querejeta} {et~al.}(2021){Querejeta}, {Schinnerer}, {Meidt}, {Sun},
  {Leroy}, {Emsellem}, {Klessen}, {Mu{\~n}oz-Mateos}, {Salo}, {Laurikainen},
  {Be{\v{s}}li{\'c}}, {Blanc}, {Chevance}, {Dale}, {Eibensteiner}, {Faesi},
  {Garc{\'\i}a-Rodr{\'\i}guez}, {Glover}, {Grasha}, {Henshaw}, {Herrera},
  {Hughes}, {Kreckel}, {Kruijssen}, {Liu}, {Murphy}, {Pan}, {Pety}, {Razza},
  {Rosolowsky}, {Saito}, {Schruba}, {Usero}, {Watkins}, \&
  {Williams}}]{querejeta21}
{Querejeta}, M., {Schinnerer}, E., {Meidt}, S., {et~al.} 2021, \aap, 656, A133,
  \dodoi{10.1051/0004-6361/202140695}

\bibitem[{{Rampazzo} {et~al.}(2013){Rampazzo}, {Panuzzo}, {Vega}, {Marino},
  {Bressan}, \& {Clemens}}]{rampazzo13}
{Rampazzo}, R., {Panuzzo}, P., {Vega}, O., {et~al.} 2013, \mnras, 432, 374,
  \dodoi{10.1093/mnras/stt475}

\bibitem[{{Rigopoulou} {et~al.}(2021){Rigopoulou}, {Barale}, {Clary}, {Shan},
  {Alonso-Herrero}, {Garc{\'\i}a-Bernete}, {Hunt}, {Kerkeni},
  {Pereira-Santaella}, \& {Roche}}]{rigopoulou21}
{Rigopoulou}, D., {Barale}, M., {Clary}, D.~C., {et~al.} 2021, \mnras, 504,
  5287, \dodoi{10.1093/mnras/stab959}

\bibitem[{{Rigopoulou} {et~al.}(2024){Rigopoulou}, {Donnan},
  {Garc{\'\i}a-Bernete}, {Pereira-Santaella}, {Alonso-Herrero}, {Davies},
  {Hunt}, {Roche}, \& {Shimizu}}]{rigopoulou24}
{Rigopoulou}, D., {Donnan}, F.~R., {Garc{\'\i}a-Bernete}, I., {et~al.} 2024,
  arXiv e-prints, arXiv:2406.11415, \dodoi{10.48550/arXiv.2406.11415}

\bibitem[{{Robitaille} {et~al.}(2020){Robitaille}, {Deil}, \&
  {Ginsburg}}]{reproject_2020}
{Robitaille}, T., {Deil}, C., \& {Ginsburg}, A. 2020, {reproject: Python-based
  astronomical image reprojection}, Astrophysics Source Code Library, record
  ascl:2011.023.
\newblock \doeprint{2011.023}

\bibitem[{{Sales} {et~al.}(2010){Sales}, {Pastoriza}, \& {Riffel}}]{sales10}
{Sales}, D.~A., {Pastoriza}, M.~G., \& {Riffel}, R. 2010, \apj, 725, 605,
  \dodoi{10.1088/0004-637X/725/1/605}

\bibitem[{{Sandstrom} {et~al.}(2023{\natexlab{a}}){Sandstrom}, {Chastenet},
  {Sutter}, {Leroy}, {Egorov}, {Williams}, {Bolatto}, {Boquien}, {Cao}, {Dale},
  {Lee}, {Rosolowsky}, {Schinnerer}, {Barnes}, {Belfiore}, {Bigiel},
  {Chevance}, {Grasha}, {Groves}, {Hassani}, {Hughes}, {Klessen}, {Kruijssen},
  {Larson}, {Liu}, {Lopez}, {Meidt}, {Murphy}, {Sormani}, {Thilker}, \&
  {Watkins}}]{sandstrom23a}
{Sandstrom}, K.~M., {Chastenet}, J., {Sutter}, J., {et~al.} 2023{\natexlab{a}},
  \apjl, 944, L7, \dodoi{10.3847/2041-8213/acb0cf}

\bibitem[{{Sandstrom} {et~al.}(2023{\natexlab{b}}){Sandstrom}, {Koch}, {Leroy},
  {Rosolowsky}, {Emsellem}, {Smith}, {Egorov}, {Williams}, {Larson}, {Lee},
  {Schinnerer}, {Thilker}, {Barnes}, {Belfiore}, {Bigiel}, {Blanc}, {Bolatto},
  {Boquien}, {Cao}, {Chastenet}, {Chevance}, {Chiang}, {Dale}, {Faesi},
  {Glover}, {Grasha}, {Groves}, {Hassani}, {Henshaw}, {Hughes}, {Kim},
  {Klessen}, {Kreckel}, {Kruijssen}, {Lopez}, {Liu}, {Meidt}, {Murphy}, {Pan},
  {Querejeta}, {Saito}, {Sardone}, {Sormani}, {Sutter}, {Usero}, \&
  {Watkins}}]{sandstrom23b}
{Sandstrom}, K.~M., {Koch}, E.~W., {Leroy}, A.~K., {et~al.} 2023{\natexlab{b}},
  \apjl, 944, L8, \dodoi{10.3847/2041-8213/aca972}

\bibitem[{{Santoro} {et~al.}(2022){Santoro}, {Kreckel}, {Belfiore}, {Groves},
  {Congiu}, {Thilker}, {Blanc}, {Schinnerer}, {Ho}, {Kruijssen}, {Meidt},
  {Klessen}, {Schruba}, {Querejeta}, {Pessa}, {Chevance}, {Kim}, {Emsellem},
  {McElroy}, {Barnes}, {Bigiel}, {Boquien}, {Dale}, {Glover}, {Grasha}, {Lee},
  {Leroy}, {Pan}, {Rosolowsky}, {Saito}, {Sanchez-Blazquez}, {Watkins}, \&
  {Williams}}]{santoro22}
{Santoro}, F., {Kreckel}, K., {Belfiore}, F., {et~al.} 2022, \aap, 658, A188,
  \dodoi{10.1051/0004-6361/202141907}

\bibitem[{{Sellgren} {et~al.}(1983){Sellgren}, {Werner}, \&
  {Dinerstein}}]{sellgren83}
{Sellgren}, K., {Werner}, M.~W., \& {Dinerstein}, H.~L. 1983, \apjl, 271, L13,
  \dodoi{10.1086/184083}

\bibitem[{{Shaya} {et~al.}(2017){Shaya}, {Tully}, {Hoffman}, \&
  {Pomar{\`e}de}}]{shaya17}
{Shaya}, E.~J., {Tully}, R.~B., {Hoffman}, Y., \& {Pomar{\`e}de}, D. 2017,
  \apj, 850, 207, \dodoi{10.3847/1538-4357/aa9525}

\bibitem[{{Smith} {et~al.}(2007){Smith}, {Draine}, {Dale}, {Moustakas},
  {Kennicutt}, {Helou}, {Armus}, {Roussel}, {Sheth}, {Bendo}, {Buckalew},
  {Calzetti}, {Engelbracht}, {Gordon}, {Hollenbach}, {Li}, {Malhotra},
  {Murphy}, \& {Walter}}]{smith07}
{Smith}, J.~D.~T., {Draine}, B.~T., {Dale}, D.~A., {et~al.} 2007, \apj, 656,
  770, \dodoi{10.1086/510549}

\bibitem[{{Steglich} {et~al.}(2010){Steglich}, {J{\"a}ger}, {Rouill{\'e}},
  {Huisken}, {Mutschke}, \& {Henning}}]{steglich10}
{Steglich}, M., {J{\"a}ger}, C., {Rouill{\'e}}, G., {et~al.} 2010, \apjl, 712,
  L16, \dodoi{10.1088/2041-8205/712/1/L16}

\bibitem[{{Steidel} {et~al.}(2016){Steidel}, {Strom}, {Pettini}, {Rudie},
  {Reddy}, \& {Trainor}}]{steidel16}
{Steidel}, C.~C., {Strom}, A.~L., {Pettini}, M., {et~al.} 2016, \apj, 826, 159,
  \dodoi{10.3847/0004-637X/826/2/159}

\bibitem[{{Steidel} {et~al.}(2014){Steidel}, {Rudie}, {Strom}, {Pettini},
  {Reddy}, {Shapley}, {Trainor}, {Erb}, {Turner}, {Konidaris}, {Kulas}, {Mace},
  {Matthews}, \& {McLean}}]{steidel14}
{Steidel}, C.~C., {Rudie}, G.~C., {Strom}, A.~L., {et~al.} 2014, \apj, 795,
  165, \dodoi{10.1088/0004-637X/795/2/165}

\bibitem[{{Sutter} {et~al.}(2024){Sutter}, {Sandstrom}, {Chastenet}, {Leroy},
  {Koch}, {Williams}, {Chown}, {Belfiore}, {Bigiel}, {Boquien}, {Cao},
  {Chevance}, {Dale}, {Egorov}, {Glover}, {Groves}, {Klessen}, {Kreckel},
  {Larson}, {Oakes}, {Pathak}, {Ramambason}, {Rosolowsky}, \&
  {Watkins}}]{sutter24}
{Sutter}, J., {Sandstrom}, K., {Chastenet}, J., {et~al.} 2024, arXiv e-prints,
  arXiv:2405.15102, \dodoi{10.48550/arXiv.2405.15102}

\bibitem[{{Tielens}(2008)}]{tielens08}
{Tielens}, A.~G.~G.~M. 2008, \araa, 46, 289,
  \dodoi{10.1146/annurev.astro.46.060407.145211}

\bibitem[{{Ujjwal} {et~al.}(2024){Ujjwal}, {Kartha}, {Krishna R}, {Mathew},
  {Subramanian}, {P}, \& {Thomas}}]{ujjwal24}
{Ujjwal}, K., {Kartha}, S.~S., {Krishna R}, A., {et~al.} 2024, arXiv e-prints,
  arXiv:2401.04061, \dodoi{10.48550/arXiv.2401.04061}

\bibitem[{{Vega} {et~al.}(2010){Vega}, {Bressan}, {Panuzzo}, {Rampazzo},
  {Clemens}, {Granato}, {Buson}, {Silva}, \& {Zeilinger}}]{vega10}
{Vega}, O., {Bressan}, A., {Panuzzo}, P., {et~al.} 2010, \apj, 721, 1090,
  \dodoi{10.1088/0004-637X/721/2/1090}

\bibitem[{{Veilleux} \& {Osterbrock}(1987)}]{veilleux87}
{Veilleux}, S., \& {Osterbrock}, D.~E. 1987, \apjs, 63, 295,
  \dodoi{10.1086/191166}

\bibitem[{{V{\'e}ron-Cetty} \& {V{\'e}ron}(2010)}]{veron_cetty10}
{V{\'e}ron-Cetty}, M.~P., \& {V{\'e}ron}, P. 2010, \aap, 518, A10,
  \dodoi{10.1051/0004-6361/201014188}

\bibitem[{{Vincenzo} {et~al.}(2016){Vincenzo}, {Belfiore}, {Maiolino},
  {Matteucci}, \& {Ventura}}]{vincenzo16}
{Vincenzo}, F., {Belfiore}, F., {Maiolino}, R., {Matteucci}, F., \& {Ventura},
  P. 2016, \mnras, 458, 3466, \dodoi{10.1093/mnras/stw532}

\bibitem[{{Weingartner} \& {Draine}(2001)}]{weingartner01}
{Weingartner}, J.~C., \& {Draine}, B.~T. 2001, \apjs, 134, 263,
  \dodoi{10.1086/320852}

\bibitem[{{Whitcomb} {et~al.}(2023){Whitcomb}, {Sandstrom}, \&
  {Smith}}]{whitcomb23}
{Whitcomb}, C.~M., {Sandstrom}, K., \& {Smith}, J.-D.~T. 2023, Research Notes
  of the American Astronomical Society, 7, 38, \dodoi{10.3847/2515-5172/acc073}

\bibitem[{{Whitcomb} {et~al.}(2024){Whitcomb}, {Smith}, {Sandstrom}, {Starkey},
  {Donnelly}, {Draine}, {Skillman}, {Dale}, {Armus}, {Hensley}, {Lai}, \&
  {Kennicutt}}]{whitcomb24}
{Whitcomb}, C.~M., {Smith}, J. D.~T., {Sandstrom}, K., {et~al.} 2024, arXiv
  e-prints, arXiv:2405.09685, \dodoi{10.48550/arXiv.2405.09685}

\bibitem[{{Williams} {et~al.}(2022){Williams}, {Kreckel}, {Belfiore}, {Groves},
  {Sandstrom}, {Santoro}, {Blanc}, {Bigiel}, {Boquien}, {Chevance}, {Congiu},
  {Emsellem}, {Glover}, {Grasha}, {Klessen}, {Koch}, {Kruijssen}, {Leroy},
  {Liu}, {Meidt}, {Pan}, {Querejeta}, {Rosolowsky}, {Saito},
  {S{\'a}nchez-Bl{\'a}zquez}, {Schinnerer}, {Schruba}, \&
  {Watkins}}]{williams22}
{Williams}, T.~G., {Kreckel}, K., {Belfiore}, F., {et~al.} 2022, \mnras, 509,
  1303, \dodoi{10.1093/mnras/stab3082}

\bibitem[{{Williams} {et~al.}(2024){Williams}, {Lee}, {Larson}, {Leroy},
  {Sandstrom}, {Schinnerer}, {Thilker}, {Belfiore}, {Egorov}, {Rosolowsky},
  {Sutter}, {DePasquale}, {Pagan}, {Anand}, {Barnes}, {Bigiel}, {Boquien},
  {Cao}, {Chastenet}, {Chevance}, {Chown}, {Dale}, {Eibensteiner}, {Emsellem},
  {Faesi}, {Glover}, {Grasha}, {Hannon}, {Hassani}, {Henshaw},
  {Jim{\'e}nez-Donaire}, {Kim}, {Klessen}, {Koch}, {Li}, {Liu}, {Meidt},
  {M{\'e}ndez-Delgado}, {Murphy}, {Neumann}, {Neumann}, {Neumayer}, {Oakes},
  {Pathak}, {Pety}, {Pinna}, {Querejeta}, {Ramambason}, {Romanelli}, {Sormani},
  {Stuber}, {Sun}, {Teng}, {Usero}, {Watkins}, \& {Weinbeck}}]{williams24}
{Williams}, T.~G., {Lee}, J.~C., {Larson}, K.~L., {et~al.} 2024, arXiv
  e-prints, arXiv:2401.15142, \dodoi{10.48550/arXiv.2401.15142}

\bibitem[{{Zhang} {et~al.}(2022){Zhang}, {Ho}, \& {Li}}]{zhang22}
{Zhang}, L., {Ho}, L.~C., \& {Li}, A. 2022, \apj, 939, 22,
  \dodoi{10.3847/1538-4357/ac930f}

\end{thebibliography}

\appendix

\onecolumngrid

\section{Additional information about the {\sc pca} visualization}\label{app:pca_3_and_4}
\setcounter{figure}{0}

\begin{figure*}
	\centering
\includegraphics[width=1\textwidth]{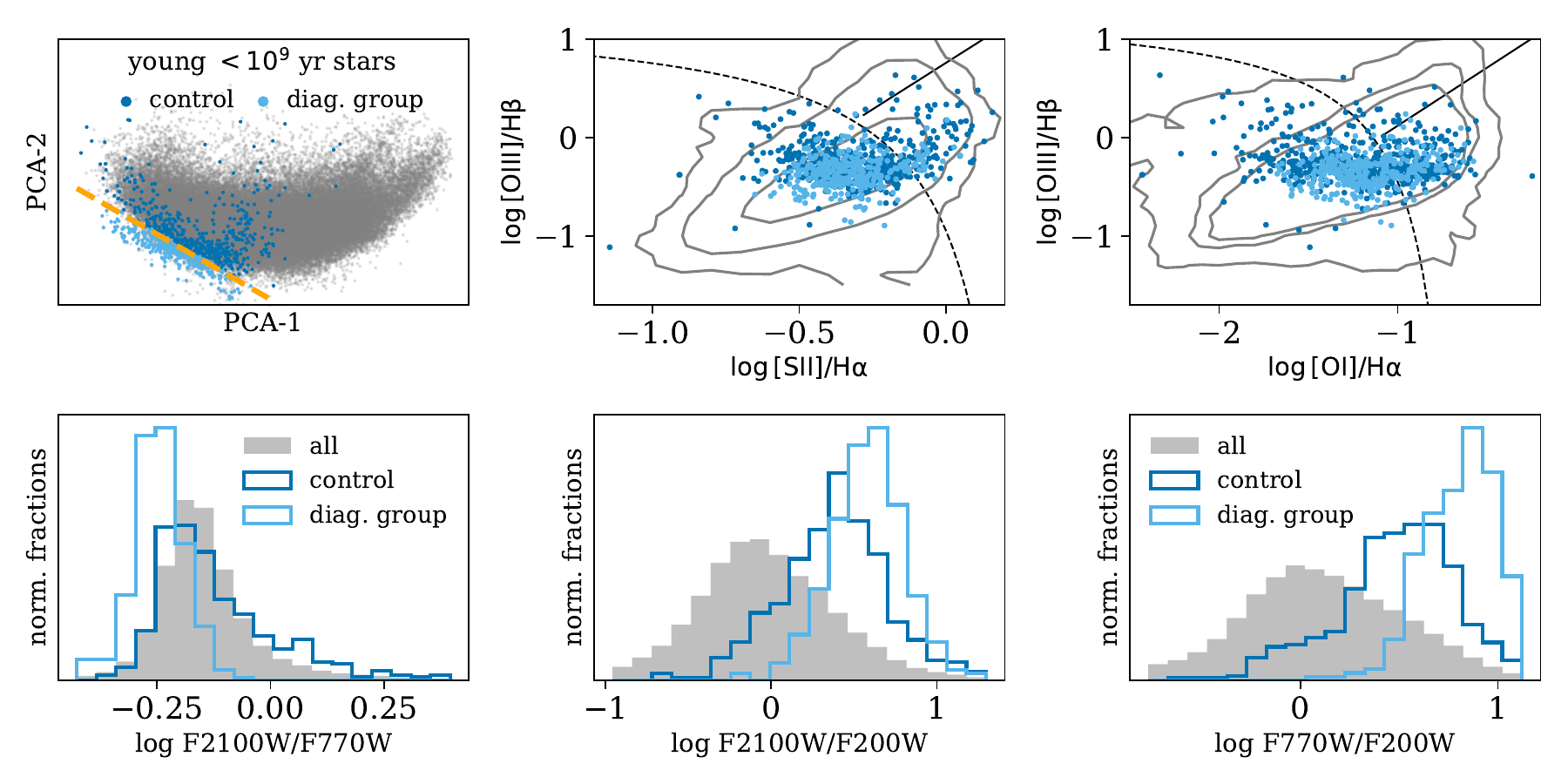}
\caption{\textbf{Observed optical and mid-infrared properties of the diagonal group that does not follow the color gradients seen in Figure \ref{f:feature_display}.} The top left panel shows the 2D distribution of the dataset in the space defined by the first two principal components (gray). Pixels that correspond to younger stellar populations are marked with blue markers, where we isolate the diagonal group as pixels that fall below the orange line. The stellar ages of the diagonal group (light blue) and of the control group (dark blue) are similar to each other. The next two panels in the top row show the distribution of pixels from the diagonal group and the control group in the \siihalpha and \oihalpha line diagnostic diagrams. The bottom panels show the distribution of all the pixels (gray), pixels from the diagonal group (light blue), and pixels from the control group (dark blue) in several JWST colors, as indicated by the labels. 
}
\label{f:diagnoal_group_additional_info}
\end{figure*}

Figure \ref{f:feature_display} in the main text reveals a group of pixels that form a diagonal stripe in the bottom left corner of the 2D distribution in the first two principal components. The group does not follow the color gradient trends seen for the rest of the pixels, especially in the optical line ratios \siihalpha and \oihalpha. Its pixels trace primarily younger stellar populations with $\log \mathrm{Age/yr} < 9$. Figure \ref{f:diagnoal_group_additional_info} presents the optical and infrared properties of this group. We identify pixels in this group (hereafter diagonal group) using their location with the principal components space (orange dashed line) and build a control group of pixels with similarly younger stellar ages $\log \mathrm{Age/yr} < 9$. Comparing the diagonal and control groups, we find comparable stellar ages and locations in the \siihalpha and \oihalpha line diagnostic diagrams. The diagonal group stands out in its JWST mid-infrared emission with respect to its stellar emission. It shows higher F2100W/F200W ratios compared to the control group and to the entire dataset. It shows significantly higher F770W/F200W ratios compared to the control group and the rest of the population. Figure \ref{f:diagnoal_group_additional_info} therefore suggests that this group is characterized by brighter dust continuum mid-infrared emission, and exceptionally bright PAH 7.7 \mic emission, with respect to the stellar emission at 2 \mic. 

\begin{figure*}
	\centering
\includegraphics[width=1\textwidth]{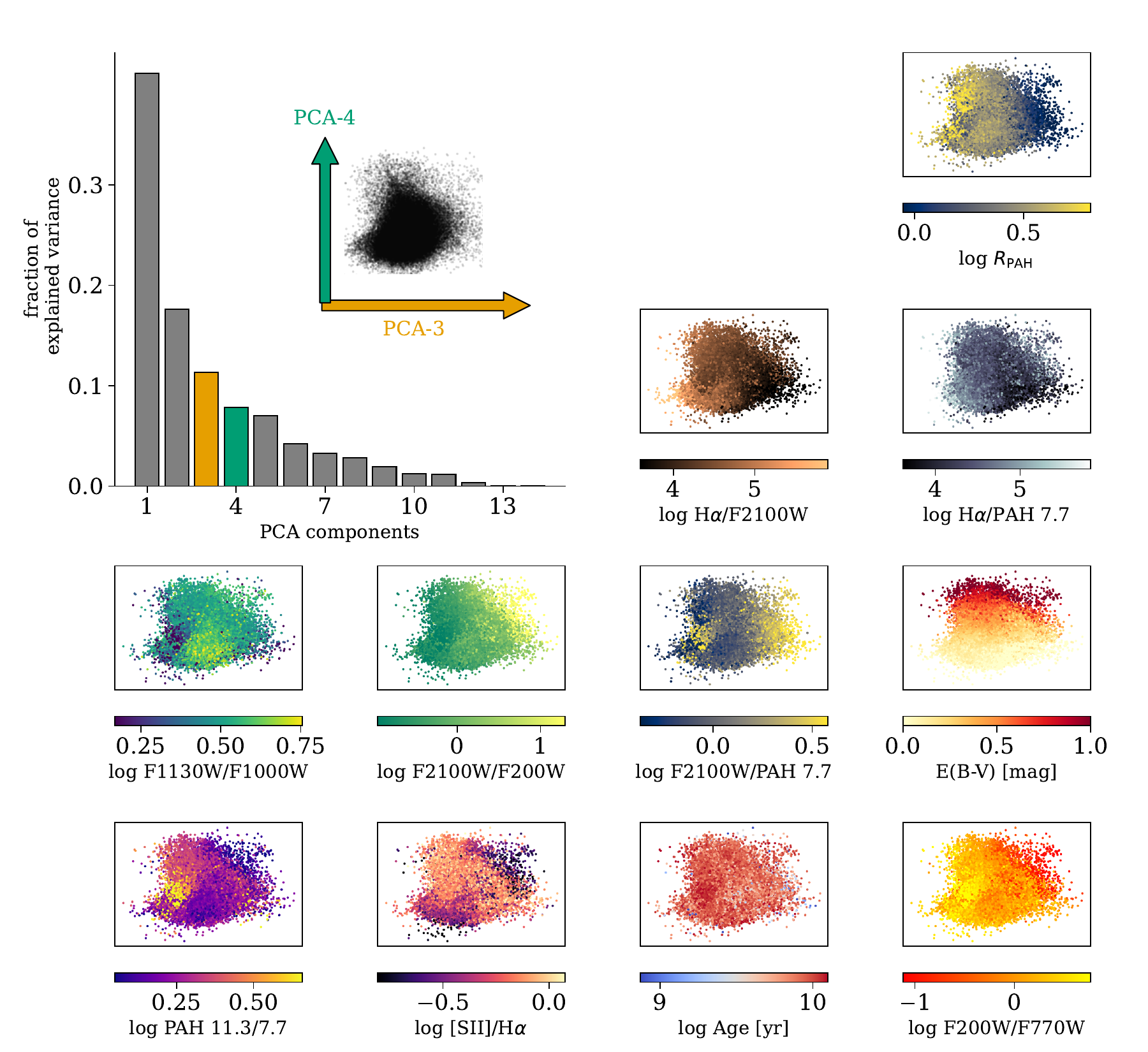}
\caption{\textbf{Distribution of the $\sim$100\,000 150 pc-scale pixels from the 19 PHANGS galaxies in the space defined by the third and fourth {\sc pca} components.} The top left panel shows the result of the {\sc pca} decomposition applied to 108\,403 spatially independent pixels that trace different optical and infrared features measured over a 150 pc scale. The bars represent the fraction of explained variance by each of the principal components. The third component accounts for 1.1\% of the total variance, while the fourth accounts for 0.78\%. The inset shows the location of the pixels in the two-dimensional plane spanned by these third and fourth components. In the rest of the panels, the distribution of the pixels in this 2D plane is color-coded by different features of interest.
 }
\label{f:feature_display_PCA3_PCA4}
\end{figure*}

Figure \ref{f:feature_display_PCA3_PCA4} shows the 2D distribution of the 150 pc-scale dataset of the 19 PHANGS galaxies in the space defined by the third and fourth principal components. They account for 1.1\% and 0.78\% of the total variance. The color gradients in F2100W/F200W and $R_{\mathrm{PAH}}$ approximately align with PCA-3. The dust reddening towards the optical line-emitting gas, $\mathrm{E}(B-V)$, aligns with PCA-4.

\section{$\log$PAH(11.3/7.7) versus \siihalpha relations in individual galaxies on a uniform 150 pc scale}\label{app:pah_gas_corr_150pc}

\begin{figure*}
\includegraphics[width=0.87\textwidth]{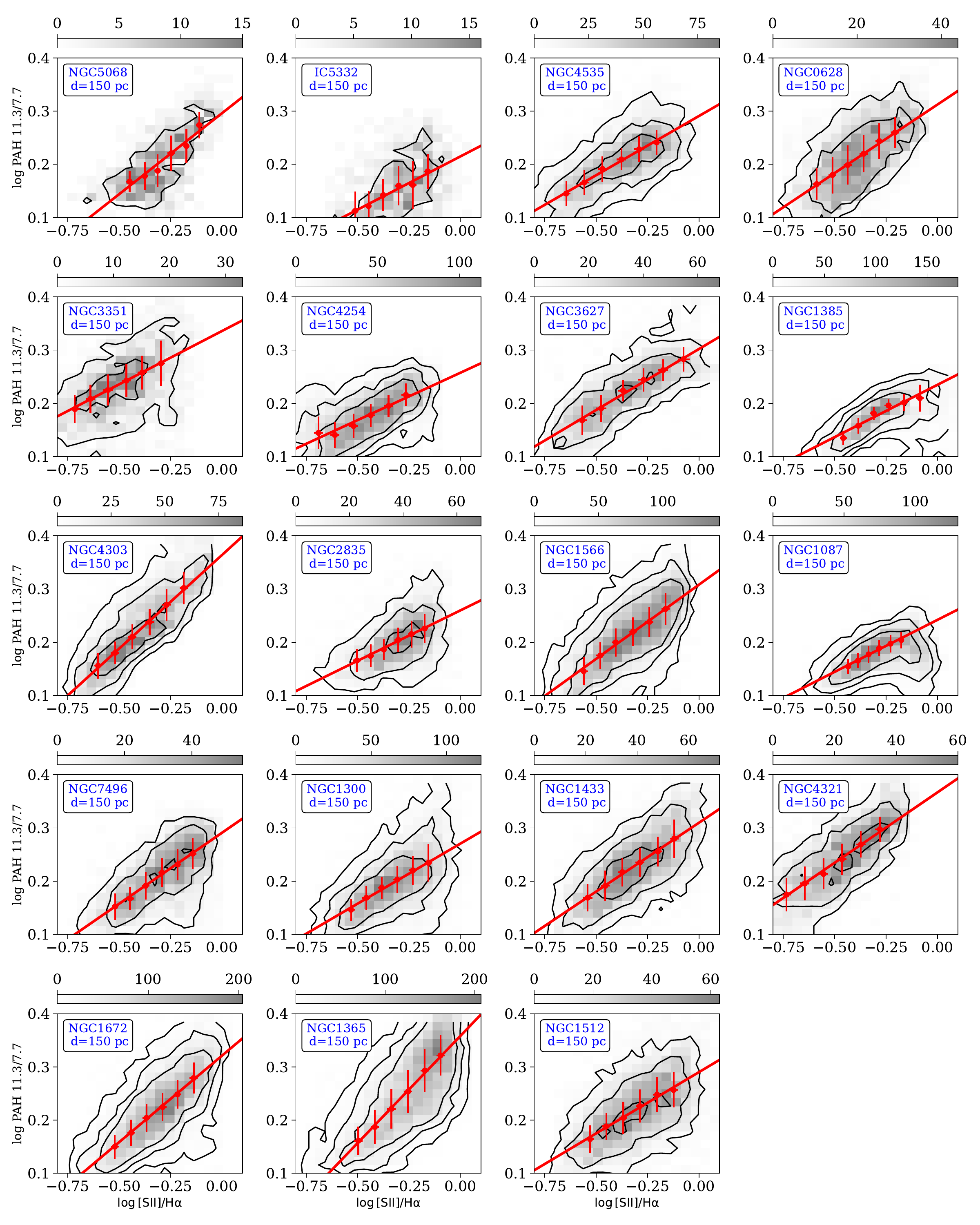}
\caption{\textbf{The $\log$PAH(11.3/7.7) versus \siihalpha relations across individual PHANGS galaxies on a uniform scale of 150 pc.} Each panel shows the 2D distribution of the $\log$PAH(11.3/7.7) band ratio versus the \siihalpha optical line ratio across a single PHANGS galaxy. The gray color-coding represent the number of pixels with the corresponding PAH the \siihalpha ratios. The black contours encompass the regions within which the counts are 5, 20, and 50. The light-blue error bars represent 6 bins in \siihalpha and their medians and median absolute deviations of the $\log$PAH(11.3/7.7) ratio in the bin. The red solid lines represent the best-fitting linear relations between $\log$PAH(11.3/7.7) and \siihalpha~.} 
\label{f:correlations_sep_150pc}
\end{figure*}

\begin{figure}
\includegraphics[width=3.5in]{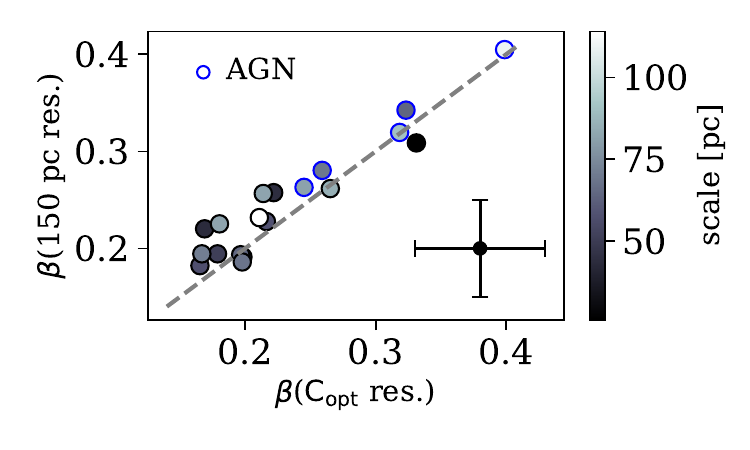}
\caption{\textbf{Comparison of the PAH-gas best-fitting slopes obtained for the $C_{\mathrm{opt}}$ resolution versus a uniform 150 pc resolution.} Each point represents a pair of best-fitting slopes of the $\log$PAH(11.3/7.7) versus \siihalpha relation for a single galaxy, once using the $C_{\mathrm{opt}}$ resolution maps and once with the 150 pc scale maps. The points are color-coded according the scale, in parsec, that the $C_{\mathrm{opt}}$ resolution probes. AGN hosts galaxies, which are systems with identified Seyfert nuclei, we marked with blue edges. The error bar in the bottom right of the panel represents the typical uncertainty of the derived slopes. The best-fitting slopes for the 150 pc resolution are comparable to those obtained for the $C_{\mathrm{opt}}$ resolution, in particular within the estimated uncertainties, though they tend to be larger than the $C_{\mathrm{opt}}$-based slopes, especially for low $\beta$ values.} 
\label{f:comparison_of_slopes_Copt_vs_150pc}
\end{figure}

\begin{figure}
\includegraphics[width=3.5in]{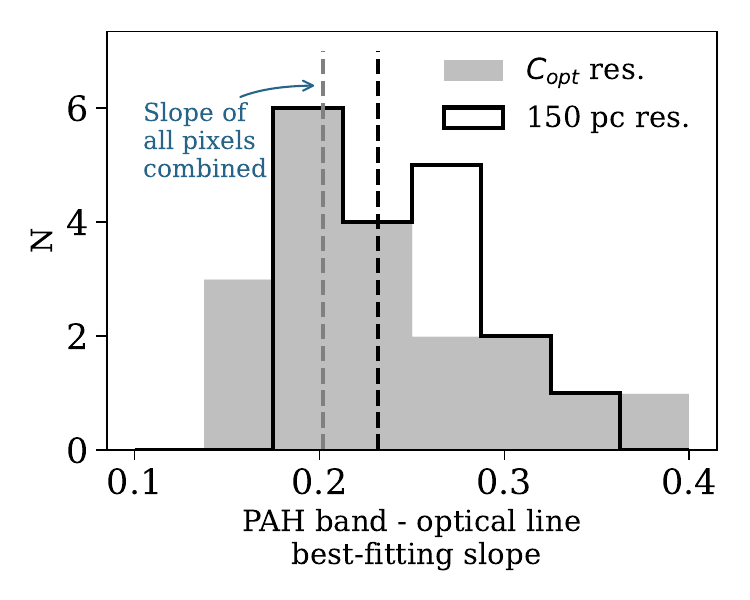}
\caption{\textbf{The distribution of best-fitting slopes obtained for the $C_{\mathrm{opt}}$ resolution versus a uniform 150 pc resolution.} The gray histogram represents the distribution of best-fitting slopes of the $\log$PAH(11.3/7.7) versus \siihalpha relation when using the $C_{\mathrm{opt}}$ maps, while the empty black histogram represents the distribution when using the 150 pc scale maps. The dashed lines represent the best-fitting slopes when combining all the pixels from all the galaxies. The distribution of best-fitting slopes on a 150 pc scale is somewhat narrower than the distribution on the smaller scales probed by $C_{\mathrm{opt}}$.} 
\label{f:beta_distribution_Copt_vs_150pc}
\end{figure}

This section summarizes the resulting best-fitting slopes of the $\log$PAH(11.3/7.7) versus \siihalpha relations on a uniform scale of 150 pc for all of the galaxies. Figure \ref{f:correlations_sep_150pc} shows the $\log$PAH(11.3/7.7) versus \siihalpha relations in individual galaxies where the PAH and optical lines ratios are derived using the 150 pc scale maps. Figure \ref{f:comparison_of_slopes_Copt_vs_150pc} compares between the best-fitting slopes derived for the 150 pc scale products and those derived at the $C_{\mathrm{opt}}$ resolution. The best-fitting slopes for the 150 pc resolution are comparable to those obtained for the $C_{\mathrm{opt}}$ resolution, in particular within the estimated uncertainties, though they tend to be larger than the $C_{\mathrm{opt}}$-based slopes, especially for low $\beta$ values. Figure \ref{f:beta_distribution_Copt_vs_150pc} compares the distribution of the best-fitting slopes for the two resolution cases. The distribution of best-fitting slopes on a 150 pc scale is somewhat narrower than the distribution on the smaller scales probed by $C_{\mathrm{opt}}$.

\section{Anomalous PAH ratios: supporting figures}\label{app:anomalous_pahs}

\begin{figure*}
	\centering
\includegraphics[width=1\textwidth]{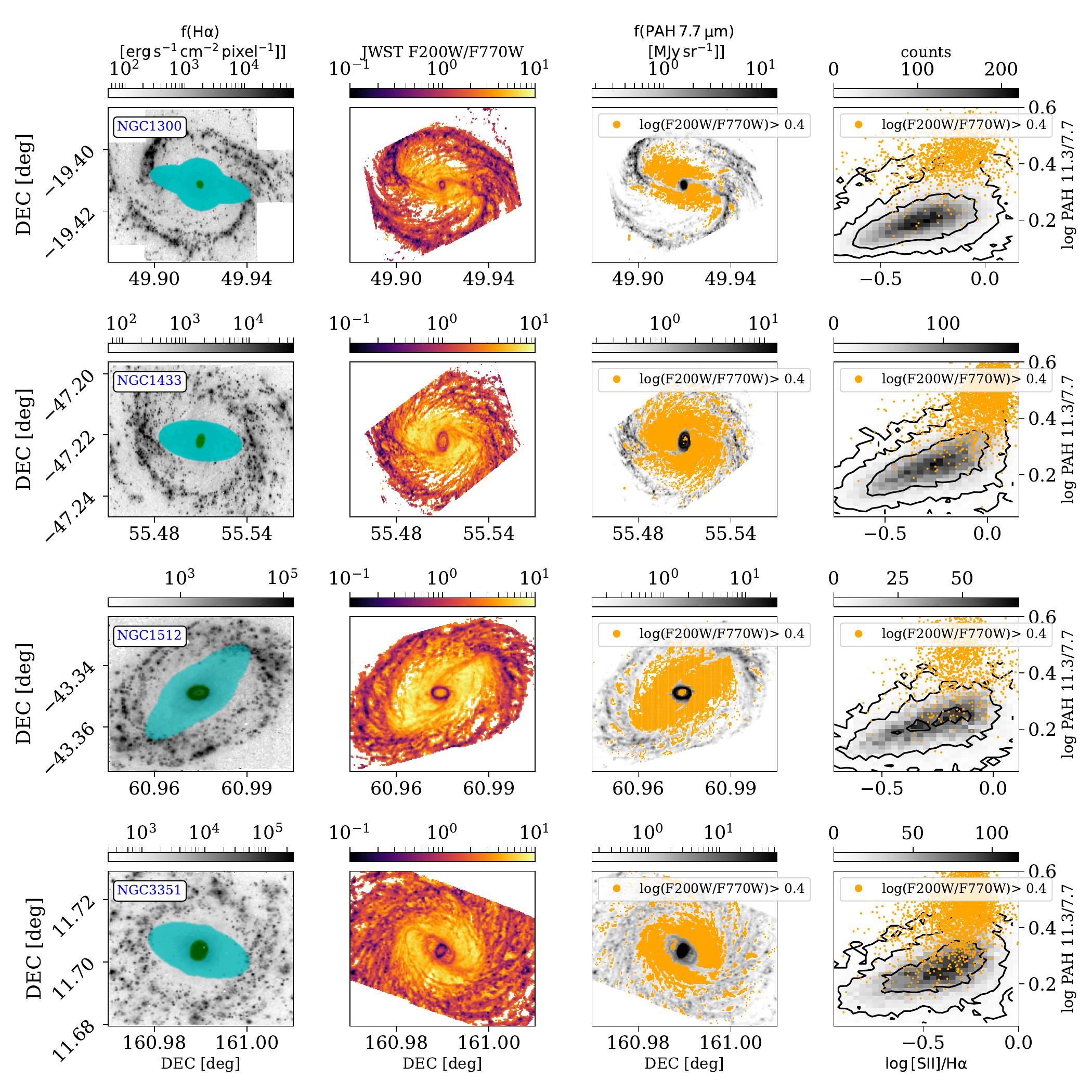}
\caption{\textbf{Anomalous $\log$PAH(11.3/7.7) regions in 4 galaxies identified using the stellar-to-mid infrared emission ratio F200W/F770W.} Each row represents a single galaxy, where we show the four galaxies where a significant number of pixels with anomalous PAH ratios have been identified. In each row, from left to right: (i) The black-white background shows the \halpha surface brightness. The cyan (green) color represents pixels that are identified as bars (centers) in the environmental maps. (2) Image showing the  F200W/F2100W ratio throughout the galaxy. (3) The black-white background shows the surface brightness of the F770W$_{\mathrm{PAH}}$ filter, and the orange points represent all pixels with $\log$(F200W/F770W)$\;>0.4$. (4) $\log$PAH(11.3/7.7) versus \siihalpha relations for pixels below the $\log$(F200W/F770W)$\;=0.4$ threshold (considered 'normal'; shown in the gray-scale color-coding and the contours), and pixels above $\log$(F200W/F770W)$\;=0.4$ threshold (orange points). We therefore the threshold $\log$(F200W/F770W)$\;>0.4$ to identify pixels that belong to the anomalous PAH group. }
\label{f:stellar_to_PAH_threshold_outlying_gals_worst}
\end{figure*}

\begin{figure*}
	\centering
\includegraphics[width=1\textwidth]{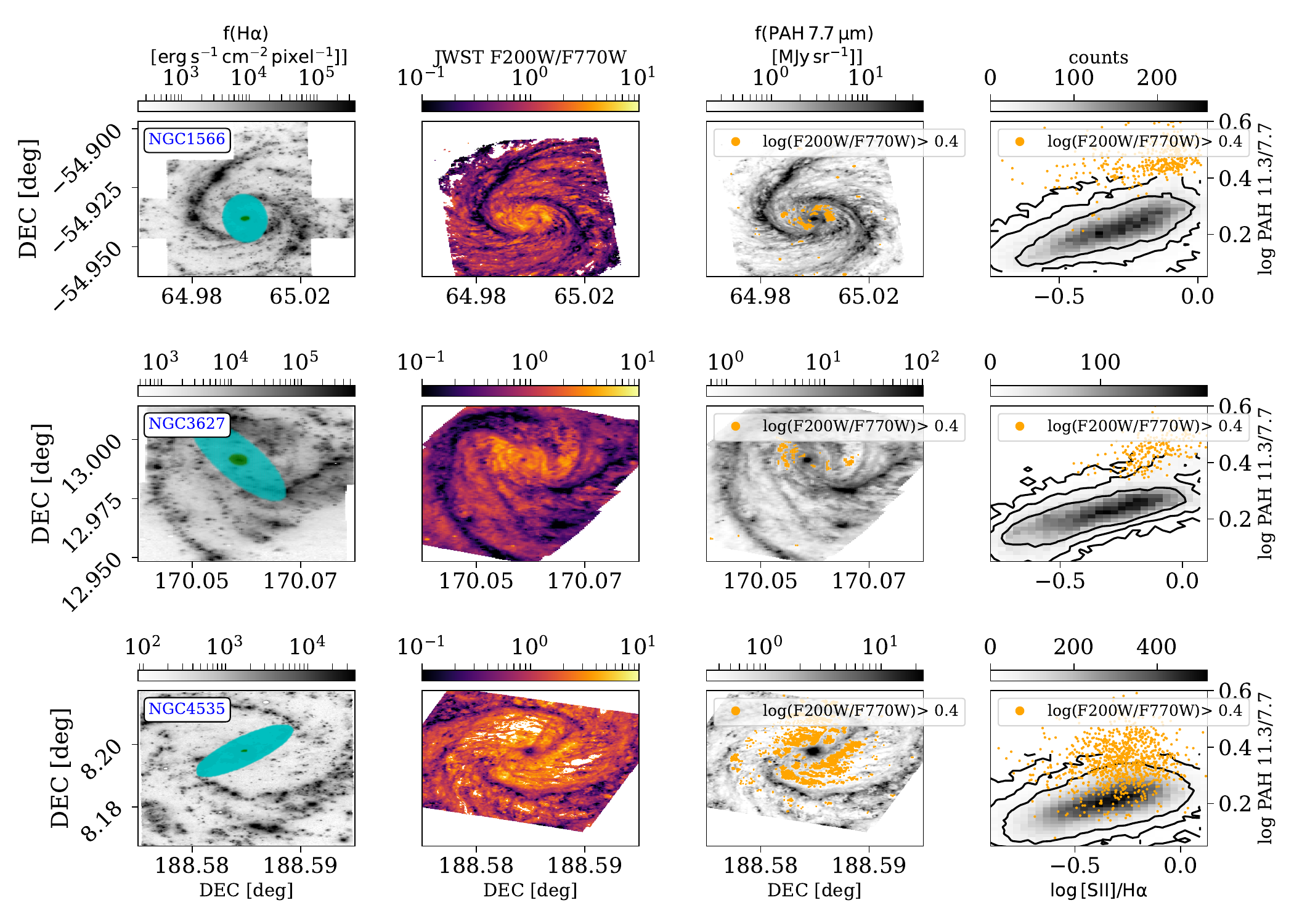}
\caption{\textbf{Anomalous $\log$PAH(11.3/7.7) regions in 3 galaxies showing more modest fractions of pixels above the $\log$(F200W/F770W)$\;>0.4$ threshold.} Similar to Figure \ref{f:stellar_to_PAH_threshold_outlying_gals_worst}.}
\label{f:stellar_to_PAH_threshold_outlying_gals_notworst}
\end{figure*}

Figure \ref{f:stellar_to_PAH_threshold_outlying_gals_worst} demonstrates our method to identify pixels with anomalous PAH ratios using the stellar-to-mid infrared emission ratio F200W/F770W. The figure focuses on the four galaxies where a significant number of pixels with anomalous PAH ratios has been identified. In Figure \ref{f:stellar_to_PAH_threshold_outlying_gals_notworst} we show the three galaxies with the more modest fraction of anomalous pixels.

\begin{figure*}
	\centering
\includegraphics[width=1\textwidth]{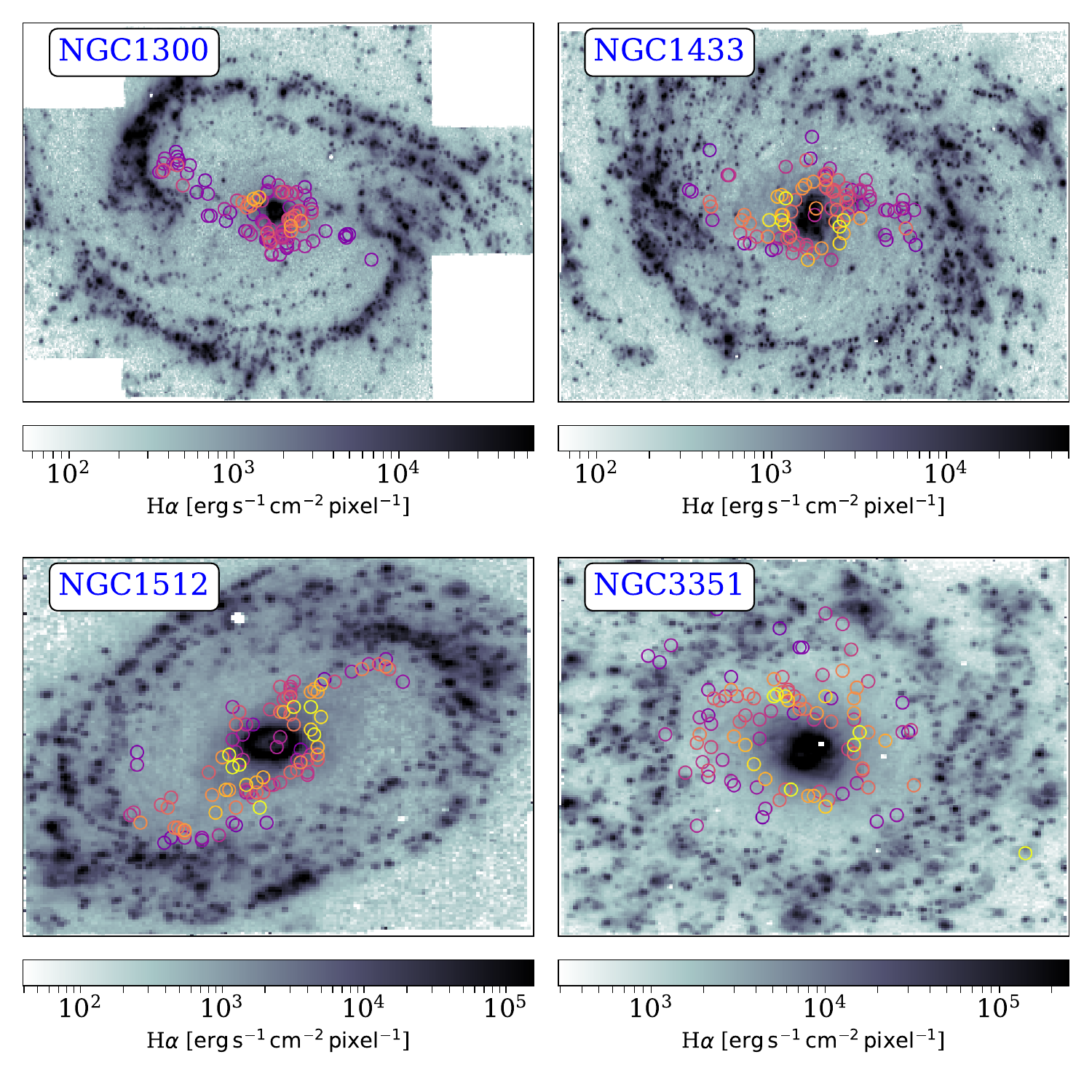}
\caption{\textbf{\halpha maps of the four galaxies that show anomalous PAH ratios.} The panels shows the \halpha surface brightness in the four galaxies showing anomalous PAH ratios. Empty circles are centered around a random subset of pixels with anomalous PAH ratios, with edge colors corresponding to those of Figure \ref{f:stacked_SEDs_weird_PAHs_fnu}. Pixels showing anomalous PAH ratios are located in regions showing very little \halpha emission.}
\label{f:Halpha_of_anomalous_PAH_ratios}
\end{figure*}

\begin{figure*}
	\centering
\includegraphics[width=1\textwidth]{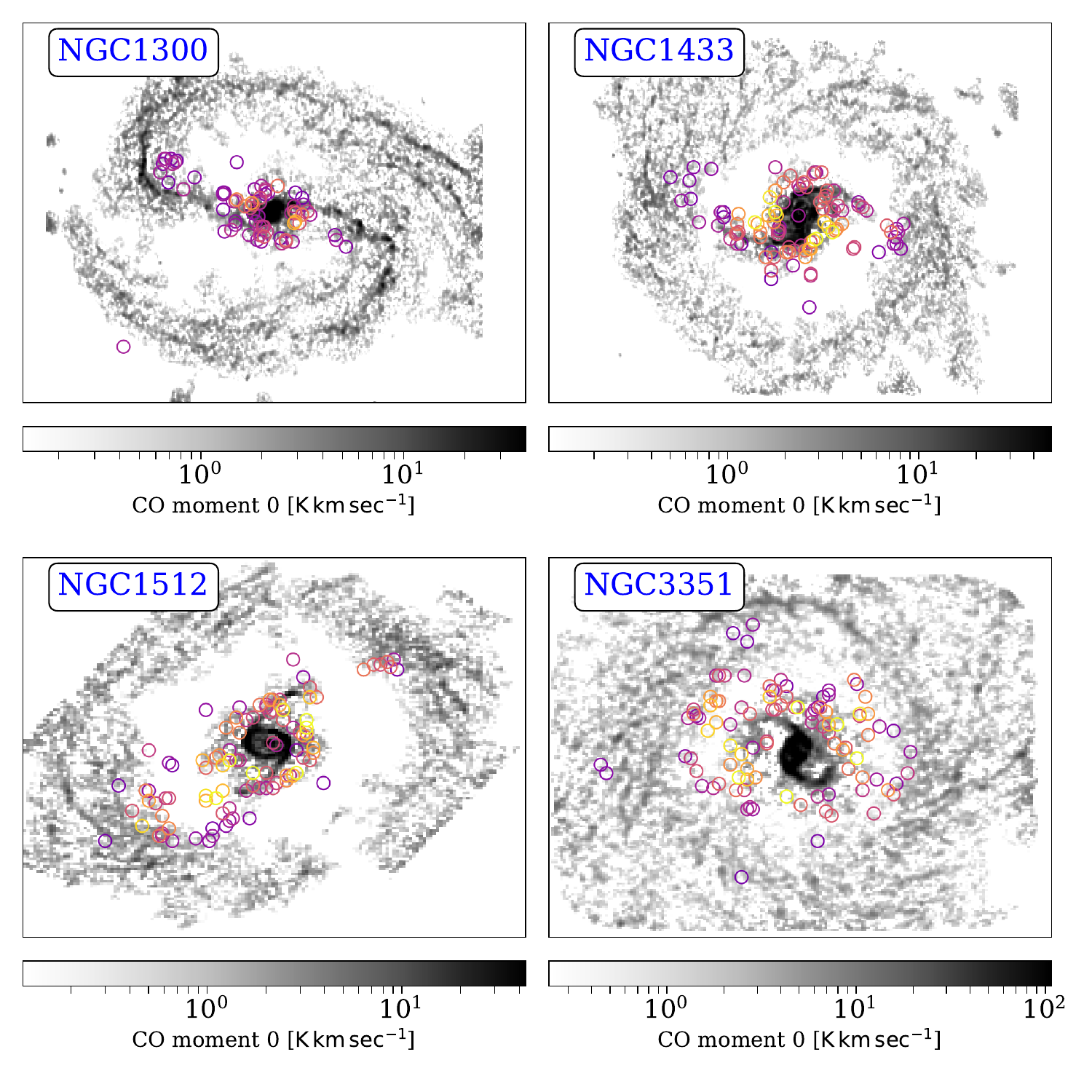}
\caption{\textbf{CO maps of the four galaxies that show anomalous PAH ratios.} The panels shows the CO moment 0 flux maps in the four galaxies showing anomalous PAH ratios. Empty circles are centered around a random subset of pixels with anomalous PAH ratios, with edge colors corresponding to those of Figure \ref{f:stacked_SEDs_weird_PAHs_fnu}. Pixels showing anomalous PAH ratios are located near the bar lanes traced by CO emission, but do not spatially coincide with them.}
\label{f:CO_of_anomalous_PAH_ratios}
\end{figure*}

In Figure \ref{f:Halpha_of_anomalous_PAH_ratios} we show the spatial distribution of a sample of anomalous pixels superimposed on the \halpha surface brightness of the galaxies. In Figure \ref{f:CO_of_anomalous_PAH_ratios} we show their distribution superimposed in the CO moment 0 maps.

To better quantify the relation between $\log$PAH(11.3/7.7) and $\log$(F200W/F2100W), we need to assess the impact of hot dust continuum contamination in the F1130W filter. The JWST broad-band observations do not allow us to disentangle between the PAH 11.3 \mic emission and the mid-infrared continuum emission by hot grains within the F1130W filter. We therefore examine two limiting cases: (1) There is no contamination of hot dust continuum to the F1130W filter, and it is dominated by PAH 11.3 \mic emission (which we implicitly assume throughout the paper), and (2) The F1000W filter is completely dominated by the hot dust continuum, with the continuum increasing linearly between 10 and 21 \mic. We consider the latter as the worst case scenario that results in the largest contamination of the F1130W filter by hot dust emission. Throughout the rest of the section, we refer to (2) as the ``aggressive hot dust subtraction''.

\begin{figure*}
	\centering
\includegraphics[width=1\textwidth]{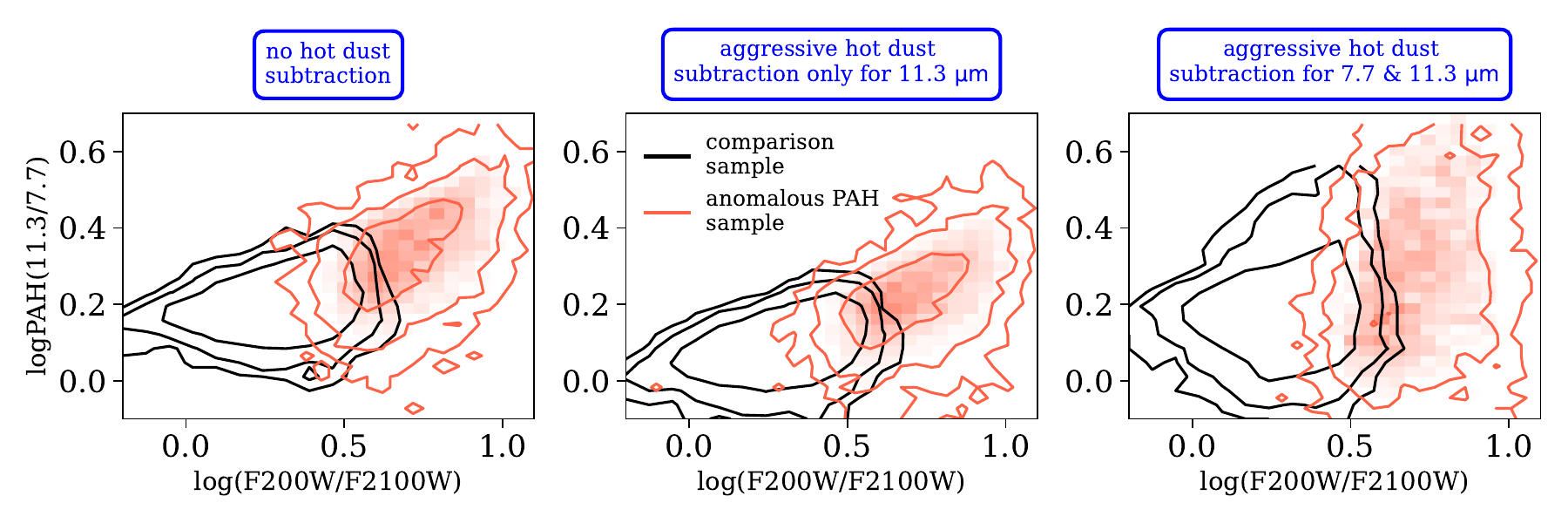}
\caption{\textbf{Relation between $\log$PAH(11.3/7.7) and the stellar-to-dust emission ratio $\log$(F200W/F2100W).} The distribution of $\log$PAH(11.3/7.7) versus $\log$(F200W/F2100W) for the comparison sample (black contours) and the anomalous PAH ratios sample (orange contours and background color). The contours encompass the regions where the counts are 2, 10, and 50. The left panel shows the relation using the standard $\log$PAH(11.3/7.7) assumed throughout the paper. The middle panel shows the relation when performing an aggressive subtraction of hot dust continuum from the F1130W filter before estimating the band ratio. The right panel shows the relation when assuming that the hot dust continuum is linear in $f_{\nu}$ between 7.7 \mic and 21 \mic, and subtracting its continuum from both F770W and F1130W.}
\label{f:log_PAH_vs_F200W_F2100W}
\end{figure*}

In Figure \ref{f:log_PAH_vs_F200W_F2100W} we show the relation between $\log$PAH(11.3/7.7) and $\log$(F200W/F2100W) for the two limiting cases. In the left panel, we show the relation using the standard $\log$PAH(11.3/7.7) assumed throughout the paper. In the middle panel, we use the F1000W and F2100W fluxes to predict the hot dust continuum flux at 11.3 \mic, assuming a linear continuum in $f_{\nu}$ units. We then estimate $\log$PAH(11.3/7.7) by subtracting from F1130W the estimated contribution from the hot dust continuum. For completeness, we also subtract the expected stellar contribution to the 11.3 \mic flux in both cases, though we find it to be negligible. We also extrapolate the linear continuum to subtract the expected hot dust contribution to the F770W filter, and show the result in the right panel. Significant correlations between $\log$PAH(11.3/7.7) and $\log$(F200W/F2100W) are observed when no hot dust continuum emission is subtracted, and when subtracting the emission from only F1130W. The correlation disappears when extrapolating the linear relation to subtract hot dust continuum from F770W. We note, however, that even in luminous infrared galaxies where the hot dust continuum emission dominates the mid-infrared, the continuum between 7.7 \mic and 21 \mic is sub-linear (e.g., \citealt{lai22}). The figure therefore illustrates the importance of obtaining spatially resolved mid-infrared spectroscopy to properly interpret the anomalous band ratios in these regions.

\section{Comparison of stacked HST+JWST SEDs in AGN and non-AGN hosts}\label{app:agn_SEDs}

\begin{figure*}
	\centering
\includegraphics[width=1\textwidth]{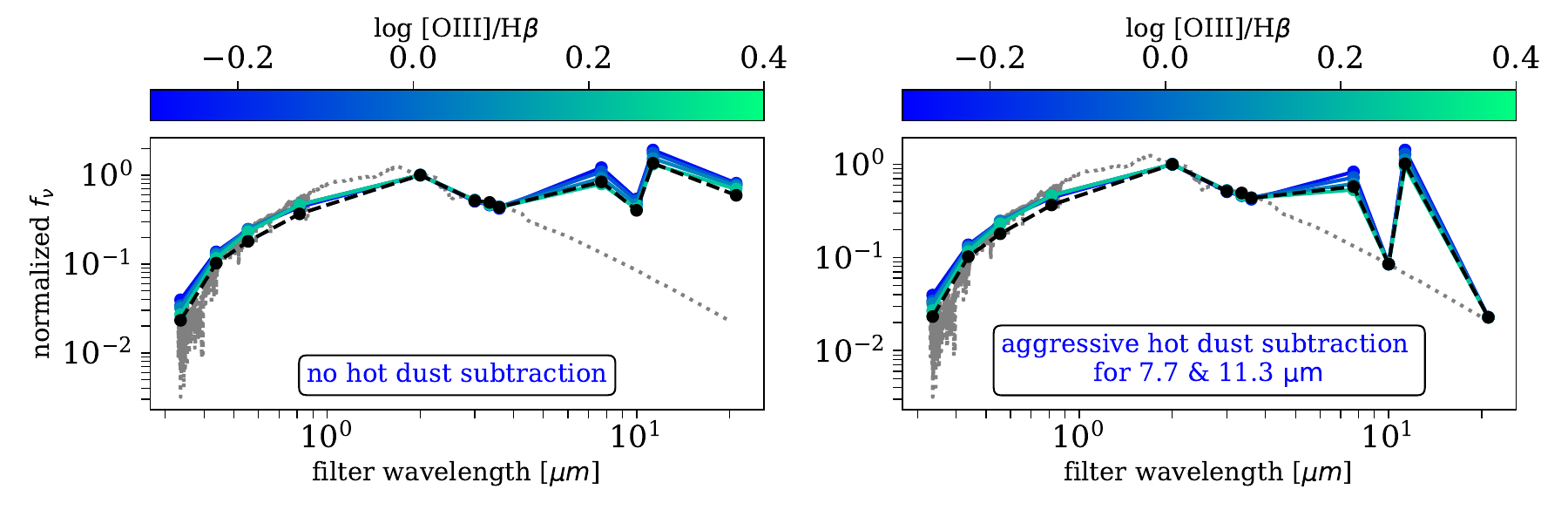}
\caption{\textbf{Stacked HST+JWST SEDs of AGN and non-AGN hosts.} The panels compare the stacked SEDs of pixels in AGN hosts (blue to green colors) to a control sample of pixels with similar \siihalpha ratios in the range [-0.2, 0] (black colors).  The gray dashed line represents a 10 Gyr SSP model by \citet{bruzual03} with a reddening of $\mathrm{E}(B-V) = 0.1\, \mathrm{mag}$. For the pixels in AGN hosts, we focus on regions with \oiiihbeta$> 0$ and consider 4 bins in \oiiihbeta, shown as different colors. The left panel shows the stacked SEDs without any hot dust continuum subtraction, while the right panel shows the result of subtracting a hot dust continuum under the assumption that the continuum is linear in $f_{\nu}$ units between 7.7 \mic and 21 \mic, and that the F1000W and F2100W filters are completely dominated by continuum, rather than PAH, emission.}
\label{f:SED_comparison_AGN_non_AGN}
\end{figure*}

Figure \ref{f:SED_comparison_AGN_non_AGN} compares the stacked HST+JWST SEDs of pixels from AGN and non-AGN groups. In both cases, we focus on pixels from the diffuse gas, with \siihalpha ratios in the range [-0.2, 0]. Pixels from regions with unusually high starlight-to-dust emission ratios are filtered out. We consider the two limiting cases examined in \ref{app:anomalous_pahs} above. (1) There is negligible contamination to the F770W and F1130W filters from hot dust continuum emission, and the ratio $\log$PAH(11.3/7.7) can be estimated in the same way as we have done throughout the paper. (2) The hot dust continuum emission is linear in $f_{\nu}$ between 7.7 \mic and 21 \mic, and dominates the filters F1000W and F2100W. In the latter case, which we call the `aggressive hot dust subtraction', the continuum is modeled with the observed F1000W and F2100W filters, and then subtracted from the F770W and F1130W filters before estimating $\log$PAH(11.3/7.7). 

First, the figure shows that the AGN and non-AGN stacks have comparable stellar continua in ultraviolet-optical wavelengths, suggesting that the increase in $\log$PAH(11.3/7.7) ratios is not due to a variation in the stellar population properties. Using the aggressive hot dust subtraction as the worst case scenario, the stacked SEDs suggest that the modest increase in $\log$PAH(11.3/7.7) observed in AGN hosts is not due to an increased contribution of hot dust continuum emission to the F1130W filter, but rather due to a small change in the PAH fluxes. Indeed, applying the aggressive hot dust continuum subtraction, we find an increase of $\sim$0.05 dex in $\log$PAH(11.3/7.7) between non-AGN and AGN hosts, and a smooth variation as a function of \oiiihbeta within the AGN group. 

\end{document}